\documentclass[preprint,12pt,nopreprintline]{elsarticle}

\usepackage{amsmath, amsfonts, amssymb, bm, dsfont}
\usepackage{algorithm}
\usepackage{algorithmic}
\usepackage{booktabs}
\usepackage{multirow}
\usepackage{threeparttable}
\usepackage{array}
\usepackage{arydshln}
\usepackage{colortbl}
\usepackage[table]{xcolor}
\usepackage{diagbox}
\usepackage{graphicx}
\usepackage{subcaption}
\usepackage{color}
\usepackage{url}
\usepackage{indentfirst}
\usepackage{textcomp}
\usepackage{stfloats}
\usepackage{verbatim}
\usepackage{txfonts}
\usepackage{microtype}
\usepackage{enumitem}
\usepackage{soul}
\usepackage{caption}
\usepackage{hyperref}
\newcommand{\PreserveBackslash}[1]{\let\temp=\\#1\let\\=\temp}
\newcolumntype{C}[1]{>{\PreserveBackslash\centering}p{#1}}
\newcolumntype{R}[1]{>{\PreserveBackslash\raggedleft}p{#1}}
\newcolumntype{L}[1]{>{\PreserveBackslash\raggedright}p{#1}}

\begin{document}
\begin{frontmatter}

\title{Deep Learning-Based Predictive Fixed-Filter Active Noise Control for Dynamic Noises}
% Predictive Fixed-Filter Active Noise Control (PFANC) Using Convolutional Recurrent Neural Networks for Dynamic Noises

\author[inst1]{Zhengding Luo}
\author[inst1]{Haowen Li}
\author[inst2]{Haozhe Ma}
\author[inst3]{Dongyuan Shi}
\author[inst3]{Wen Zhang}
\author[inst1]{Woon-Seng Gan}

\affiliation[inst1]{organization={School of Electrical and Electronic Engineering},
            addressline={Nanyang Technological University},
            country={Singapore}}
\affiliation[inst2]{organization={School of Computing, National University of Singapore},
            country={Singapore}}
\affiliation[inst3]{organization={Center of Intelligent Acoustics and Immersive Communications},
            addressline={Northwestern Polytechnical University},
            country={China}}

\begin{abstract}
The existing Generative Fixed-Filter Active Noise Control (GFANC) method generates a suitable control filter based on the current noise frame. This reactive design aims to estimate a control filter that is optimal for the present frame rather than the upcoming one. Consequently, it suffers from an inherent tracking lag and lacks the predictive capability to handle rapidly varying noises. To address this limitation, we propose the Predictive Fixed-Filter Active Noise Control (PFANC) method with a proactive control paradigm in this paper. In the PFANC method, multiple consecutive noise frames are processed by a Convolutional Recurrent Neural Network (CRNN) to predict the next-frame control filter. By utilizing temporal correlations across noise frames to anticipate the control filter in advance, the PFANC method can effectively track dynamic noise changes. Furthermore, the theoretical analysis based on a higher-order Markov formulation shows that incorporating multiple noise frames enhances the estimation of the control filter. Numerical simulations with linear and logarithmic chirp signals, as well as real-world dynamic noises, validate the effectiveness of the PFANC method and its superiority over GFANC and its variations. The PFANC method also exhibits good transferability across different acoustic paths.\footnotetext[1]{The code will be accessible at \href{https://github.com/Luo-Zhengding/Predictive-ANC}{https://github.com/Luo-Zhengding/Predictive-ANC}}
\end{abstract}

\begin{keyword}
Active Noise Control (ANC) \sep Predictive ANC \sep Proactive Control Paradigm \sep Convolutional Recurrent Neural Network \sep Dynamic Noises
\end{keyword}

\end{frontmatter}

\section{Introduction}
The growing use of industrial equipment such as engines, fans, and compressors has intensified acoustic noise problems \cite{ANC-Review,UnderstandingANC,KajikawaANC}. Prolonged exposure to excessive noise can cause both physical and psychological harm, including fatigue, loss of concentration, and stress \cite{YangJunANC,HuangGongping,Zhangxiaolei}. Traditional passive noise control methods, such as enclosures and barriers, are often bulky, costly, and ineffective at low frequencies, where the acoustic wavelength is large relative to the silencer size. Most of the noise encountered in industrial and daily life is dominated by low-frequency components, making Active Noise Control (ANC) particularly suitable \cite{CheerJordan,ElliottANC,ZhangHongwei}. ANC systems employ a secondary source to generate an anti-noise signal with equal amplitude and opposite phase to the primary noise, achieving cancellation through superposition \cite{ShiChuang,ZhangJihui}. This approach is particularly effective for low-frequency attenuation and serves as a powerful complement to traditional passive methods.

An ANC system is generally composed of three key components: microphones to acquire the primary noise and form reference signals, loudspeakers to produce the anti-noise signal, and a Digital Signal Processor (DSP) to implement the control algorithm \cite{LuJing,Marek,Denmark}. Among various ANC strategies, adaptive ANC algorithms are most widely employed due to their ability to track variations in real-world noise environments. These algorithms, such as the Filtered-x Least Mean Square (FxLMS) and its variants, iteratively update the coefficients of the control filter to minimize the error signal detected by the error microphone \cite{TaoJiancheng,ChangChengYuan-MSSP}. Due to their reliance on the feedback error signal, adaptive algorithms exhibit several limitations. Their convergence is typically slow, making it difficult to track non-stationary or dynamic noise \cite{ChrisFuller,VaryPeter,SFANC-FxNLMSLuo}. In addition, improper step-size selection may lead to instability or even divergence during adaptation, which poses challenges for robust deployment in practical ANC applications \cite{Simon,lusurvey-nonlinear}.

To improve response speed and stability, some applications have adopted fixed-filter ANC \cite{gao2026active-passive, Boxiang-SFANC}. In this approach, the control filter is pre-trained offline and then implemented in the real-time controller. Owing to their simplicity and robustness, fixed-filter ANC methods have been widely deployed in headphones, ducts, windows, and room environments \cite{FixedBhanwindow,Fixed-FilterHeadphone,miyoshifixedroom}. However, the pre-trained control filter is usually tailored to a specific noise type, resulting in degraded performance when the incoming noise differs significantly from the training noise \cite{Kajikawa-SFANC,Xiruo-SFANC,FPGA-SFANC,SFANC-Yangjun}. This limitation motivates the exploration of advanced fixed-filter ANC techniques that can handle diverse noise types. Leveraging the strong representation capability of deep learning, the Selective Fixed-Filter ANC (SFANC) method \cite{DYSFANCCNN,KangJian2025subjective} was proposed to reduce a wide range of noises.

The SFANC method utilizes a Convolutional Neural Network (CNN) to select appropriate pre-trained control filters for different primary noises. However, the limited number of pre-trained control filters restricts its noise reduction performance, especially when the incoming noise deviates significantly from the filter-training noises. To obtain more appropriate and comprehensive control filters, the Generative Fixed-Filter ANC (GFANC) method \cite{luoGFANCMssp} was proposed. It generates the control filter through a weighted combination of sub control filters, with the weight vector automatically provided by a CNN conditioned on the input noise. Despite its flexibility, the GFANC method relies only on the current noise frame to generate the control filter. To incorporate additional temporal information, two smoothing-based extensions of GFANC, namely GFANC with CNN-Bayesian Filtering (GFANC-Bayes) and GFANC with CNN-Kalman Filtering (GFANC-Kalman), were proposed \cite{LuoBayes,LuoKalman}. They smooth the newly generated control filter with that of the previous frame, thus exploiting the temporal correlations between adjacent noise frames.

The GFANC-Bayes and GFANC-Kalman methods utilize adjacent frame correlations to refine the generated control filter; however, like GFANC, they are unable to predict the underlying evolution of the noise, leading to poor tracking of rapidly varying or dynamic noises. Moreover, both the Bayesian and Kalman filtering modules require manual parameter tuning, and inappropriate parameter settings may degrade the noise control performance of GFANC-Bayes and GFANC-Kalman. In practice, rapidly varying noises exhibit frequency evolution across consecutive frames, which provides predictive cues for upcoming noise characteristics \cite{ZhangHao-ASLP,CanadaNN,DNoiseNet}. This observation highlights the potential of deep neural networks to learn such temporal context for prediction \cite{ZhangHao2021,ANCautoencoder,SeoulNN,xiang2023combined}. Motivated by this, we propose the Predictive Fixed-Filter Active Noise Control (PFANC) method in this paper, which adopts a proactive control paradigm to predict the upcoming control filter.

%------------------------------------------------------------------------
\begin{table}[!t]
\caption{Comparison between the proposed proactive PFANC and existing reactive GFANC methods.}\vspace*{-0.2cm}
\centering
\resizebox{\textwidth}{!}{
\begin{tabular}{|>{\raggedright\arraybackslash}m{2.2cm}|>{\raggedright\arraybackslash}m{2.2cm}|>{\raggedright\arraybackslash}m{6cm}|>{\raggedright\arraybackslash}m{1.9cm}|>{\raggedright\arraybackslash}m{1.2cm}|}
\hline
\rowcolor{gray!20}
\textbf{Approach} & \textbf{Control Paradigm} & \textbf{Learning Target} & \textbf{Temporal Context} & \textbf{Model} \\
\hline
GFANC \cite{luoGFANCMssp} & Reactive & Current-frame optimal control filter & Single frame & CNN \\
\hline
GFANC-Bayes \cite{LuoBayes} & Reactive & Current-frame optimal control filter & Single frame & CNN \\
\hline
GFANC-Kalman \cite{LuoKalman} & Reactive & Current-frame optimal control filter & Single frame & CNN \\
\hline
\textbf{PFANC} & Proactive & Next-frame optimal control filter & Multi-frame & CRNN \\
\hline
\end{tabular}
}
\label{Table 1}
\end{table}
%----------------------------------------------------------------------

In the PFANC method, a Convolutional Recurrent Neural Network (CRNN) processes multiple consecutive noise frames to predict the next-frame weight vector, which is then used to combine the sub control filters and generate the upcoming control filter. A comparison between PFANC and representative GFANC-based ANC approaches is summarized in Table~\ref{Table 1}. Existing GFANC-based methods follow a reactive control paradigm, estimating a control filter optimized for the current noise frame. Consequently, they suffer from an inherent tracking lag and lack the predictive capability to handle rapidly varying noises. In contrast, the PFANC method adopts a proactive control paradigm that leverages multi-frame temporal context to predict the control filter for the upcoming frame. Accordingly, the next-frame optimal control filter, rather than the current-frame optimal filter, is used as the learning target. By jointly exploiting multiple historical noise frames and the current frame, the CRNN can capture the temporal evolution of the noise and anticipate its future variation. This predictive capability is particularly beneficial for tracking and controlling dynamic noises.

In addition, as a fully data-driven approach, the PFANC method avoids the manual parameter tuning required by the previous GFANC-Bayes and GFANC-Kalman methods. To further justify the proactive control paradigm, a theoretical analysis based on a higher-order Markov formulation is provided, showing that incorporating multiple noise frames improves the predictive capability of control-filter generation. Numerical simulations using linear and logarithmic chirp signals as well as real-world dynamic noises demonstrate that PFANC consistently outperforms GFANC, GFANC-Bayes, and GFANC-Kalman, while introducing only a slight increase in network parameters. Moreover, PFANC exhibits good transferability across different acoustic paths, indicating its suitability for a wide range of practical ANC scenarios.

The remainder of this paper is organized as follows. Section~\ref{Preliminaries} introduces the vanilla GFANC method and discusses its limitations. To address these challenges, Section~\ref{The PFANC Method} presents the proposed PFANC method. Section~\ref{Theoretical Analysis of Filter Prediction based on Multiple Noise Frames} provides a theoretical analysis showing why incorporating multiple noise frames improves control filter prediction. Section~\ref{Numerical Simulation} presents numerical simulations that evaluate the effectiveness of PFANC in handling different dynamic noises. Finally, Section~\ref{Conclusion} concludes the paper.

\section{Preliminaries}\label{Preliminaries}
This section first introduces the fundamentals of ANC systems, and then presents the existing GFANC method~\cite{luoGFANCMssp} along with its limitations.

\subsection{ANC Systems}\label{ANC Systems}
ANC aims to attenuate unwanted sound by generating a secondary anti-noise signal that destructively interferes with the disturbance at the error microphone. A single-channel feedforward ANC system generally consists of a reference microphone, a controller, a secondary source (e.g., loudspeaker), and an error microphone~\cite{UnderstandingANC}. The reference microphone captures the primary noise as the reference signal \(\mathbf{x}(n)\). This signal is passed through the control filter to generate the control signal $y(n)$:
%---------------------------------------------------------------------------------
\begin{equation}
y(n) = \mathbf{w}^\mathrm{T}(n) \mathbf{x}(n),
\end{equation}
%---------------------------------------------------------------------------------
where \(n\) and \(\mathrm{T}\) denote the time index and the transpose operation, respectively. The control signal $y(n)$ drives the secondary source, and the generated control sound propagates through the secondary path. The corresponding secondary-path output $y^{\prime}(n)$ at the error microphone is expressed as
%---------------------------------------------------------------------------------
\begin{equation}
\setlength{\abovedisplayskip}{2pt}
\setlength{\belowdisplayskip}{2pt}
\begin{aligned}
y^{\prime}(n) &= y(n) * s(n)\\
&= \mathbf{w}^{\mathrm{T}}(n)\mathbf{x}(n) * s(n),
\end{aligned}
\end{equation}
%---------------------------------------------------------------------------------
where $*$ denotes linear convolution, and $s(n)$ represents the impulse response of the secondary path. The anti-noise component at the error microphone is represented by $-y^{\prime}(n)$, where the negative sign denotes the phase inversion relative to the disturbance $d(n)$ required for destructive acoustic interference~\cite{ANC-Review}. The resulting residual noise captured by the error microphone, i.e., the error signal $e(n)$, is therefore given by
%---------------------------------------------------------------------------------
\begin{equation}
\setlength{\abovedisplayskip}{2pt}
\setlength{\belowdisplayskip}{2pt}
\begin{aligned}
e(n) &= d(n) - y^{\prime}(n)\\
&= d(n) - \mathbf{w}^{\mathrm{T}}(n)\mathbf{x}(n) * s(n).
\end{aligned}
\end{equation}
%---------------------------------------------------------------------------------
Under ideal cancellation, $y^{\prime}(n)$ approximates $d(n)$, resulting in $e(n)\approx 0$.

%---------------------------------------------------------
\begin{figure}[!t]
\centering
\includegraphics[width=0.38\linewidth, height=3.2cm]{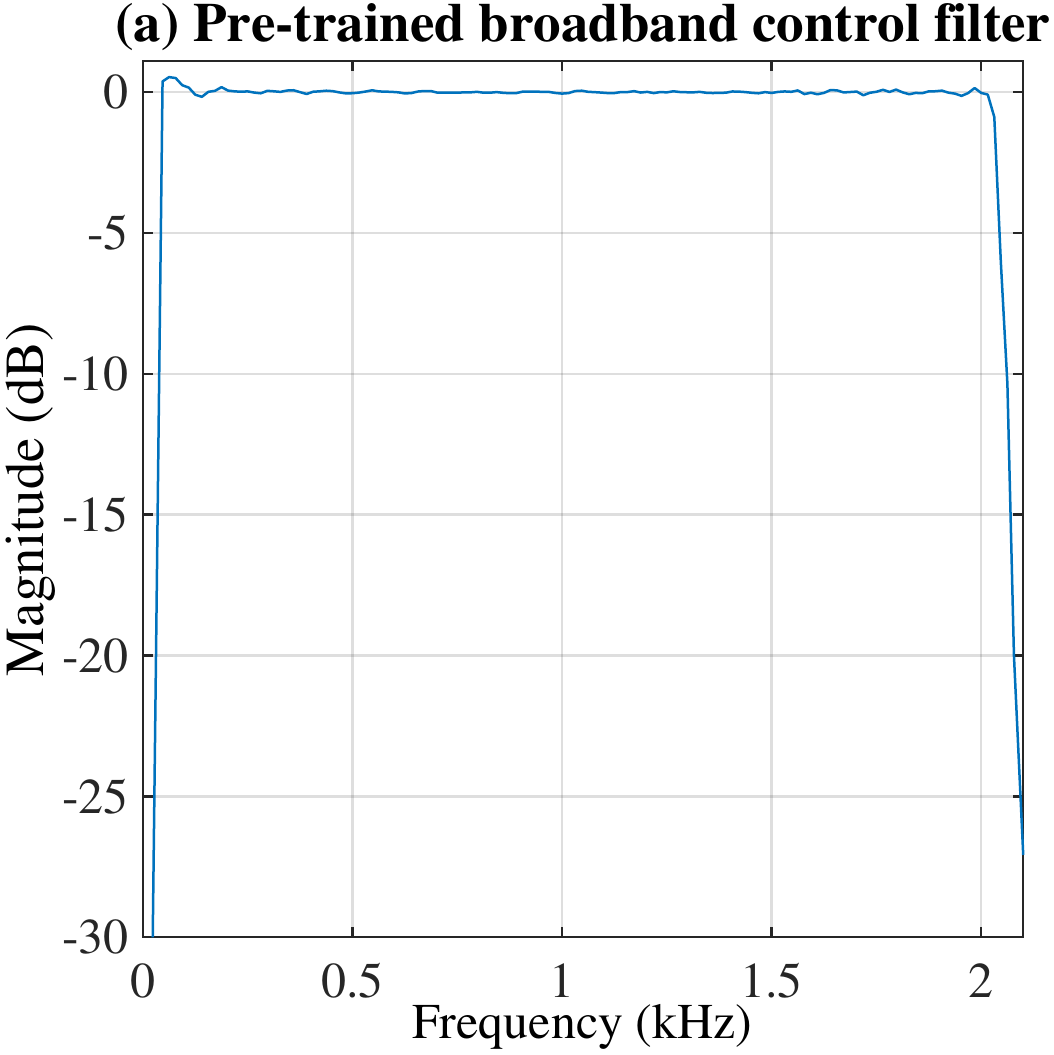}
\includegraphics[width=0.37\linewidth, height=3.2cm]{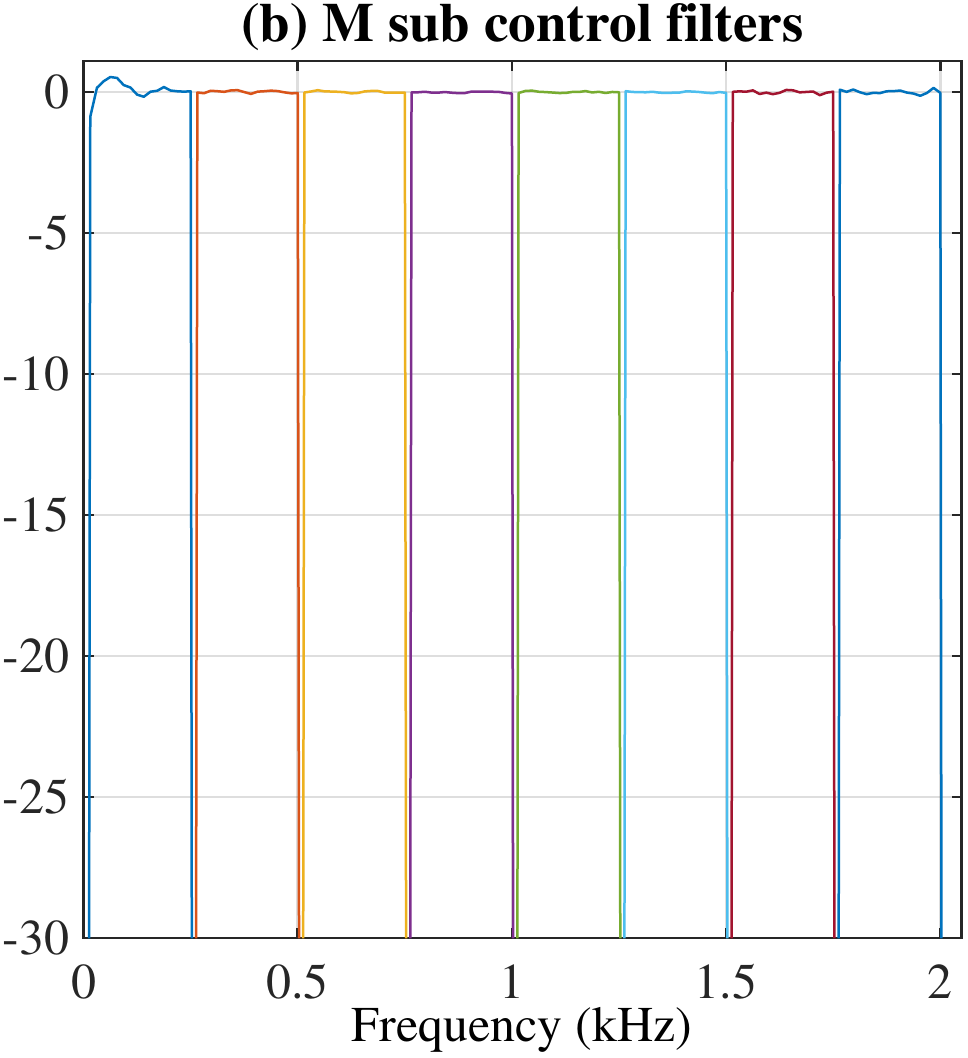}
\caption{Frequency spectrum of (a) the pre-trained broadband control filter and (b) the $M$ sub control filters.}
\label{Fig Sub Control Filters}
\end{figure}
%--------------------------------------------------------

\subsection{The GFANC Method}\label{The GFANC Method}
Traditional adaptive ANC algorithms, such as the FxLMS algorithm and its variants, update the control filter coefficients based on the feedback error signal. In contrast, the GFANC method~\cite{luoGFANCMssp} automatically generates suitable control filters by combining sub control filters, with the combination weights output by a CNN. GFANC does not rely on error-signal-based filter updates during noise control. Hence, it avoids the feedback mechanism and alleviates the risk of divergence.

\subsubsection{Sub Control Filters}\label{Sub Control Filters}
The construction of sub control filters is a crucial step in the GFANC method. First, the target ANC system is used to cancel a broadband noise covering the frequency components of interest. The optimal control filter is referred to as the pre-trained broadband control filter and serves as the sole prior information required by the GFANC method. Following the approach in \cite{LuoBayes}, the pre-trained broadband control filter is decomposed into $M$ sub control filters. The frequency spectrum of the pre-trained broadband control filter and its sub control filters on synthetic acoustic paths are shown in Figure~\ref{Fig Sub Control Filters}. The $M$ sub control filters form the sub control filter matrix $\mathbf{C}$, defined as
%----------------------------------------------------
\begin{equation}
\setlength{\abovedisplayskip}{2pt}
\setlength{\belowdisplayskip}{2pt}
\mathbf{C} = \left[\mathbf{c}_1,\ldots,\mathbf{c}_m,\ldots,\mathbf{c}_M \right]^\mathrm{T} \in \mathbb{R}^{M\times L},
\end{equation}
%----------------------------------------------------
where $\mathbf{c}_m \in \mathbb{R}^{1\times L}$ represents the impulse response of the $m$-th sub control filter of length $L$. These sub control filters serve as the orthogonal basis for generating various control filters.

\subsubsection{Generation of Control Filters}
In the GFANC method, the CNN generates a weight vector for each noise frame. Specifically, the $t$-th noise frame $\mathbf{x}_t$ is fed into the CNN, which outputs a weight vector $\mathbf{g}$ for combining the sub control filters. This corresponds to a regression task, where the elements of $\mathbf{g}$ take values between $0$ and $1$. The weight vector $\mathbf{g}$ is obtained by
%----------------------------------------------------
\begin{equation}
    \mathbf{g} = \textit{CNN}(\mathbf{x}_t;\, \Theta^{*}), \quad
    \mathbf{g} = \left[g_{1}, \ldots, g_{m}, \ldots, g_{M} \right],
    \label{Equation CNN in GFANC}
\end{equation}
%----------------------------------------------------
where $g_m$ denotes the weight associated with the $m$-th sub control filter $\mathbf{c}_m$. The function $\textit{CNN}(\cdot;\Theta^{*})$ represents the trained CNN model with parameters $\Theta^{*}$. The weight vector $\mathbf{g}$ is then transmitted to the real-time controller, which forms the control filter $\mathbf{w}$ by taking the weighted combination of the sub control filters as
%----------------------------------------------------
\begin{equation}
    \mathbf{w}
    = \mathbf{g}\mathbf{C}
    = \sum_{m=1}^{M} g_{m}\mathbf{c}_m
    \in \mathbb{R}^{1\times L}.
    \label{Equation Generate Control Filter in GFANC}
\end{equation}
%----------------------------------------------------
The generated control filter $\mathbf{w}$ of length $L$ is subsequently used for sample-by-sample noise cancellation, as described in Section~\ref{ANC Systems}.

%-------------------------------------------------------------------------
\begin{figure}[!t]
\centering
\includegraphics[width=0.8\linewidth]{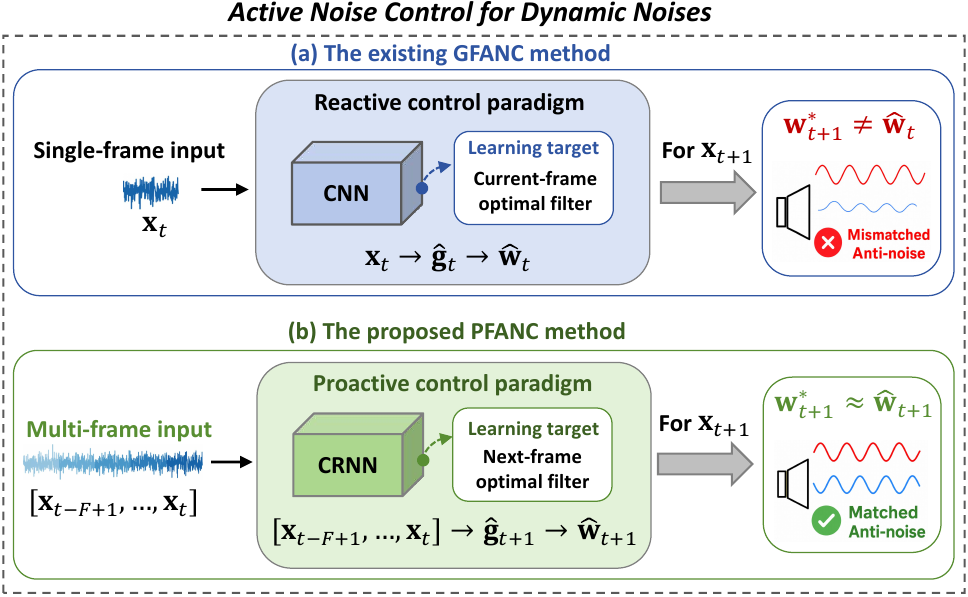}
\caption{Comparison of the existing GFANC and the proposed PFANC methods for dynamic noises. GFANC performs reactive control based on a single-frame input, while PFANC employs multi-frame temporal information to proactively predict the next-frame optimal filter.}
\label{Fig Compare GFANC PFANC}
\end{figure}
%-------------------------------------------------------------------------

\subsubsection{Limitations of the GFANC Method}\label{Limitations of the GFANC Method}
In the GFANC method, the training noise dataset, consisting of $N$ noise instances each containing $T$ frames, is denoted as $\{\mathbf{x}^{i}_t, \mathbf{g}^{*i}_t\}_{t=1, i=1}^{T, N}$, where $\mathbf{x}^{i}_t$ represents the $t$-th frame of the $i$-th noise instance. An adaptive labeling mechanism~\cite{LuoBayes} is employed to automatically assign each noise frame $\mathbf{x}^{i}_{t}$ its optimal weight vector $\mathbf{g}^{*i}_{t}$. The CNN is trained to map the current noise frame $\mathbf{x}^{i}_t$ to its corresponding optimal weight vector $\mathbf{g}^{i}_t$, with the training loss defined as
%-------------------------------------------------------------------------
\begin{equation}
    \mathcal{L}_{\text{CNN-train}}
    = \frac{1}{N T}\sum_{i=1}^{N}\sum_{t=1}^{T}
    \mathcal{L}_{\mathrm{MSE}}\Big(
    \textit{CNN}(\mathbf{x}^{i}_t;\, \Theta),
    \mathbf{g}^{*i}_t
    \Big),
    \label{GFANC-CNN-train loss}
\end{equation}
%-------------------------------------------------------------------------
where $\mathcal{L}_{\text{MSE}}(\cdot)$ denotes the Mean Squared Error (MSE) loss function used to quantify the difference between the predicted weight vector and the target label.

However, this training scheme inherently limits the predictive capability of GFANC. Since the CNN uses only the current noise frame as input and the current-frame optimal weight vector as the learning target, the model learns to estimate the control filter that is optimal for the present frame rather than for the upcoming one. This design implicitly assumes that the control filter varies slowly over time, which holds for stationary or slowly varying noises but leads to tracking lag and significant performance degradation for rapidly changing or dynamic noises. In this work, dynamic noises refer to noises with time-varying frequency characteristics, which exhibit spectral evolution across consecutive frames. Moreover, the GFANC method ignores the temporal correlations across consecutive frames, limiting the neural network’s ability to capture the temporal evolution of noise and anticipate its future variations.

As extensions of the GFANC method, although the GFANC-Bayes and GFANC-Kalman \cite{LuoBayes,LuoKalman} methods can utilize the correlations of adjacent noise frames, they only smooth the generated filter for the current frame rather than predicting the upcoming control filter. Thus, both GFANC and its extensions are limited by the lack of predictive capability, leading to poor tracking of rapidly varying noises. To address this issue, we propose the PFANC method, in which the CRNN processes the current and multiple historical noise frames and uses the next-frame optimal weight vector as the learning target. This proactive design enables PFANC to capture the temporal dynamics of noise and predict the optimal control filter for the subsequent frame. A comparison between the existing GFANC method and the proposed PFANC method is illustrated in Figure~\ref{Fig Compare GFANC PFANC}. Consequently, the PFANC method can predict a control filter that produces anti-noise better matched to the upcoming disturbance, thereby improving noise attenuation.

%-------------------------------------------------------------------------
\begin{figure}[tp]
\centering
\includegraphics[width=\linewidth]{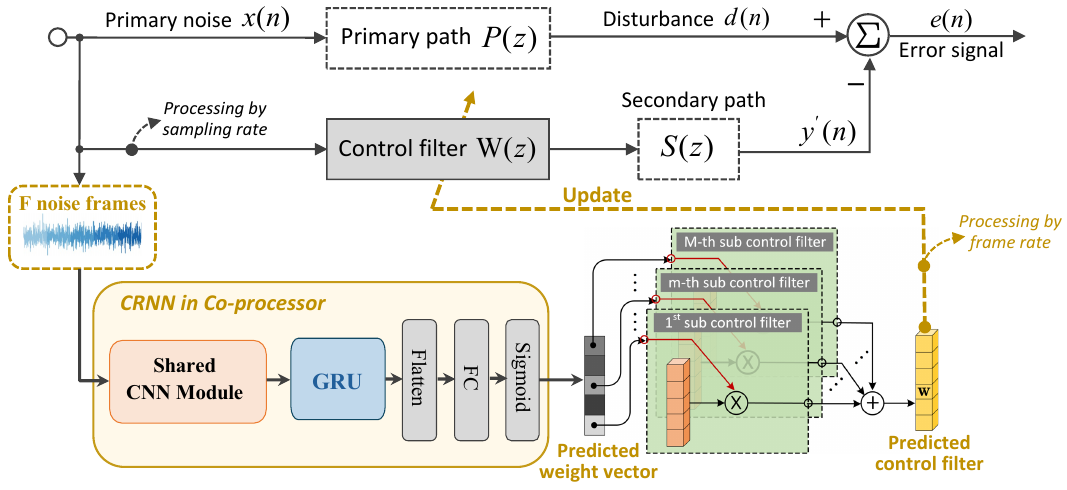}
\caption{Schematic of the PFANC method, where the CRNN runs on a co-processor to predict the next-frame weight vector from every $F$ noise frames, while the real-time controller simultaneously performs noise control at the sampling rate.}
\label{Fig PFANC}
\end{figure}
%-------------------------------------------------------------------------

%-------------------------------------------------------------------------
\begin{figure}[tp]
\centering
\includegraphics[width=\linewidth]{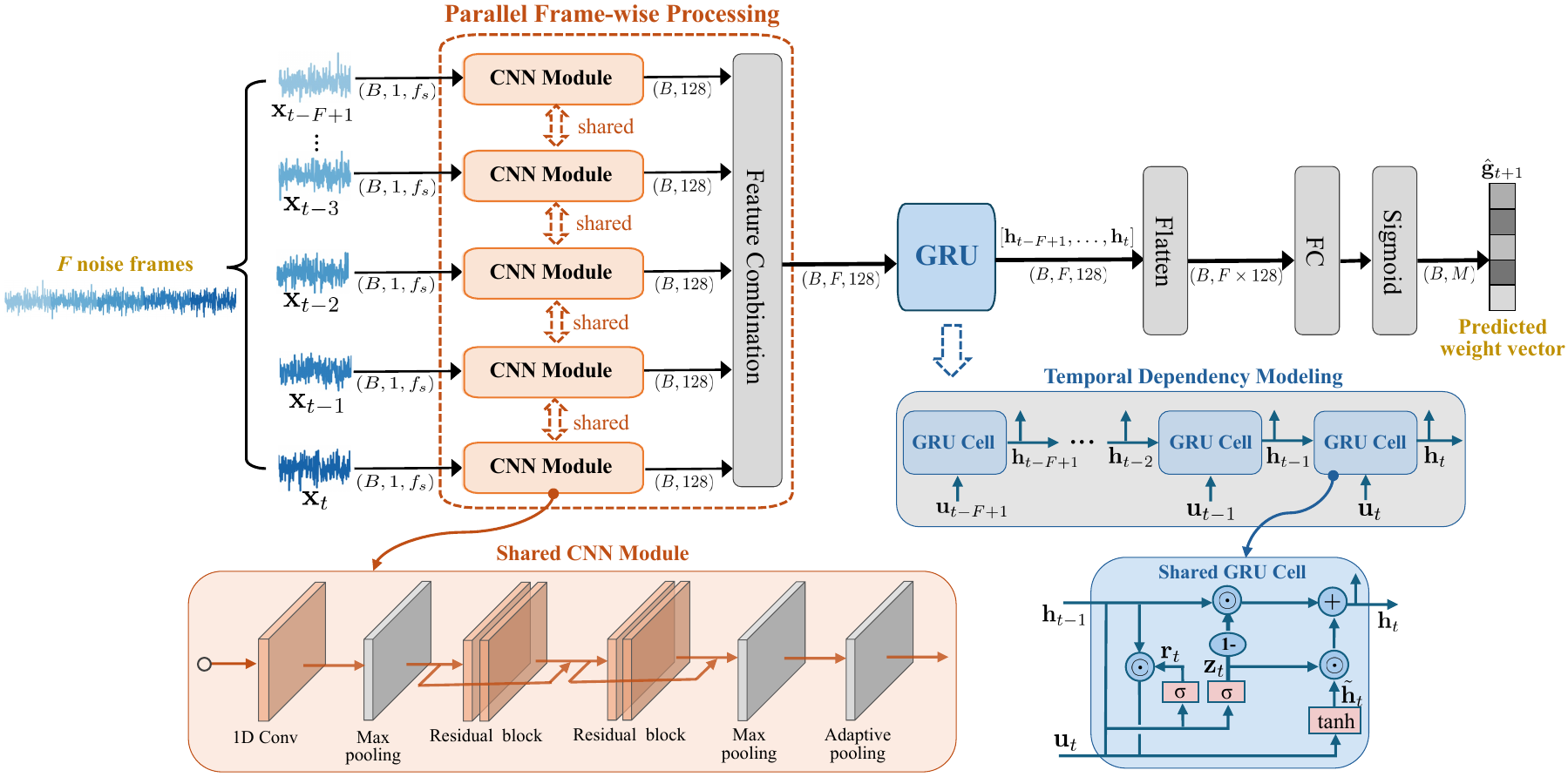}
\caption{Block diagram of the CRNN model, which learns dynamics from $F$ consecutive noise frames to predict the next-frame weight vector. It incorporates a shared CNN module for parallel frame-wise feature extraction and a GRU module for temporal dependency modeling. Especially, when $1 \le t < F$, zero-padding is applied to maintain the input length $F$.}
\label{Fig CRNN}
\end{figure}
%-------------------------------------------------------------------------

\section{The Proposed PFANC Method}\label{The PFANC Method}
The schematic of the proposed PFANC method is shown in Figure~\ref{Fig PFANC}. In this method, the CRNN operates on a co-processor to predict the weight vector for combining sub control filters at the frame rate, while the real-time controller performs noise control at the sampling rate in parallel. Unlike the GFANC method, which exploits only the current noise frame to generate a control filter, the PFANC method leverages multiple historical noise frames in addition to the current frame to achieve next-frame control filter prediction. The block diagram of the CRNN is illustrated in Figure~\ref{Fig CRNN}, which consists of a shared CNN module for frame-wise feature extraction and a Gated Recurrent Unit (GRU) module for temporal dependency modeling. The detailed CRNN architecture and training paradigm are described in the following subsection.

\subsection{Convolutional Recurrent Neural Network}
The proposed CRNN architecture illustrated in Figure~\ref{Fig CRNN} primarily consists of a CNN module, a GRU module, a flatten layer, a fully connected (FC) layer, and a Sigmoid layer. First, $F$ consecutive noise frames are fed into the shared CNN module for parallel extraction of frame-wise features. Each noise frame $\mathbf{x}_t \in \mathbb{R}^{(B, 1, f_s)}$ represents a waveform segment containing $f_s$ sampling points, where $f_s$ denotes the sampling rate, and $B$ is the batch size during training. By sharing parameters across frames, the CNN module ensures consistent feature extraction and enhances parameter efficiency. The shared CNN module adopts the same architecture as in the GFANC method~\cite{luoGFANCMssp} and directly operates on raw one-dimensional (1D) noise waveforms without converting them into spectrogram representations. Specifically, the shared CNN module comprises a 1D convolutional layer, two residual blocks, and pooling layers to perform frame-wise feature extraction and generate compact feature embeddings $\mathbf{u}_t \in \mathbb{R}^{(B, 128)}$. Subsequently, the features from $F$ consecutive frames are combined as $\mathbb{R}^{(B, F, 128)}$ to serve as the input to the subsequent GRU module.

Following the CNN-based feature extraction, a GRU module is employed to model temporal dependencies among the extracted features from $F$ consecutive noise frames. The GRU, a type of Recurrent Neural Network (RNN), utilizes two gating mechanisms, the reset gate and the update gate, to adaptively regulate information flow across time steps \cite{gru2015,RNNreview,lhw-sharedCRNN}. The GRU features a simple architecture with few parameters, resulting in fast training and low computational cost. Owing to these advantages, the GRU is well suited for temporal modeling in the proposed CRNN framework. The GRU module is composed of a series of cascaded GRU cells that share the same parameters across time steps. Each GRU cell receives the input feature at the current frame and the hidden state from the previous frame to update its internal state. The internal operations of the GRU cell in the PFANC method can be formulated as follows:
\begin{itemize}
    \item \textbf{Inputs:} The current input noise feature $\mathbf{u}_t$ and the previous hidden state $\mathbf{h}_{t-1}$ are fed into the GRU cell.

    \item \textbf{Reset Gate:}
    \begin{equation}
        \mathbf{r}_t = \sigma(\mathbf{W}_r \mathbf{u}_t + \mathbf{U}_r \mathbf{h}_{t-1} + \mathbf{b}_r),
    \end{equation}
    where $\sigma(\cdot)$ denotes the element-wise sigmoid function. The reset gate controls how much of the previous hidden state $\mathbf{h}_{t-1}$ should be considered when generating the candidate hidden state $\tilde{\mathbf{h}}_t$.

    \item \textbf{Update Gate:}  
    \begin{equation}
        \mathbf{z}_t = \sigma(\mathbf{W}_z \mathbf{u}_t + \mathbf{U}_z \mathbf{h}_{t-1} + \mathbf{b}_z),
    \end{equation}
    which controls the trade-off between retaining the previous hidden state $\mathbf{h}_{t-1}$ and incorporating the newly computed candidate state $\tilde{\mathbf{h}}_t$.

    \item \textbf{Candidate Hidden State:}
    \begin{equation}
        \tilde{\mathbf{h}}_t = \tanh(\mathbf{W}_h \mathbf{u}_t + \mathbf{U}_h (\mathbf{r}_t \odot \mathbf{h}_{t-1}) + \mathbf{b}_h),
    \end{equation}
    where $\odot$ denotes element-wise multiplication. The reset gate $\mathbf{r}_t$ modulates the contribution of the previous hidden state before computing the candidate activation. $\mathbf{W}_*, \mathbf{U}_*, \mathbf{b}_*$ represent the weight matrices and bias vectors of the corresponding gates.

    \item \textbf{Final Hidden State:}  
    \begin{equation}
        \mathbf{h}_t = (1 - \mathbf{z}_t) \odot \mathbf{h}_{t-1} + \mathbf{z}_t \odot \tilde{\mathbf{h}}_t,
    \end{equation}
    which performs a gated interpolation between the previous hidden state $\mathbf{h}_{t-1}$ and the candidate hidden state $\tilde{\mathbf{h}}_t$, effectively balancing memory preservation and update.

    \item \textbf{Output:} 
    The updated hidden state $\mathbf{h}_t$ is propagated to the next time step or to subsequent layers.
\end{itemize}
In summary, the GRU cell performs the following operations sequentially at each time step:
\begin{enumerate}
    \item Compute the reset and update gates, $\mathbf{r}_t$ and $\mathbf{z}_t$, from $\mathbf{u}_t$ and $\mathbf{h}_{t-1}$;
    \item Use $\mathbf{r}_t$ to regulate the contribution of $\mathbf{h}_{t-1}$ when forming the candidate state $\tilde{\mathbf{h}}_t$;
    \item Combine $\tilde{\mathbf{h}}_t$ and $\mathbf{h}_{t-1}$ through the update gate $\mathbf{z}_t$;
    \item Output the updated hidden state $\mathbf{h}_t$.
\end{enumerate}
The GRU produces a sequence of hidden states $\mathbf{H} = [\mathbf{h}_{t-F+1}, \dots, \mathbf{h}_t] \in \mathbb{R}^{(B, F, 128)}$ across $F$ time steps, which captures the temporal dependencies among $F$ noise frames. Subsequently, the sequence $\mathbf{H}$ aggregates the temporal cues across $F$ noise frames and is flattened into a feature vector of size $(B, F\times128)$. This feature vector is then processed by a FC layer followed by a Sigmoid activation to produce the predicted weight vector $\hat{\mathbf{g}}_{t+1} \in \mathbb{R}^{(B, M)}$.

\subsection{CRNN Training Strategy}\label{Training of the CRNN}
As discussed in Section~\ref{Limitations of the GFANC Method}, the GFANC method uses the current noise frame as input and trains the CNN with the optimal weight vector of the current frame as the label. Differently, the PFANC method takes multiple historical noise frames together with the current frame as input and uses the next-frame optimal weight vector as the label, enabling the CRNN to learn the temporal dynamics of the noise for filter prediction.

The training noise dataset for CRNN is identical to that described in Section~\ref{Limitations of the GFANC Method}. For each noise instance $\mathbf{x}^{i}$, the sequence from the $(t-F+1)$-th frame to the $t$-th frame, denoted as $[\mathbf{x}^{i}_{t-F+1}, \dots, \mathbf{x}^{i}_t]$, is used as the input to the CRNN, which predicts the weight vector for every $F$ consecutive noise frames. The index $t$ satisfies $1 \le t < T$. Especially, when $1 \le t < F$, zero-padding is applied to the preceding frames to maintain the input length $F$. The corresponding label of $[\mathbf{x}^{i}_{t-F+1}, \dots, \mathbf{x}^{i}_t]$ is the optimal weight vector $\mathbf{g}^{*i}_{t+1}$ associated with the next noise frame $\mathbf{x}^{i}_{t+1}$.

The objective of training is to minimize the discrepancy between the CRNN-predicted and target weight vectors over the entire training dataset. The optimal parameters of the CRNN in the PFANC method, denoted by $\Theta^{*}$, are obtained by minimizing the following training loss:
%-------------------------------------------------------------------------
\begin{equation}
    \begin{split}
       \mathcal{L}_{\text{CRNN-train}} = &\frac{1}{N (T-1)}\cdot \\
        &\sum_{i=1}^{N} \sum_{t=1}^{T-1}\mathcal{L}_{\text{MSE}}\Big(
        \textit{CRNN}([\mathbf{x}^{i}_{t-F+1}, \dots, \mathbf{x}^{i}_t]; \Theta),\,
        \mathbf{g}^{*i}_{t+1}
        \Big).
    \end{split}
\end{equation}
%-------------------------------------------------------------------------
Notably, the training loss differs from that used in the GFANC method in Eq.~(\ref{GFANC-CNN-train loss}). The PFANC method differs from GFANC in that it takes multiple consecutive noise frames $[\mathbf{x}^i_{t-F+1}, \dots, \mathbf{x}^i_t]$ as input instead of a single current frame $\mathbf{x}^i_t$, and uses the next-frame optimal weight vector $\mathbf{g}^{*i}_{t+1}$ as the training label rather than $\mathbf{g}^{*i}_t$. This training scheme enables the CRNN to leverage the temporal correlations among consecutive noise frames to predict the upcoming weight vector, thereby enabling a proactive control paradigm in PFANC.

%---------------------------------------------------------
\begin{table}[!t]
\renewcommand{\arraystretch}{1.2}
\centering
\caption{Pseudo-code of online noise control in the PFANC method.}\vspace*{-0.1cm}
\resizebox{\linewidth}{!}{
\begin{tabular}{|l|}
\hline
\textbf{Description:} The weight vector is predicted by the CRNN in the co-processor based \\on $F$ noise frames, while real-time noise control is performed at the sampling rate.\\
\hline
\textbf{Initialization:} $F$ consecutive noise frames are denoted as $[\mathbf{x}_{t-F+1}, \dots, \mathbf{x}_t]$.\\
The sub control filter matrix is $\mathbf{C} = \left[\mathbf{c}_1,\ldots,\mathbf{c}_m,\ldots,\mathbf{c}_M \right]^\mathrm{T}$. \\
The weight vector is denoted as $\mathbf{g} = \left[g_{1}, \ldots, g_{m}, \ldots, g_{M} \right]$.\\
Initialize the weight vector as a zero vector, i.e., $\mathbf{g}=\mathbf{0}$.\\
$\textit{CRNN}(\cdot;\Theta^{*})$ represents the trained CRNN with parameters $\Theta^{*}$.\\
\hline
\textbf{\# Noise control in the real-time controller:}\\
\textbf{for} each sample of the reference signal \textbf{do}\\
~~~$\mathbf{w} = \mathbf{g} \mathbf{C} = \sum_{m=1}^{M} g_{m}\mathbf{c}_m$~~~~$\triangleright$ Generate the control filter.\\
~~~$e(n) = d(n) - \mathbf{w}^\mathrm{T}(n) \mathbf{x}(n) * s(n)$~~~~$\triangleright$ Real-time noise control.\\
\textbf{end for}\\
\textbf{\# Weight vector prediction in the co-processor:}\\
\textbf{for} $F$ frames of the reference signal \textbf{do}\\
~~~$\hat{\mathbf{g}}_{t+1} = \textit{CRNN}([\mathbf{x}_{t-F+1}, \dots, \mathbf{x}_t]; \Theta^{*})$~~$\triangleright$ Predict the next-frame weight vector.\\
~~~\textbf{\# Update the used weight vector:}\\
~~~\textbf{if}~~~$\mathbf{g}~\neq~\hat{\mathbf{g}}_{t+1}$~~\textbf{do}~~~~$\triangleright$ If $\hat{\mathbf{g}}_{t+1}$ is different from the currently used weight vector.\\
~~~~~~~~$\mathbf{g} \leftarrow \hat{\mathbf{g}}_{t+1}$~~~~$\triangleright$ Transmit $\hat{\mathbf{g}}_{t+1}$ to the real-time controller to update $\mathbf{g}$.\\
\textbf{end for}\\
\hline
\end{tabular}
}
\label{Table PFANC}
\end{table}
%---------------------------------------------------------

\subsection{Online Noise Control in PFANC}
The online noise control procedure of the PFANC method is summarized in Table~\ref{Table PFANC}. After training, the CRNN with its optimal parameters $\Theta^{*}$ is used to predict a weight vector from every $F$ noise frames in the PFANC method. The predicted weight vector $\hat{\mathbf{g}}_{t+1}$ is obtained as  
\begin{equation}
\hat{\mathbf{g}}_{t+1} = \textit{CRNN}\left([\mathbf{x}_{t-F+1}, \dots, \mathbf{x}_t];\, \Theta^{*}\right),
\end{equation}
which differs from the CNN in GFANC (Eq.~\ref{Equation CNN in GFANC}), where the weight vector is generated solely from the current frame $\mathbf{x}_t$. Thus, compared with GFANC, the proposed PFANC method leverages temporal dependencies across consecutive noise frames, enabling it to capture cross-frame correlations in the reference signal and anticipate upcoming noise variations.

When the newly predicted weight vector $\hat{\mathbf{g}}_{t+1}$ differs from the one currently in use $\mathbf{g}$, it is transmitted to the real-time controller to update $\mathbf{g}$. A new control filter $\mathbf{w}$ is then generated by combining sub control filters using $\mathbf{g}$. The control filter $\mathbf{w}$ is applied for real-time noise control, which operates at the sampling rate as described in Section~\ref{ANC Systems}. By adopting a dual-rate architecture, the co-processor and real-time controller operate in parallel, ensuring that the CRNN’s processing latency does not interfere with real-time noise control. This coordinated operation enables delayless noise attenuation, making the PFANC system suitable for practical deployment. Potential applications of the PFANC method include vehicle and railway cabins and noise-control systems for variable-speed engines, motors, fans, and compressors, where low-frequency noise characteristics evolve with operating speed and load.

%-------------------------------------------------------------------
\begin{figure}
    \centering
    \includegraphics[width=0.65\linewidth]{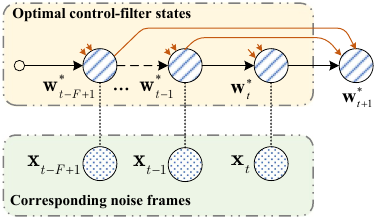}
    \caption{Higher-order Markov formulation of optimal control-filter evolution. The upcoming state $\mathbf{w}_{t+1}^{*}$ depends on the preceding $F$ filter states, while the dotted lines indicate the temporal correspondence between the filter states and the associated noise frames.}
    \label{Fig Higher-order Markov}
\end{figure}
%-------------------------------------------------------------------

\section{Theoretical Analysis of Filter Prediction based on Multiple Noise Frames}\label{Theoretical Analysis of Filter Prediction based on Multiple Noise Frames}

This section presents a theoretical analysis of how incorporating multiple noise frames influences the estimation of the control filter. We model the temporal evolution of the optimal control filter using a higher-order Markov formulation to capture dependencies across multiple preceding states. Based on the Minimum Mean Square Error (MMSE) property of conditional expectation, we show that incorporating additional historical observations cannot increase the minimum achievable prediction error. We further establish the theoretical connection between this optimal predictor and the CRNN employed in the proposed PFANC method.

\subsection{From Single-Frame to Multi-Frame Temporal Modeling}\label{From Single-Frame to Multi-Frame Temporal Modeling}
In the previous GFANC method \cite{luoGFANCMssp}, the optimal control filter was assumed to evolve slowly over time under stationary or slowly varying noise conditions, i.e.,
\begin{equation}\label{eq:w slow}
    \mathbf{w}_{t+1}^{*} \approx \mathbf{w}_{t}^{*}.
\end{equation}
Under the GFANC formulation, the optimal control filter for the current frame is estimated from the current noise observation through an approximate mapping
\begin{equation}
    \mathbf{w}_{t}^{*} = \varphi(\mathbf{x}_{t}),
\end{equation}
where $\varphi(\cdot)$ denotes the mapping from the current reference frame to its corresponding optimal control filter. When the noise varies slowly, the current-frame optimal filter $\mathbf{w}_{t}^{*}$ remains close to the upcoming optimal filter $\mathbf{w}_{t+1}^{*}$ and can therefore provide effective control for the subsequent frame.

To further exploit the temporal correlation between adjacent frames, GFANC-Bayes \cite{LuoBayes} and GFANC-Kalman \cite{LuoKalman} recursively refine the current-frame filter estimate using information from the preceding frame, which can be abstracted as
\begin{equation}
    \mathbf{w}_{t}^{*}
    =
    \mathcal{F}\big(\varphi(\mathbf{x}_{t}),\,\mathbf{w}_{t-1}^{*}\big),
\end{equation}
where $\mathcal{F}(\cdot)$ represents the corresponding Bayesian- or Kalman-based filtering operation. These GFANC-based methods therefore primarily exploit short-term temporal correlations while still relying on the slowly varying behavior of the optimal control filter. For dynamic noises with rapidly varying frequency characteristics, however, the upcoming optimal filter may differ substantially from the current one, leading to an inherent tracking lag when the current-frame filter is applied to the subsequent frame.

For dynamic noises with rapidly varying frequency characteristics, the slowly varying-filter assumption may no longer hold, and the optimal control filter for the upcoming frame can differ substantially from that of the current frame. In this case, multiple preceding noise frames may provide additional information about the temporal evolution of the noise. To account for such temporal dependencies, we consider a higher-order Markov formulation of the optimal control-filter evolution, as illustrated in Figure~\ref{Fig Higher-order Markov}. The upcoming optimal filter $\mathbf{w}_{t+1}^{*}$ depends on the $F$ most recent filter states, i.e.,
\begin{equation}
    p\!\left(\mathbf{w}_{t+1}^{*} \mid \mathbf{w}_{1:t}^{*}\right)
    =
    p\!\left(\mathbf{w}_{t+1}^{*} \mid \mathbf{w}_{t-F+1:t}^{*}\right).
    \label{eq:markov}
\end{equation}
The slowly varying-filter assumption in Eq.~(\ref{eq:w slow}) adopted by the GFANC method corresponds to the special case of $F=1$. Since the optimal filter states are determined by the corresponding noise frames, the same order-$F$ temporal dependency is assumed to hold in the observation domain.

In the proposed PFANC method, however, the optimal filter states are not directly observable. Instead, the prediction is performed from the available noise observations. Let
\begin{equation}
    \mathbf{x}_{t-F+1:t}
    =
    [\mathbf{x}_{t-F+1},\ldots,\mathbf{x}_{t}]
\end{equation}
denote the sequence of $F$ consecutive noise frames available at time $t$. We compare the minimum achievable prediction error when only the current frame $\mathbf{x}_{t}$ is available with that obtained when the complete multi-frame observation $\mathbf{x}_{t-F+1:t}$ is used.

Under the Mean Square Error (MSE) criterion, the optimal predictor of the upcoming control filter based on the current noise frame is given by the conditional expectation
\begin{equation}
    \widetilde{\mathbf{w}}_{t+1}^{(1)}
    =
    \mathbb{E}
    \left[
        \mathbf{w}_{t+1}^{*}
        \mid
        \mathbf{x}_{t}
    \right].
\end{equation}
Accordingly, the minimum achievable MSE using the current frame alone is
\begin{equation}
\label{eq:mmse_single}
    \mathrm{MMSE}_{1}
    =
    \mathbb{E}
    \left[
        \left\|
            \mathbf{w}_{t+1}^{*}
            -
            \widetilde{\mathbf{w}}_{t+1}^{(1)}
        \right\|_{2}^{2}
    \right].
\end{equation}

When the $F$ consecutive noise frames are available, the corresponding optimal predictor becomes
\begin{equation}
    \widetilde{\mathbf{w}}_{t+1}^{(F)}
    =
    \mathbb{E}
    \left[
        \mathbf{w}_{t+1}^{*}
        \mid
        \mathbf{x}_{t-F+1:t}
    \right],
\end{equation}
and the minimum achievable MSE is
\begin{equation}
\label{eq:mmse_multi}
    \mathrm{MMSE}_{F}
    =
    \mathbb{E}
    \left[
        \left\|
            \mathbf{w}_{t+1}^{*}
            -
            \widetilde{\mathbf{w}}_{t+1}^{(F)}
        \right\|_{2}^{2}
    \right].
\end{equation}

Since the current observation $\mathbf{x}_{t}$ is contained in the multi-frame observation $\mathbf{x}_{t-F+1:t}$, the latter provides an enlarged information set \cite{Reviewer1,Reviewer2}. According to the orthogonal projection property of conditional expectation \cite{StatisticalSP}, the prediction errors satisfy
\begin{equation}
\begin{aligned}
\mathrm{MMSE}_{1} = \mathrm{MMSE}_{F} + \mathbb{E}\left[\left\|
    \widetilde{\mathbf{w}}_{t+1}^{(F)}
    -
    \widetilde{\mathbf{w}}_{t+1}^{(1)}
\right\|_{2}^{2}
\right].
\end{aligned}
\label{eq:mmse_decomposition}
\end{equation}
Therefore,
\begin{equation}
\label{eq:mmse_monotonicity}
    \mathrm{MMSE}_{F}
    \le
    \mathrm{MMSE}_{1}.
\end{equation}

Eq.~(\ref{eq:mmse_monotonicity}) shows that, for the optimal MSE predictor, incorporating additional historical noise observations cannot increase the minimum achievable prediction error. Moreover, from Eq.~(\ref{eq:mmse_decomposition}), the reduction in MMSE can be expressed as
\begin{equation}
\begin{aligned}
\mathrm{MMSE}_{1}-\mathrm{MMSE}_{F}
=
\mathbb{E}
\left[
\left\|
    \mathbb{E}
    \left[
        \mathbf{w}_{t+1}^{*}
        \mid
        \mathbf{x}_{t-F+1:t}
    \right]
    -
    \mathbb{E}
    \left[
        \mathbf{w}_{t+1}^{*}
        \mid
        \mathbf{x}_{t}
    \right]
\right\|_{2}^{2}
\right]
\ge 0.
\end{aligned}
\label{eq:mmse_gain}
\end{equation}

When the historical noise frames provide no additional information for predicting the upcoming optimal control filter beyond that contained in the current frame, the two conditional expectations become identical and $\mathrm{MMSE}_{F}=\mathrm{MMSE}_{1}$. This situation approximately corresponds to stationary or slowly varying noises, for which the current noise frame can already provide sufficient information for control-filter estimation. In contrast, for dynamic noises with temporal spectral evolution, historical frames may contain additional predictive information about the upcoming noise characteristics, leading to $\mathrm{MMSE}_{F}<\mathrm{MMSE}_{1}$. Therefore, the MMSE analysis provides a theoretical justification for exploiting multi-frame temporal information in the proposed PFANC method.

\subsection{CRNN-Based Approximation of the MMSE Predictor}

The control filter is generated as a weighted combination of $M$ sub control filters. Let $\mathbf{C}\in\mathbb{R}^{M\times L}$ denote the sub control filter matrix and $\mathbf{g}_{t+1}^{*}\in\mathbb{R}^{1\times M}$ denote the optimal weight vector associated with the $(t+1)$-th noise frame. The corresponding optimal control filter is given by
\begin{equation}
\mathbf{w}_{t+1}^{*}
=
\mathbf{g}_{t+1}^{*}\mathbf{C}.
\end{equation}
As established in the previous subsection, under the MSE criterion, the optimal predictor based on the multi-frame observation $\mathbf{x}_{t-F+1:t}$ is given by the corresponding conditional expectation. Therefore, the MMSE predictor of the upcoming weight vector is
\begin{equation}
\widetilde{\mathbf{g}}_{t+1}^{(F)}
=
\mathbb{E}
\left[
\mathbf{g}_{t+1}^{*}
\mid
\mathbf{x}_{t-F+1:t}
\right].
\label{eq:mmse_g}
\end{equation}

As described in Section~\ref{Training of the CRNN}, the CRNN is trained by minimizing the MSE between its predicted weight vector and the optimal next-frame weight vector. Since the MSE-optimal regression function is the conditional expectation, the trained CRNN aims to approximate the MMSE predictor in Eq.~(\ref{eq:mmse_g}):
\begin{equation}
\hat{\mathbf{g}}_{t+1}
=
\textit{CRNN}
\left(
\mathbf{x}_{t-F+1:t};
\Theta^{*}
\right)
\approx
\widetilde{\mathbf{g}}_{t+1}^{(F)}
=
\mathbb{E}
\left[
\mathbf{g}_{t+1}^{*}
\mid
\mathbf{x}_{t-F+1:t}
\right].
\label{eq:crnn_approx_g}
\end{equation}
The MMSE predictor of the upcoming control filter is $\widetilde{\mathbf{w}}_{t+1}^{(F)}$. Since the control filter is a fixed linear transformation of the weight vector, we have
\begin{equation}
\widetilde{\mathbf{w}}_{t+1}^{(F)}
=
\widetilde{\mathbf{g}}_{t+1}^{(F)}\mathbf{C}.
\label{eq:mmse_w}
\end{equation}
Consequently, the control filter generated from the CRNN-predicted weight vector satisfies
\begin{equation}
\hat{\mathbf{w}}_{t+1}
=
\hat{\mathbf{g}}_{t+1}\mathbf{C}
\approx
\widetilde{\mathbf{g}}_{t+1}^{(F)}\mathbf{C}
=
\widetilde{\mathbf{w}}_{t+1}^{(F)}
=
\mathbb{E}
\left[
\mathbf{w}_{t+1}^{*}
\mid
\mathbf{x}_{t-F+1:t}
\right].
\label{eq:crnn_approx_w}
\end{equation}

Therefore, by training the CRNN to predict the optimal next-frame weight vector, the PFANC method approximates the MMSE predictor of the upcoming control filter from multi-frame noise observations. Combined with the MMSE analysis in the previous subsection, this establishes the theoretical connection between multi-frame temporal modeling and the CRNN-based predictive implementation in the PFANC method.

%---------------------------------------------------------
\begin{table}[!t]
\centering        
\caption{Parameter configuration in simulations.}\vspace*{-0.2cm}
\begin{tabular}{ll}
\hline
\textbf{Definition} & \textbf{Parameter} \\
\hline
Frequency range of interest & $20$–$2,000$ Hz \\
Sampling rate ($f_s$) & $16$ kHz \\
Number of frames used to predict weight vector ($F$) & $5$ \\
Each frame length & $1$ second \\
Number of sub control filters ($M$) & $8$ \\
Control filter length ($L$) & $1,024$ \\
Impulse response length of the secondary path & $256$ \\
Impulse response length of the primary path & $512$ \\
\hline
\end{tabular}
\label{Table GFANC Headphone Parameters}
\end{table}
%---------------------------------------------------------

\section{Numerical Simulation}\label{Numerical Simulation}
This section presents simulations conducted to evaluate the effectiveness of the proposed PFANC method. We first describe the training dataset used for CRNN training and examine the model’s performance after training. Subsequently, the noise control capability of the PFANC method is evaluated using linear and logarithmic chirp signals as well as real-world dynamic noises. Finally, the transferability of PFANC is examined across different acoustic paths.

\subsection{Parameter Configuration}
The parameter configurations used in the PFANC simulations are summarized in Table~\ref{Table GFANC Headphone Parameters}. The sampling rate is set to $16$~kHz, and the control filter length is chosen as $1,024$ taps. The lengths of the secondary path and primary path are $256$ and $512$ taps, respectively. Since ANC is particularly effective for low-frequency noises, the target frequency band is defined as $20$–$2,000$~Hz to cover most typical low-frequency noises. The number of sub control filters, denoted as $M$, is set to $8$, indicating that the pre-trained $20$–$2,000$~Hz broadband control filter is decomposed into $8$ sub control filters. The number of frames used to predict the weight vector ($F$) is set to $5$. In particular, when $1 \le t < 5$, zero-padding is applied to the preceding frames to ensure a fixed input of $5$ frames. Therefore, except for the first second, a predicted control filter can be generated and applied to each subsequent frame.

\subsection{Effectiveness of the CRNN}
In this subsection, the dataset used for training the CRNN is first described, and the effectiveness of the trained CRNN is then evaluated on the testing dataset.

%---------------------------------------------------------------------------------
\begin{figure}[!t]
\centering
\includegraphics[width=0.8\linewidth]{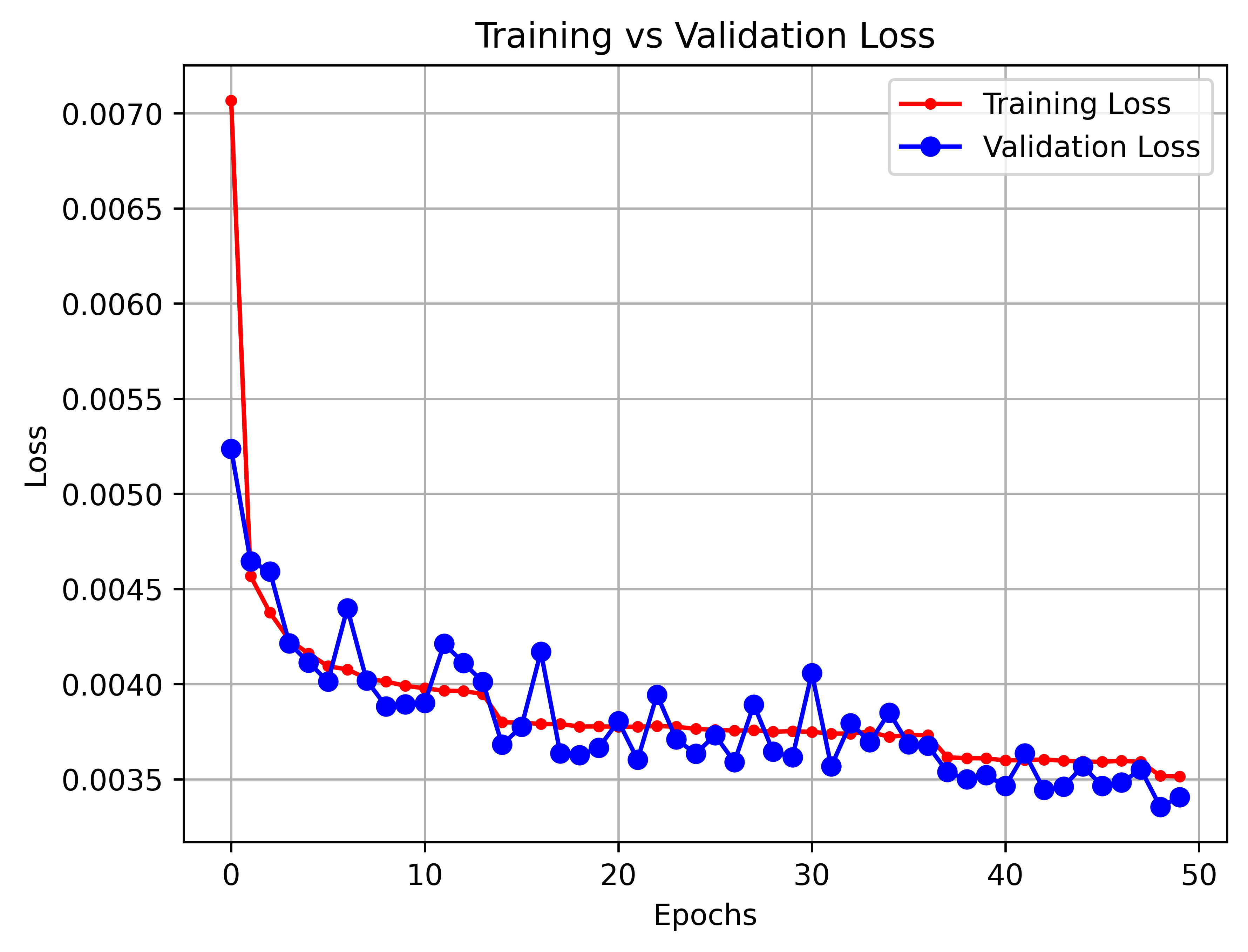}\vspace*{-0.2cm}
\caption{Training and validation loss evolution during CRNN model training.}
\label{Fig CRNN Training Loss}
\end{figure}
%---------------------------------------------------------------------------------

\subsubsection{Training dataset}
We used two synthetic noise datasets and one real noise dataset to train and evaluate the CRNN. The two synthetic datasets respectively consist of linear and logarithmic chirp signals, while the real dataset contains environmental noises. Each synthetic dataset includes $10{,}000$ noise instances for training and $1{,}000$ instances for testing. The synthetic linear and logarithmic chirp signals are generated by continuously sweeping the frequency from a random start frequency to a random end frequency within the $20$–$2{,}000$~Hz range. The real noise dataset is constructed from a subset of the \href{https://zenodo.org/records/3966543}{SONYC Urban Sound Tagging Dataset}, from which $13{,}000$ noise instances are selected for training and $2{,}000$ for testing. Each noise instance in all datasets has a duration of $10$ seconds, corresponding to $T = 10$ frames per instance.

%-------------------------------------------------
\begin{table}[!t]
\centering
\caption{Evaluation results of the CRNN on the test dataset.}\vspace*{-0.2cm}
\begin{tabular}{lc}
\hline
\textbf{Metric} & \textbf{Value} \\
\hline
Network input length & $5$ seconds\\
MSE loss of weight vector prediction  & $0.0033$ \\
Total trainable parameters & $0.31$ million \\
Execution time on GPU & $0.79$ milliseconds \\
Execution time on CPU & $7.35$ milliseconds \\
\hline
\end{tabular}
\label{Table CRNN Performance}
\end{table}
%-------------------------------------------------

\subsubsection{Training and testing of the CRNN}
The proposed CRNN was trained using the Adam optimizer with an MSE loss between the predicted and target weight vectors. Figure~\ref{Fig CRNN Training Loss} shows the convergence of the training and validation losses, which stabilized after approximately $30$ epochs, indicating good generalization and the absence of overfitting. As summarized in Table~\ref{Table CRNN Performance}, the CRNN contains approximately $0.31$ million trainable parameters, slightly higher than the CNN model used in GFANC ($0.22$ million) due to the additional GRU layer for temporal modelling. During testing, the CRNN achieved an MSE loss of $0.0033$ on the test dataset. The model input length was $5$ seconds (corresponding to $F$ frames), and its inference time was $0.79$~ms on an NVIDIA~H200 Tensor~Core GPU and $7.35$~ms on an Intel~Xeon~Platinum~8558 CPU, demonstrating real-time capability. These results confirm that the CRNN model maintains low computational overhead while effectively capturing temporal dependencies in the primary noise.

%------------------------------------------------------------------------
\begin{figure}[!t]
\centering
\includegraphics[width=0.48\linewidth, height=4.5cm]{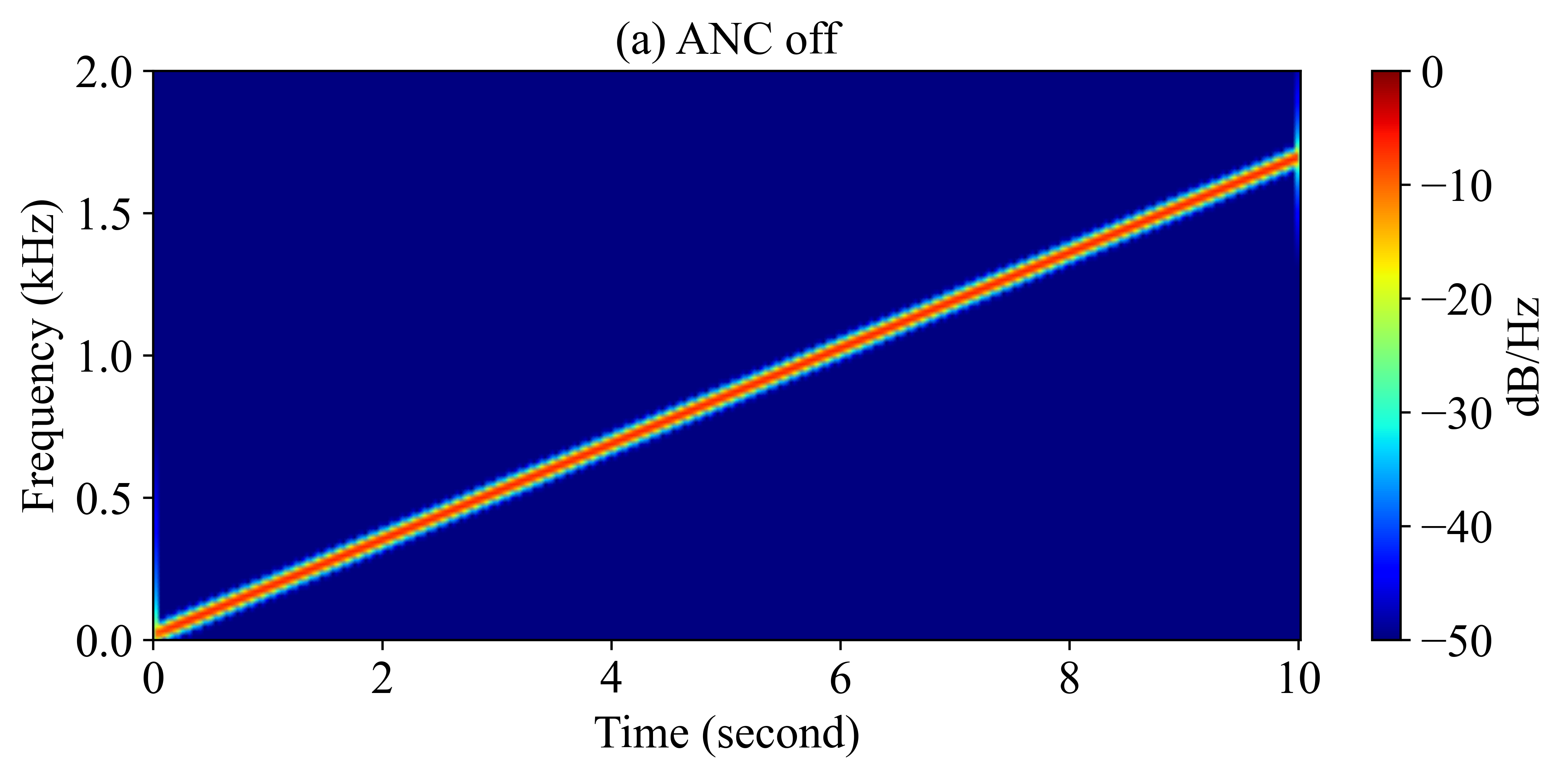}
\includegraphics[width=0.48\linewidth, height=4.5cm]{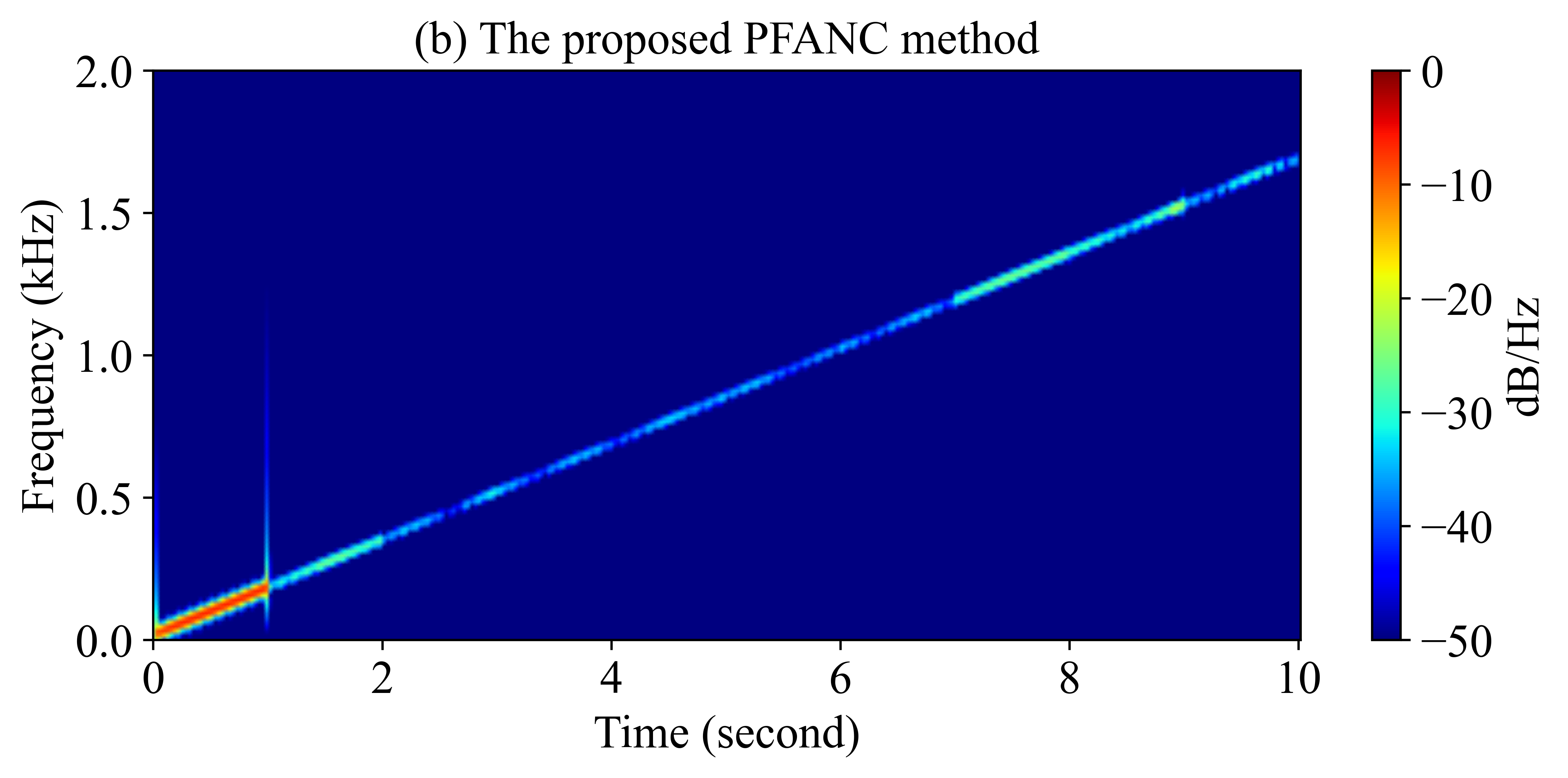}
\includegraphics[width=0.48\linewidth, height=4.5cm]{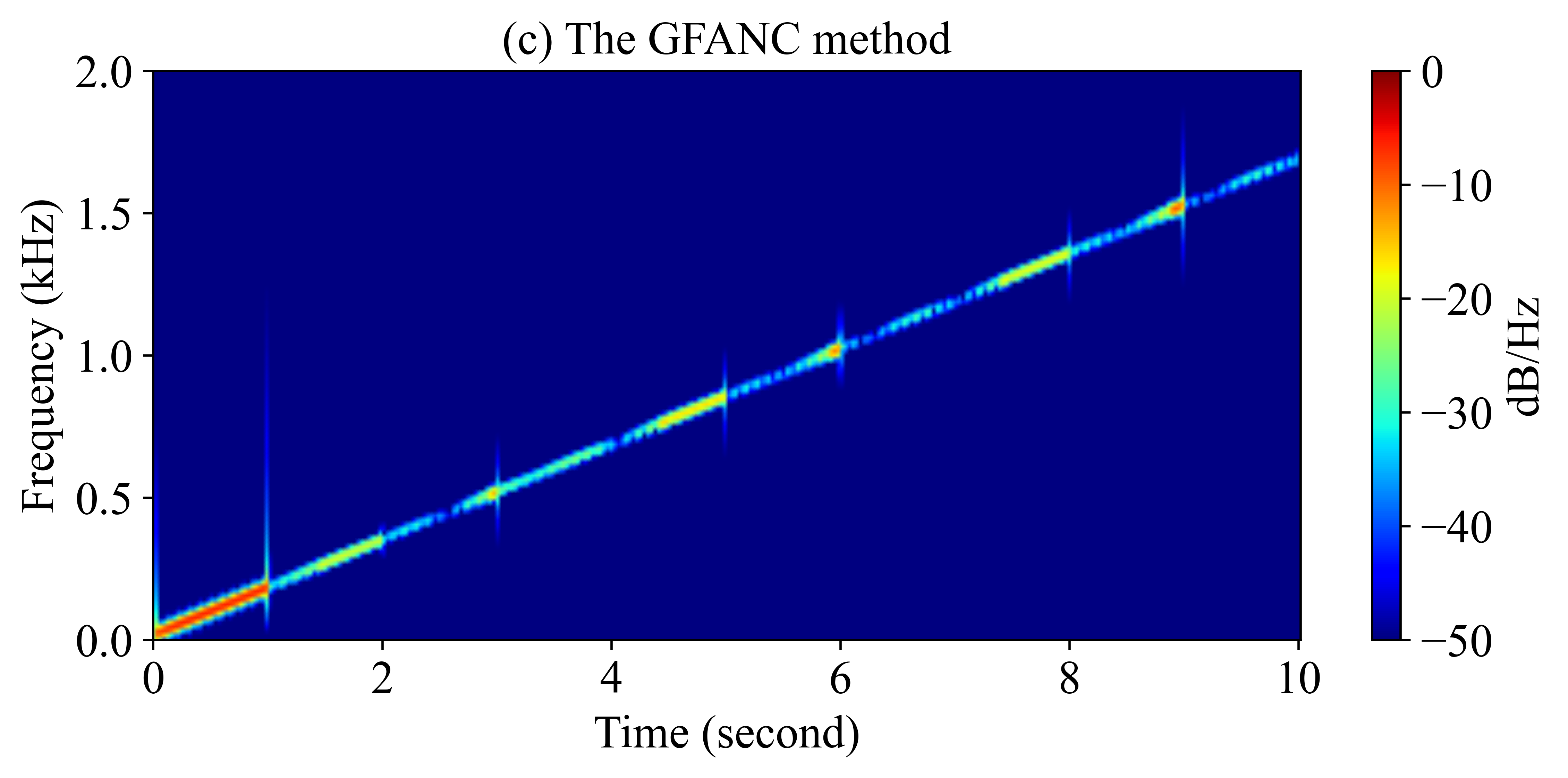}
\includegraphics[width=0.48\linewidth, height=4.5cm]{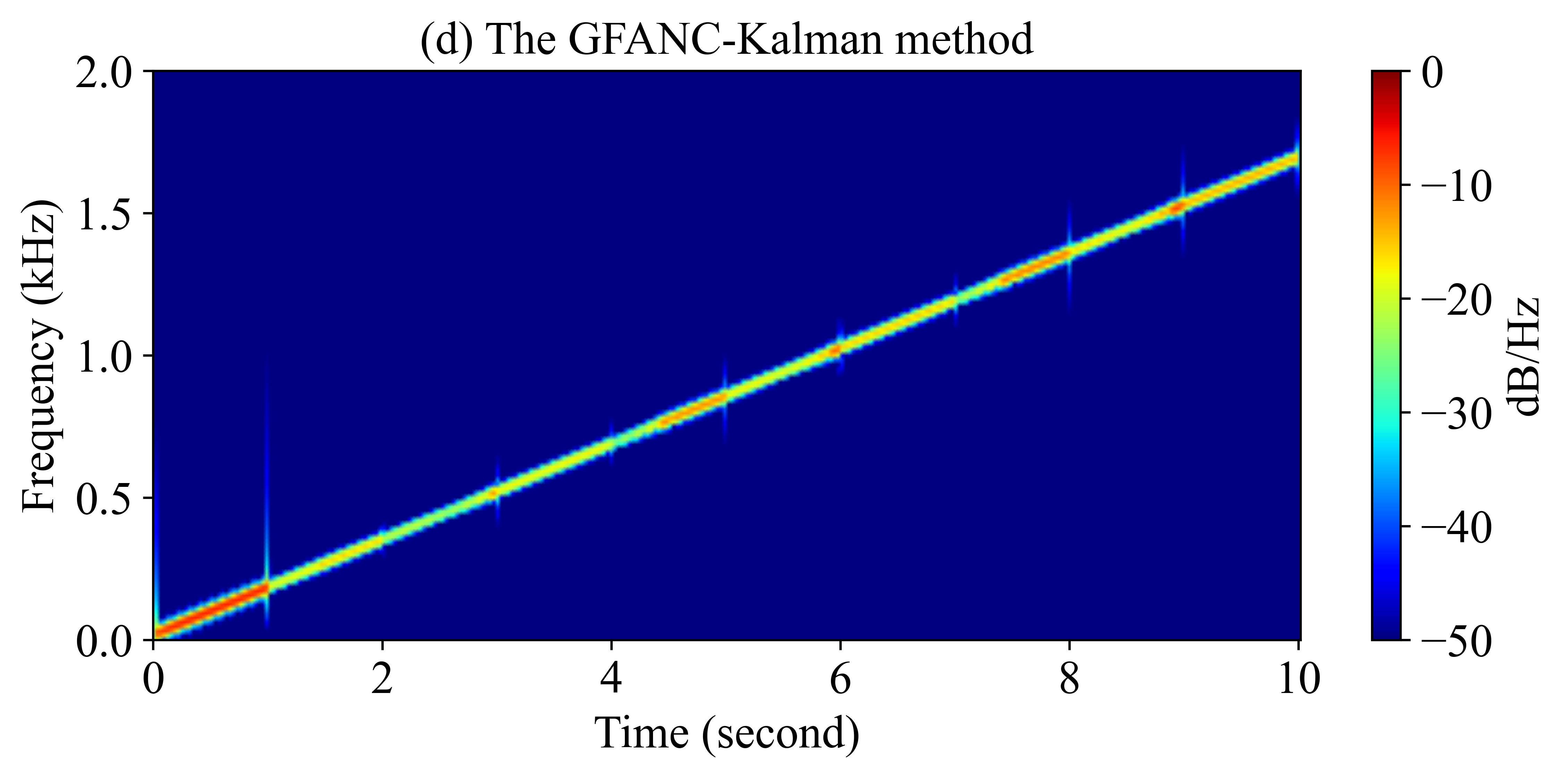}
\includegraphics[width=0.48\linewidth, height=4.5cm]{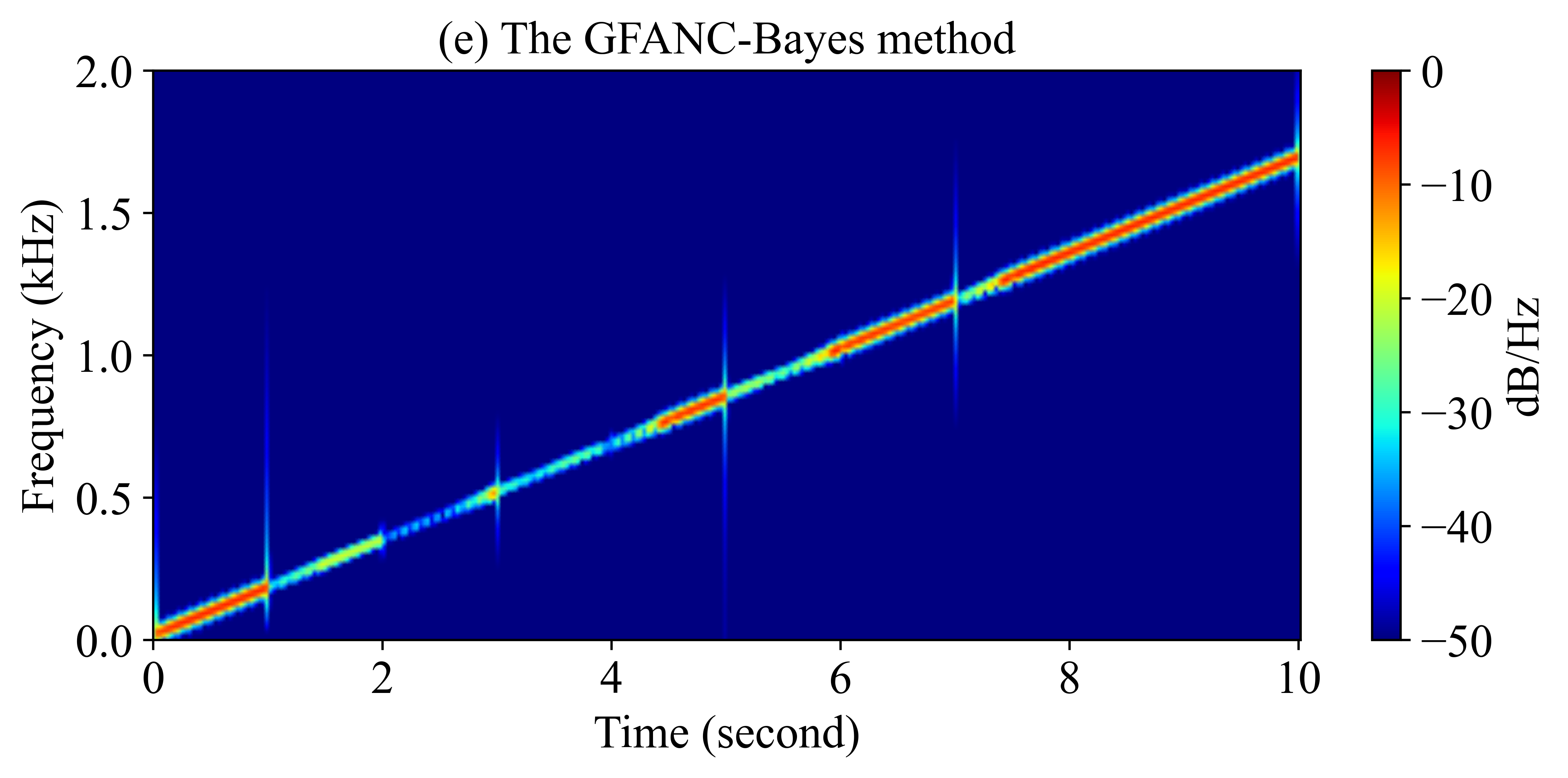}
\includegraphics[width=0.47\linewidth, height=4.5cm]{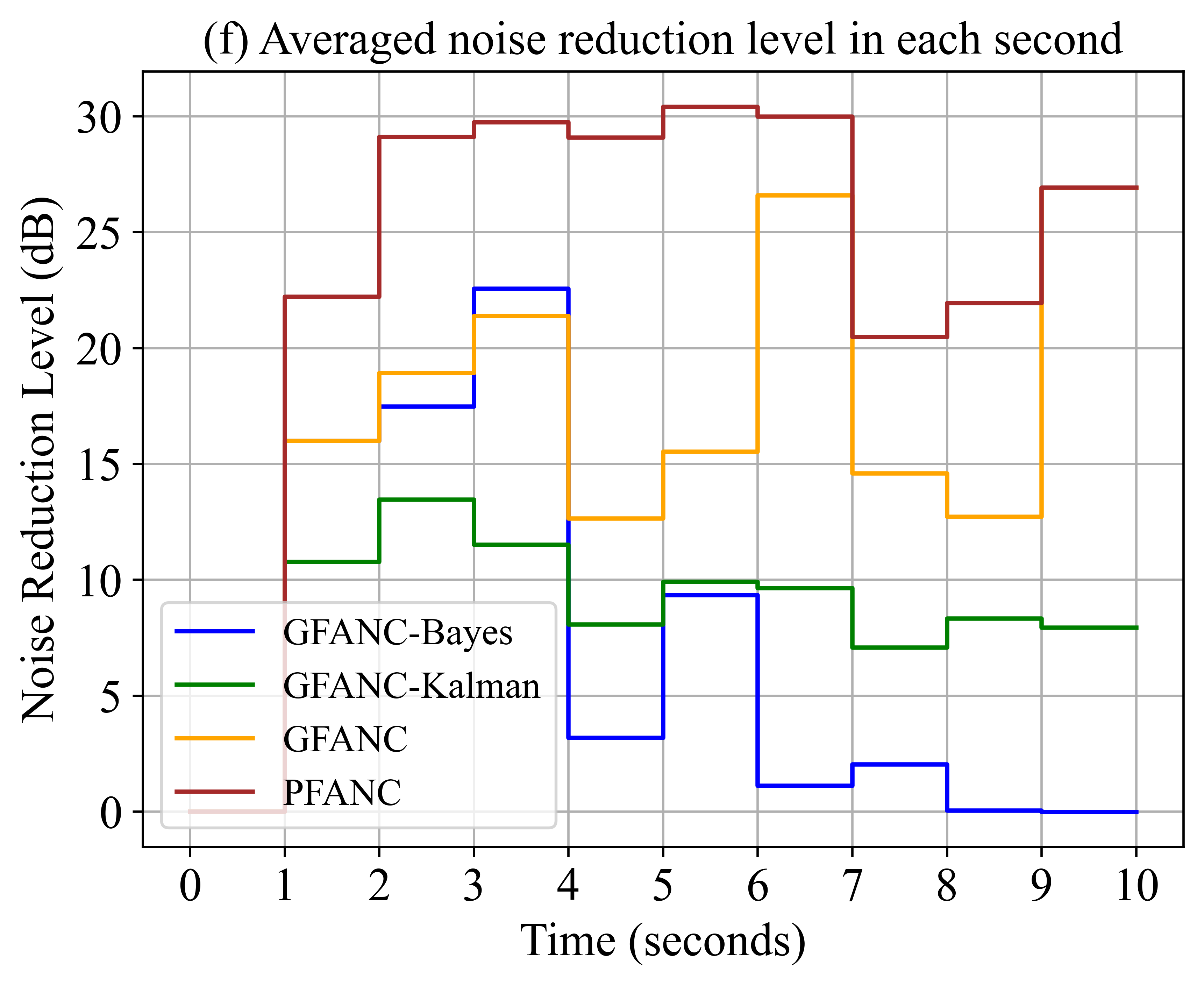}
\caption{Spectrograms of the error signal obtained (a) without ANC, and with (b) the PFANC method, (c) the GFANC method, (d) the GFANC-Kalman method, and (e) the GFANC-Bayes method, for the $20$–$1{,}700$~Hz linear chirp noise. (f) shows the averaged noise reduction level per second for the four ANC methods.}
\label{Fig Simulation linear chirp}
\end{figure}
%-------------------------------------------------------------------------

%------------------------------------------------------------------------
\begin{figure}[!t]
\centering
\includegraphics[width=0.48\linewidth, height=4.5cm]{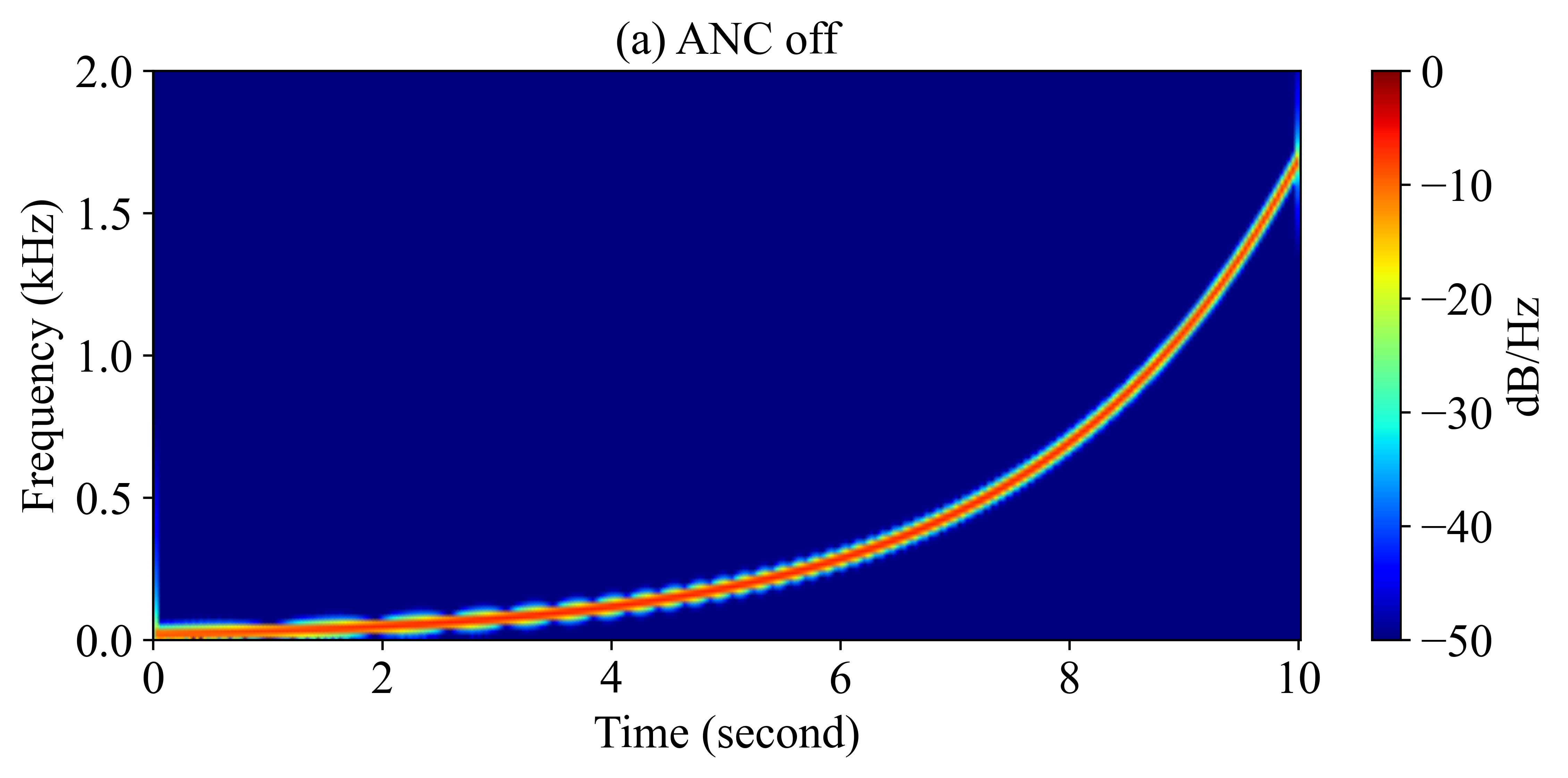}
\includegraphics[width=0.48\linewidth, height=4.5cm]{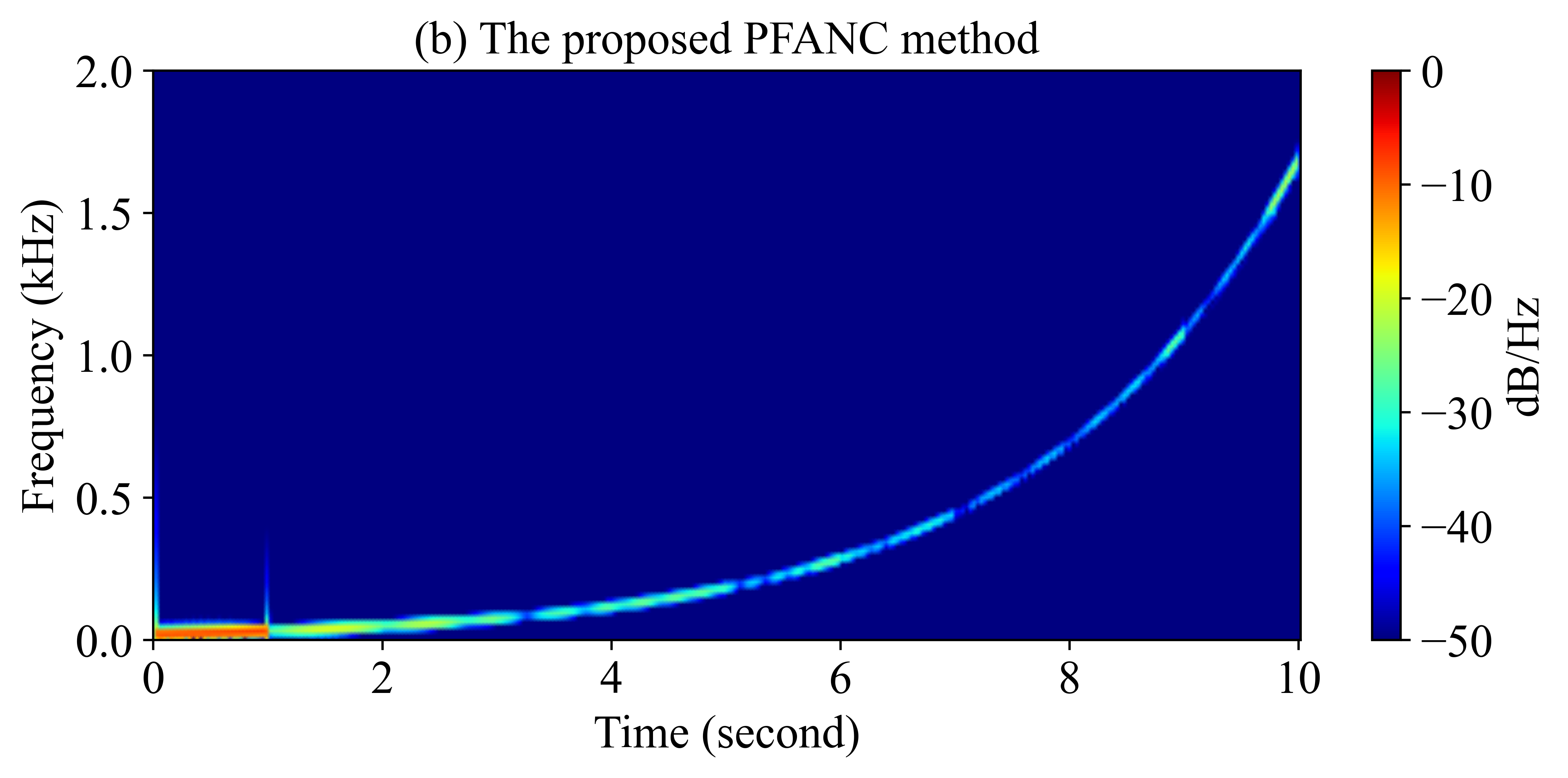}
\includegraphics[width=0.48\linewidth, height=4.5cm]{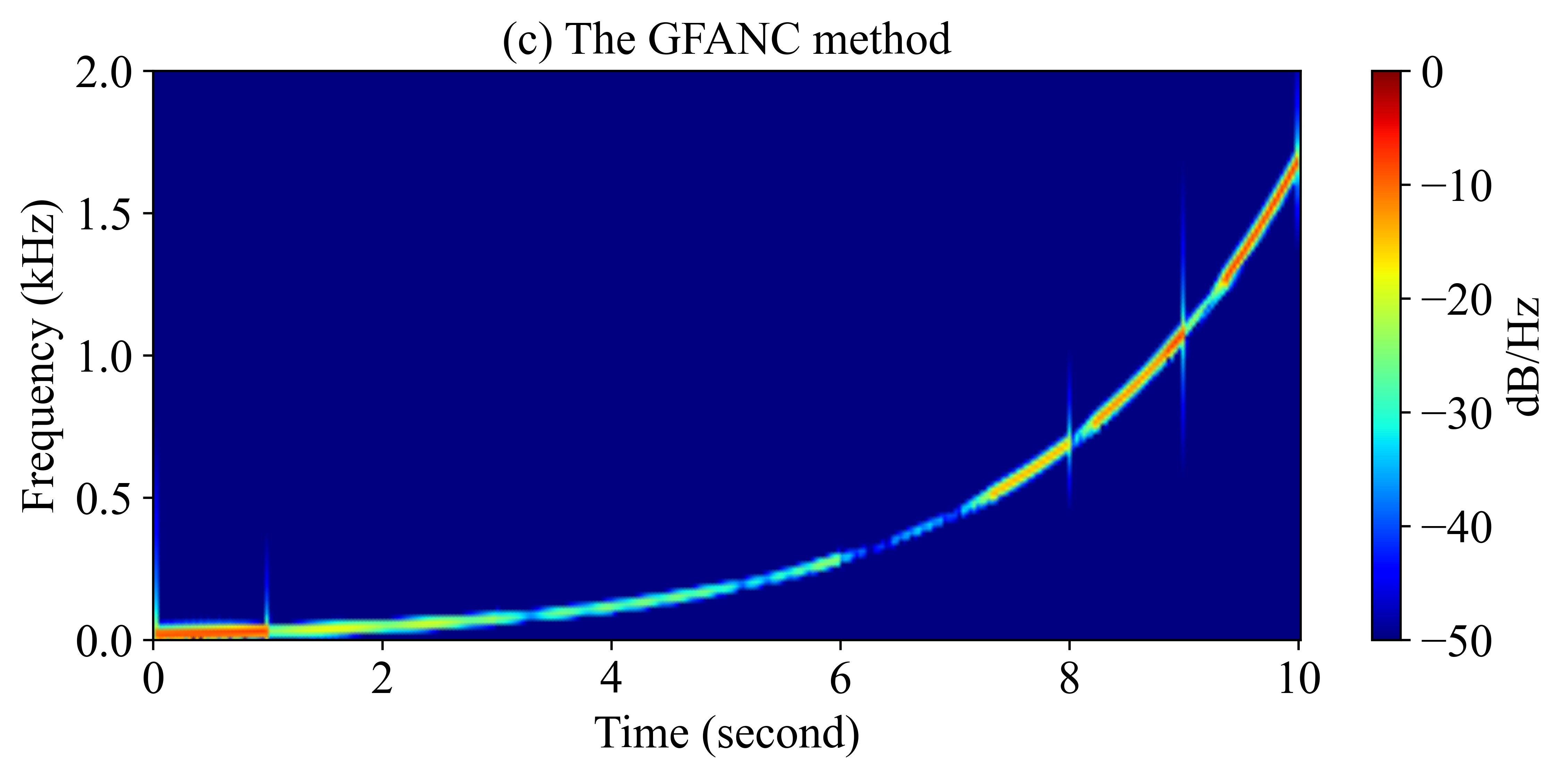}
\includegraphics[width=0.48\linewidth, height=4.5cm]{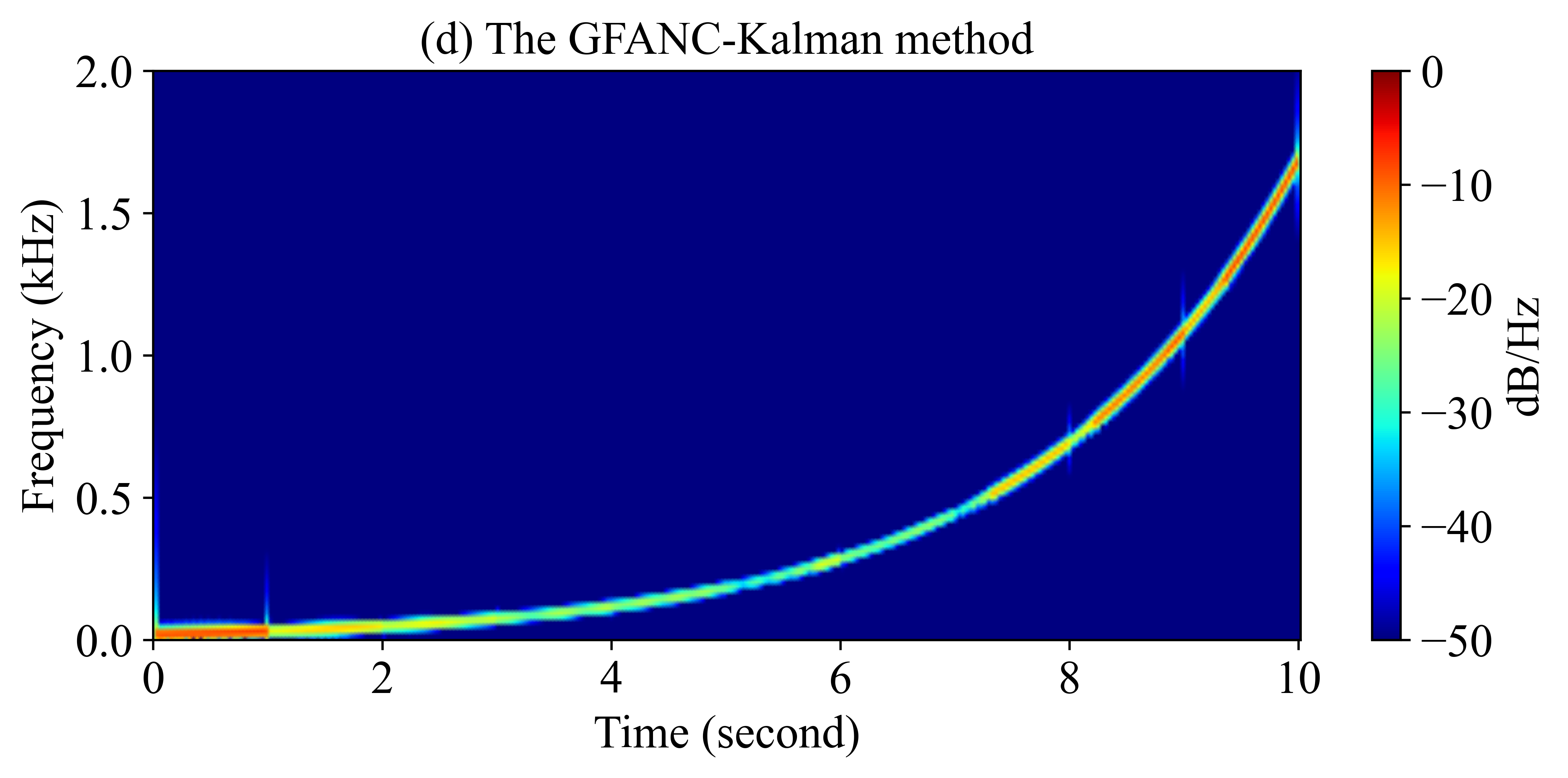}
\includegraphics[width=0.48\linewidth, height=4.5cm]{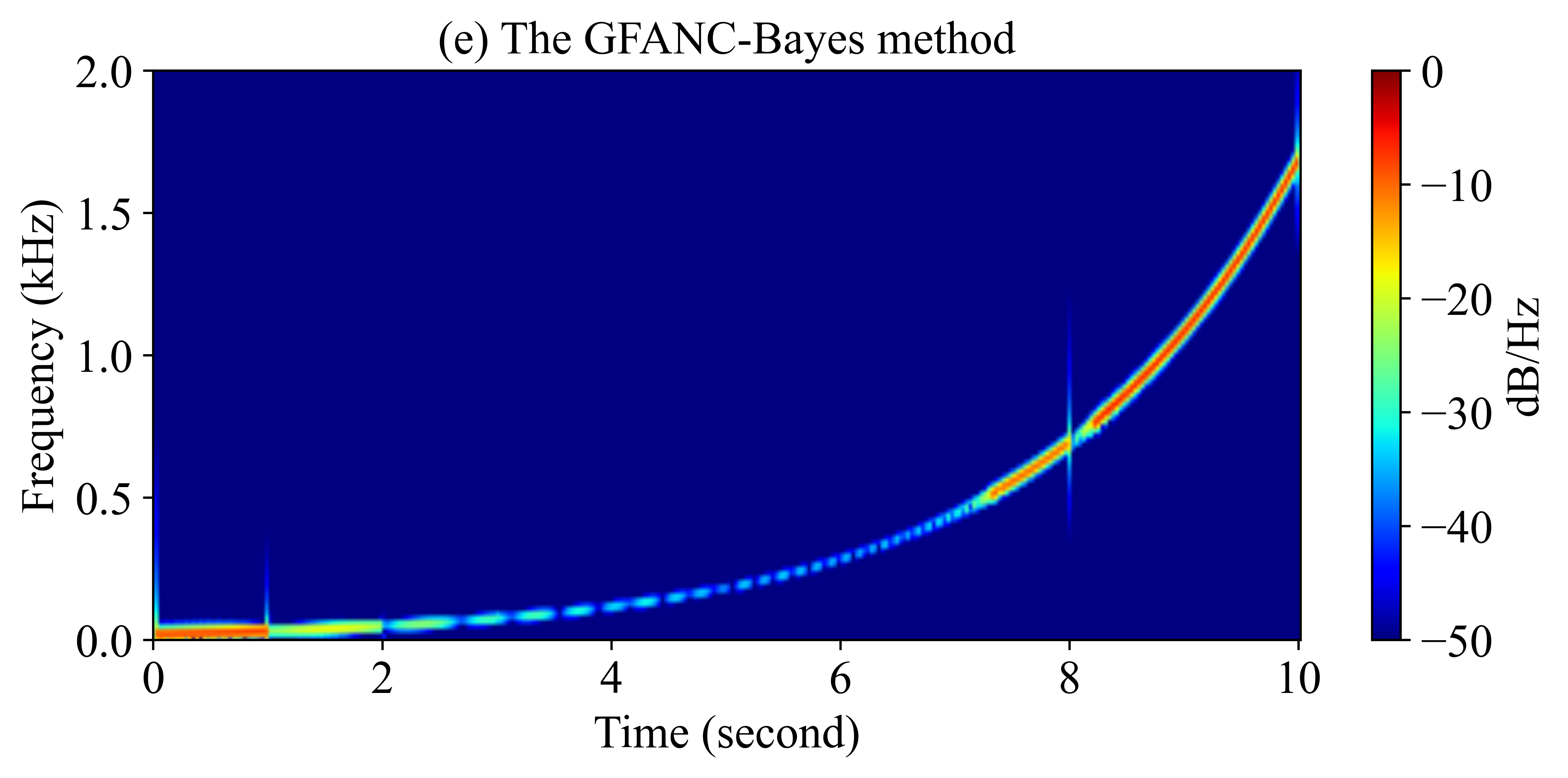}
\includegraphics[width=0.47\linewidth, height=4.5cm]{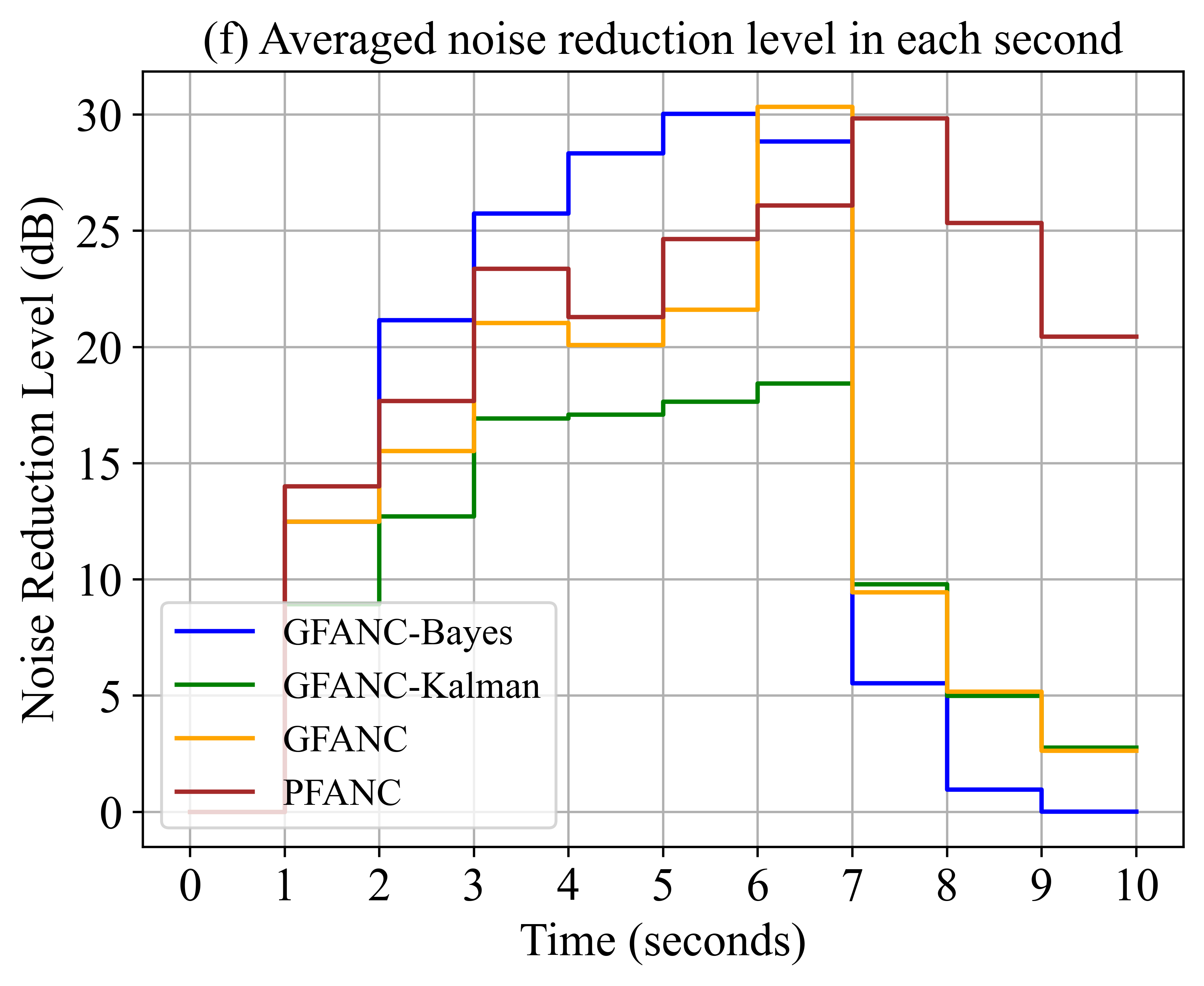}
\caption{Spectrograms of the error signal obtained (a) without ANC, and with (b) the PFANC method, (c) the GFANC method, (d) the GFANC-Kalman method, and (e) the GFANC-Bayes method, for the $20$–$1{,}700$~Hz logarithmic chirp noise. (f) shows the averaged noise reduction level per second for the four ANC methods.}
\label{Fig Simulation log chirp}
\end{figure}
%-------------------------------------------------------------------------

\subsection{Noise Control Performance of PFANC}\label{Noise Control Performance of PFANC}
In this section, the noise control performance of the PFANC method is evaluated using non-stationary noises, including linear and logarithmic chirp signals as well as real-world dynamic noises. Its performance is further compared with those of the GFANC, GFANC-Bayes, GFANC-Kalman, and FxLMS algorithms. The primary and secondary paths used in this section are synthetic acoustic paths modeled as pure acoustic delays. The next section assesses the transferability of the PFANC method in measured acoustic paths.

\subsubsection{Comparison with GFANC}
Spectrograms of the error signal and the averaged noise reduction level per second obtained by different ANC methods for the $20$–$1,700$~Hz linear chirp noise and the $20$–$1,700$~Hz logarithmic chirp noise are illustrated in Figure~\ref{Fig Simulation linear chirp} and Figure~\ref{Fig Simulation log chirp}, respectively. Noise reduction level (NR) in dB is defined as
\begin{equation}
\mathrm{NR}
=10 \log_{10} \frac{\sum_{n=1}^{l} d^{2}(n)}{\sum_{n=1}^{l} e^{2}(n)},
\end{equation}
where $l$ denotes the length of the signal segment. The averaged NR level per second is computed by setting $l = 16{,}000$ samples, corresponding to one second at a sampling rate of 16~kHz. In Figure~\ref{Fig Simulation linear chirp} and Figure~\ref{Fig Simulation log chirp}, the proposed PFANC method demonstrates clear superiority in controlling both the $20$–$1{,}700$~Hz linear chirp noise and the logarithmic chirp noise. For both signals, the PFANC method effectively suppresses the full frequency band after the first second. The lack of noise reduction during the first second is expected, as the CRNN requires multiple preceding noise frames to predict the upcoming-frame weight vector; thus, there is no previous frame before the first second.

%------------------------------------------------------------------------
\begin{figure}[!t]
\centering
\includegraphics[width=0.48\linewidth, height=4.5cm]{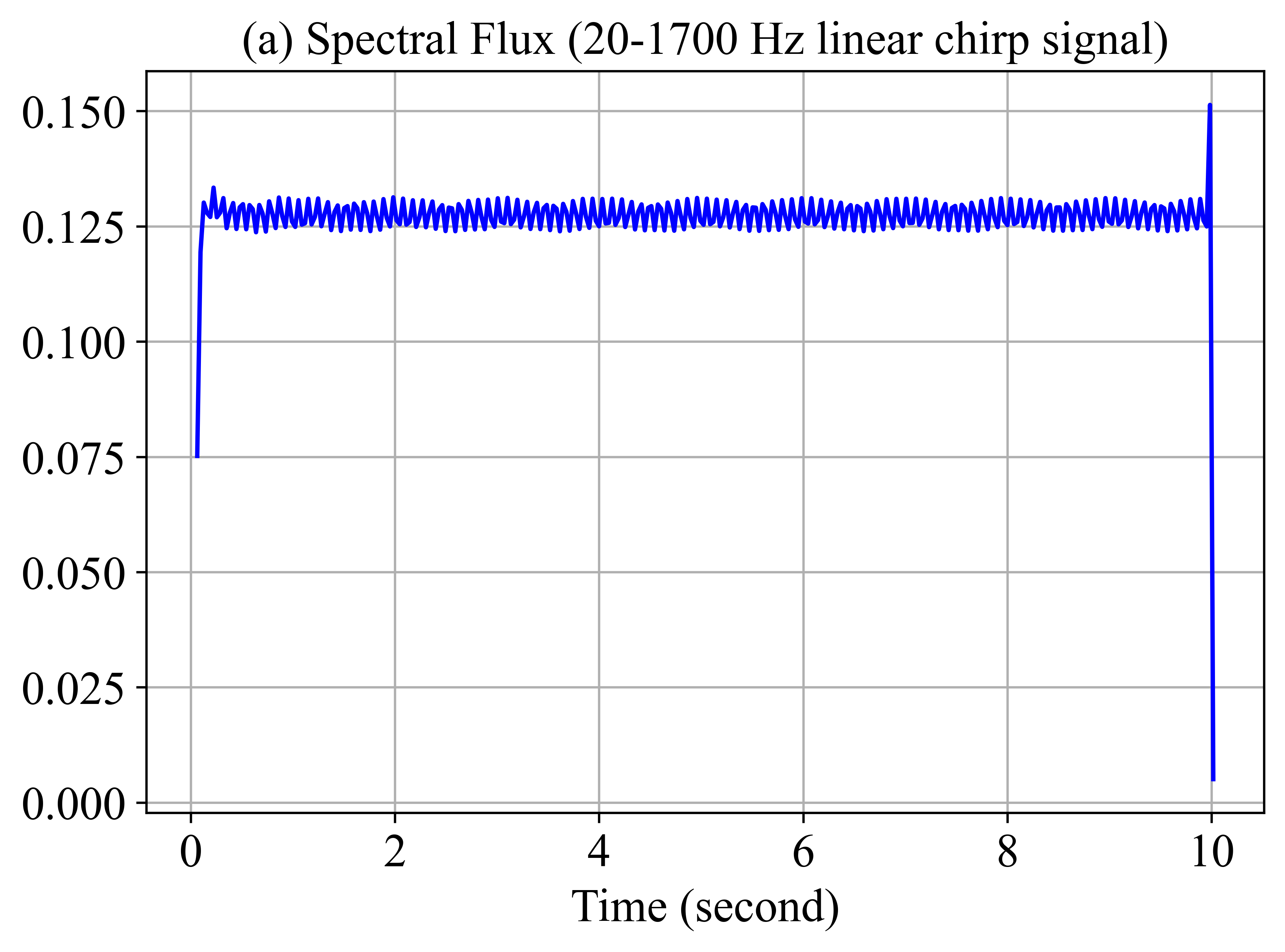}
\includegraphics[width=0.48\linewidth, height=4.5cm]{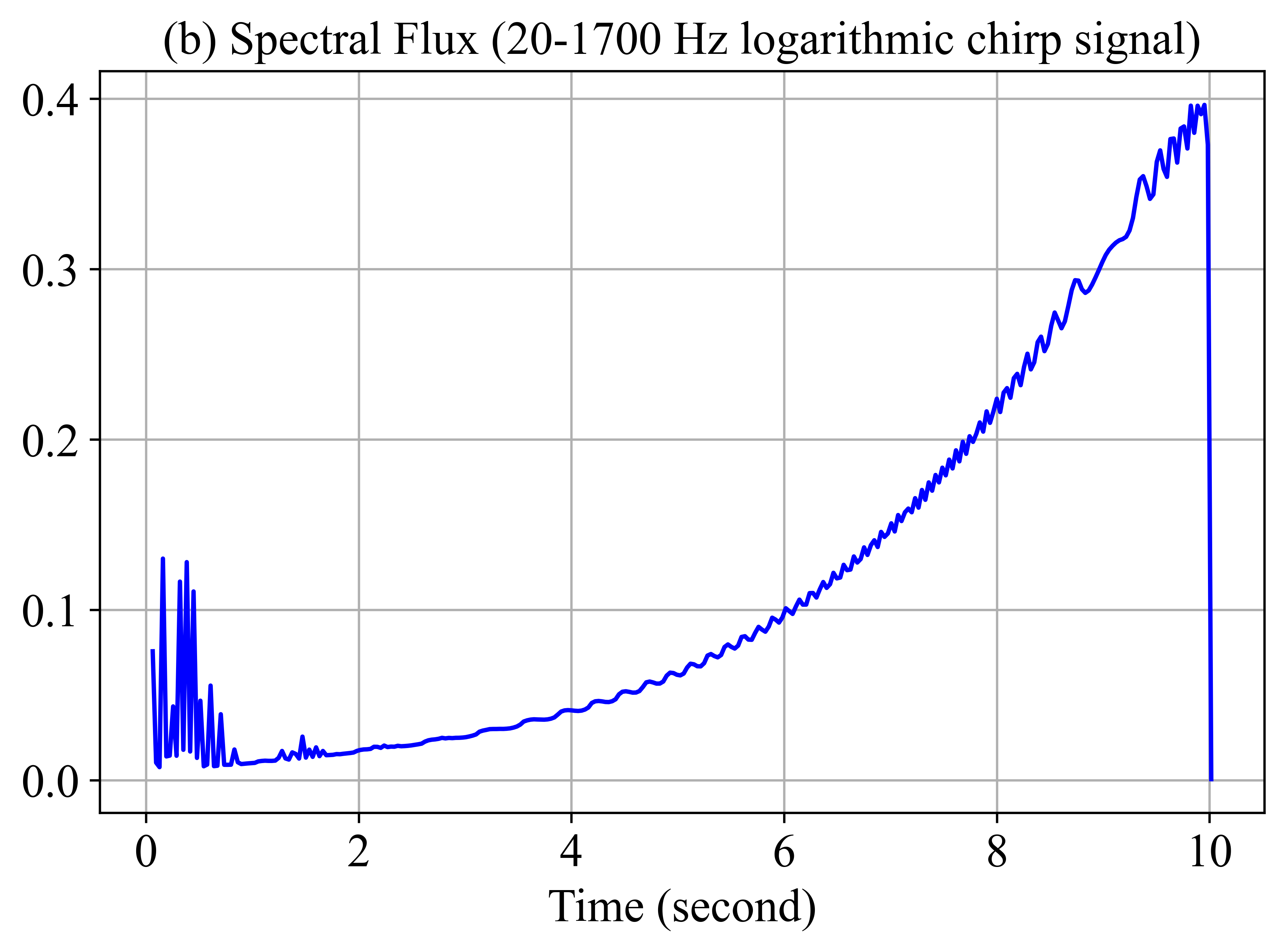}
\includegraphics[width=0.48\linewidth, height=4.5cm]{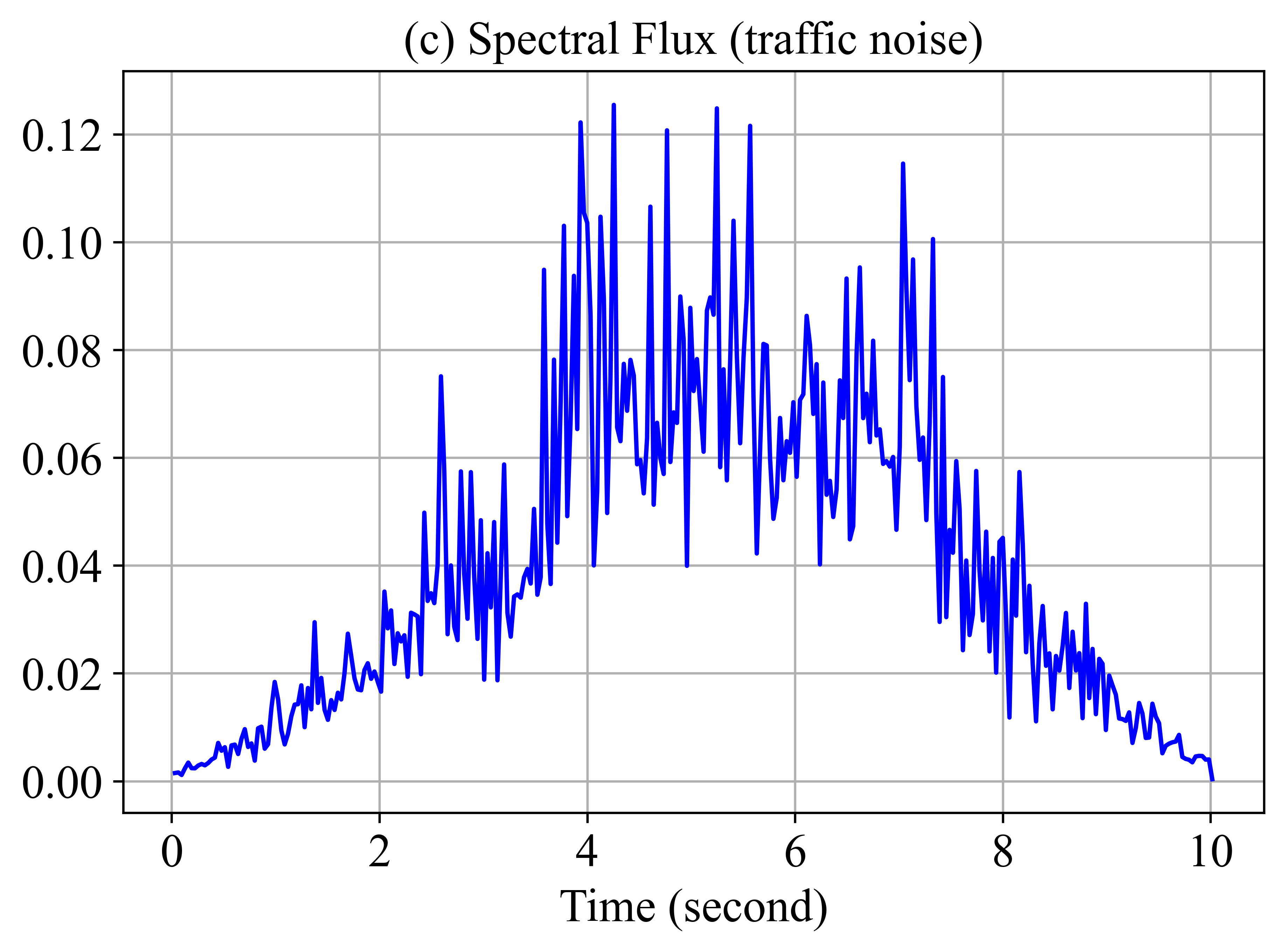}
\includegraphics[width=0.48\linewidth, height=4.5cm]{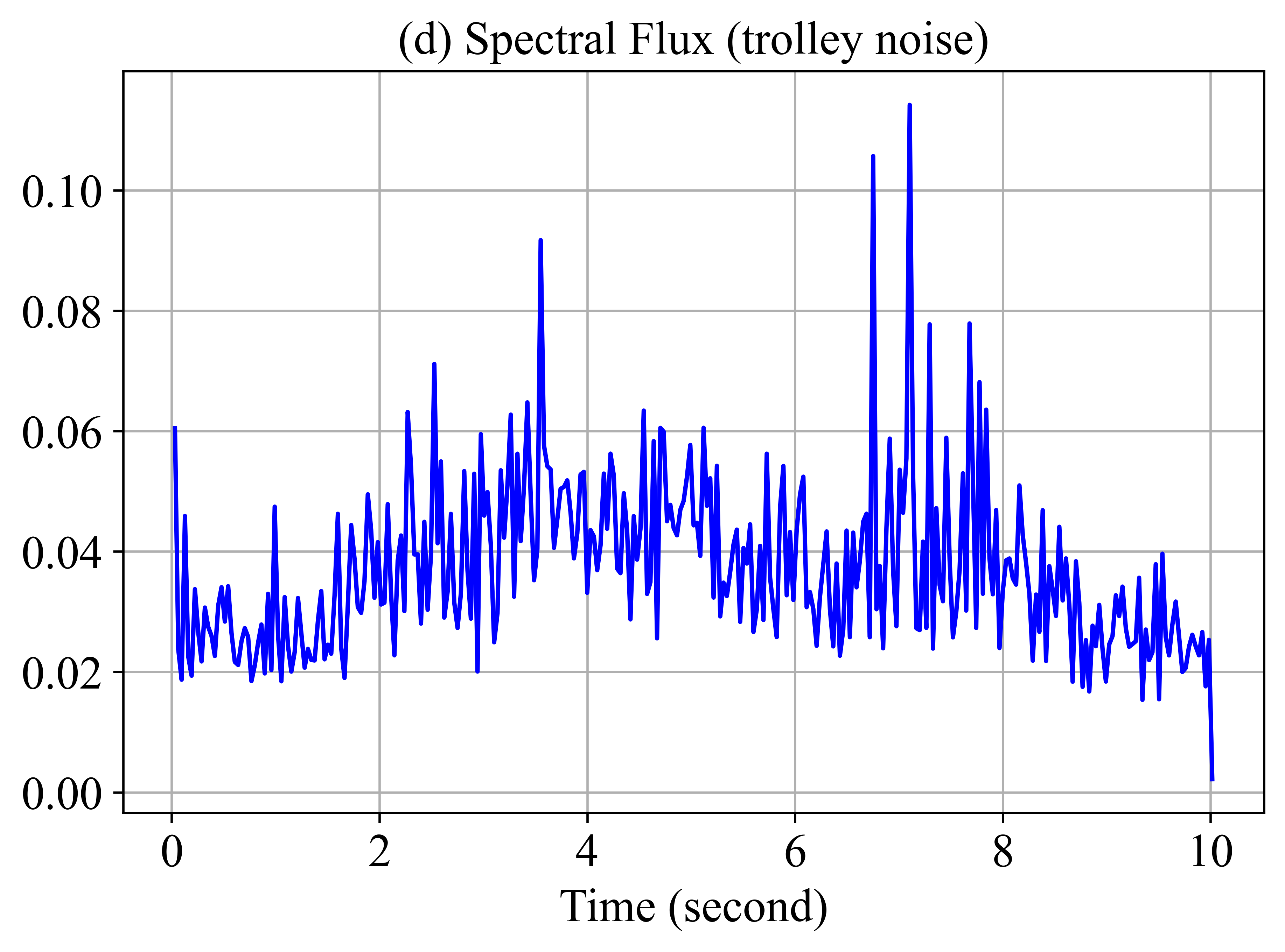}
\caption{Spectral flux of four types of noises: (a) $20$–$1{,}700$~Hz linear chirp noise, (b) $20$–$1{,}700$~Hz logarithmic chirp noise, (c) traffic noise, and (d) trolley noise.}
\label{Fig Spectral flux}
\end{figure}
%-------------------------------------------------------------------------

In contrast, the GFANC method exhibits limited capability in tracking the frequency evolution of chirp signals, particularly for the logarithmic chirp noise. Its NR level remains consistently lower than that of the PFANC method. The performance gap becomes especially pronounced during the $7-10$ second interval of the logarithmic chirp noise, where the instantaneous frequency rises rapidly. The rapid variation renders the GFANC method, which depends only on the current noise frame and lacks predictive capability, unable to track the accelerated spectral changes, resulting in poor noise attenuation. The performance comparisons between PFANC and GFANC validate the effectiveness of incorporating multiple historical frames and control filter prediction for attenuating rapidly varying noises.

\begin{figure}[!t]
\centering
\includegraphics[width=0.49\linewidth, height=4.5cm]{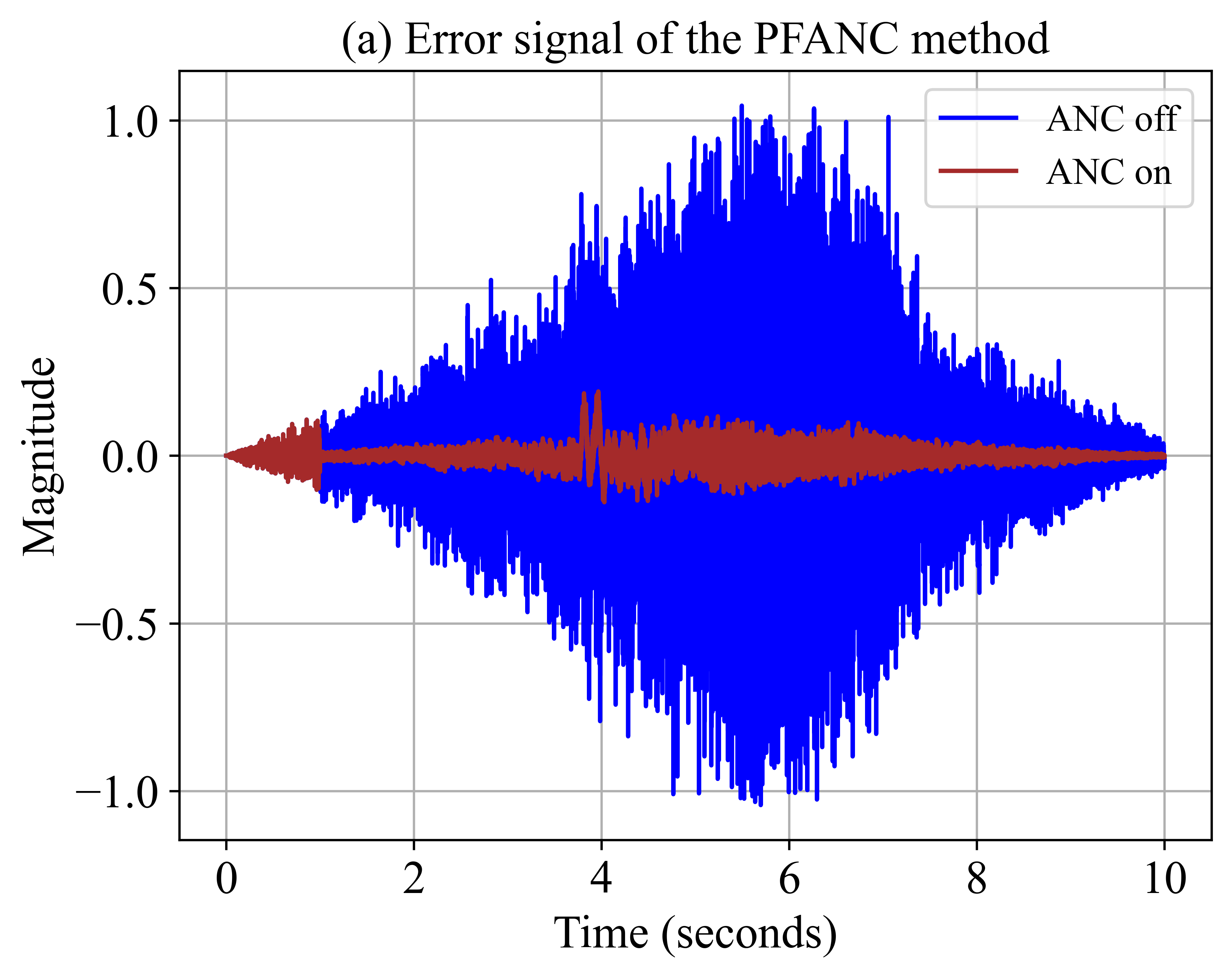}
\includegraphics[width=0.49\linewidth, height=4.5cm]{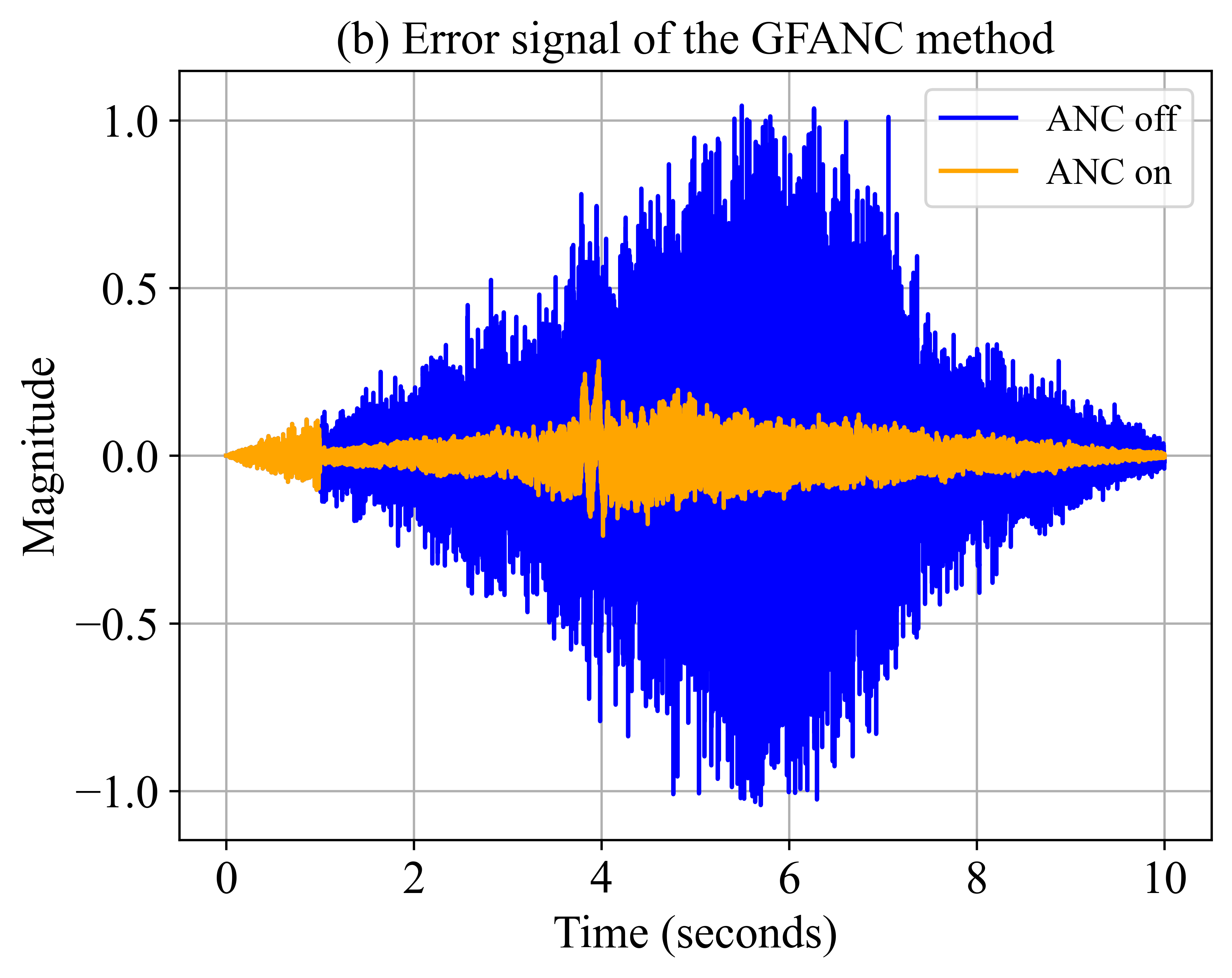}
\includegraphics[width=0.49\linewidth, height=4.5cm]{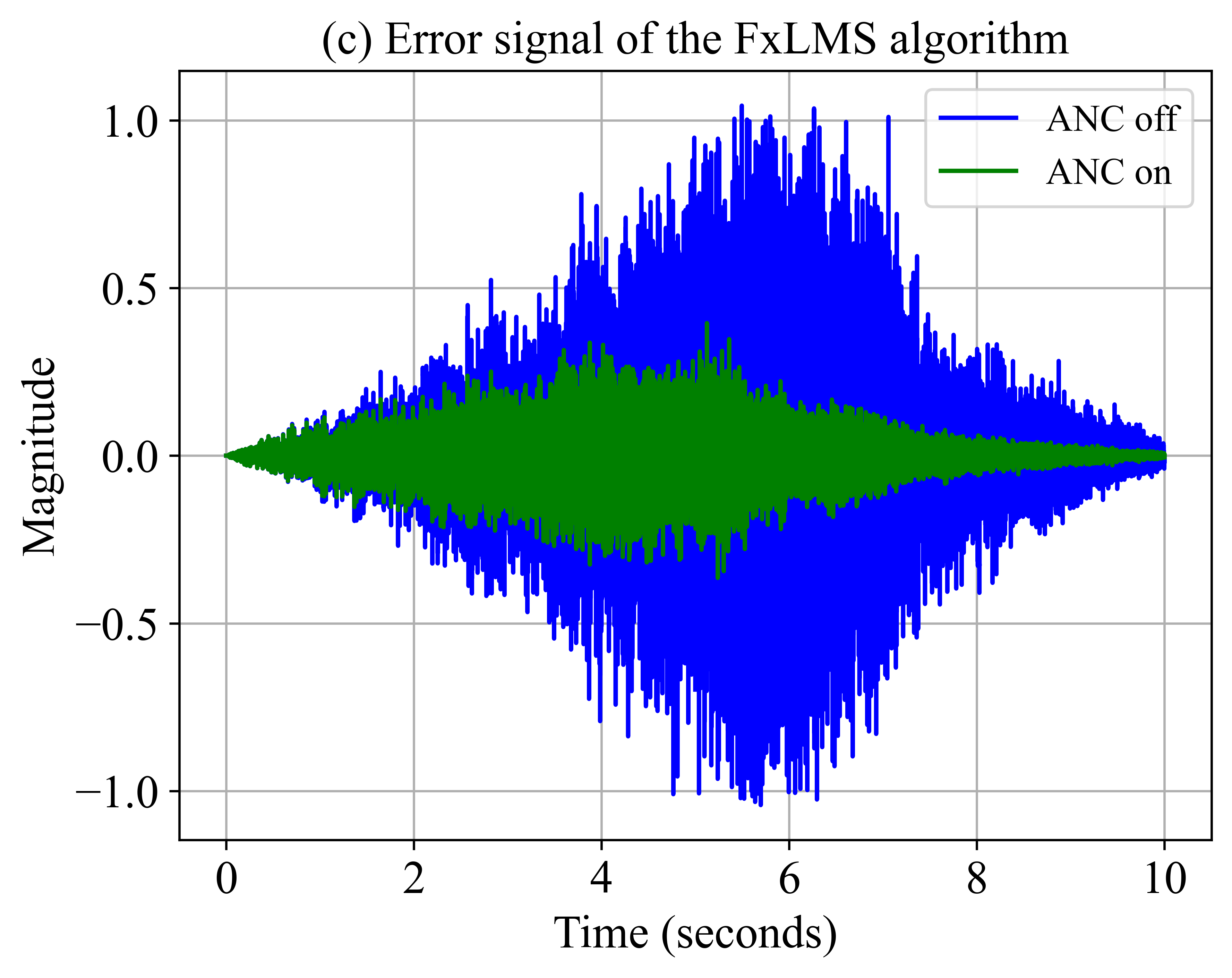}
\includegraphics[width=0.49\linewidth, height=4.5cm]{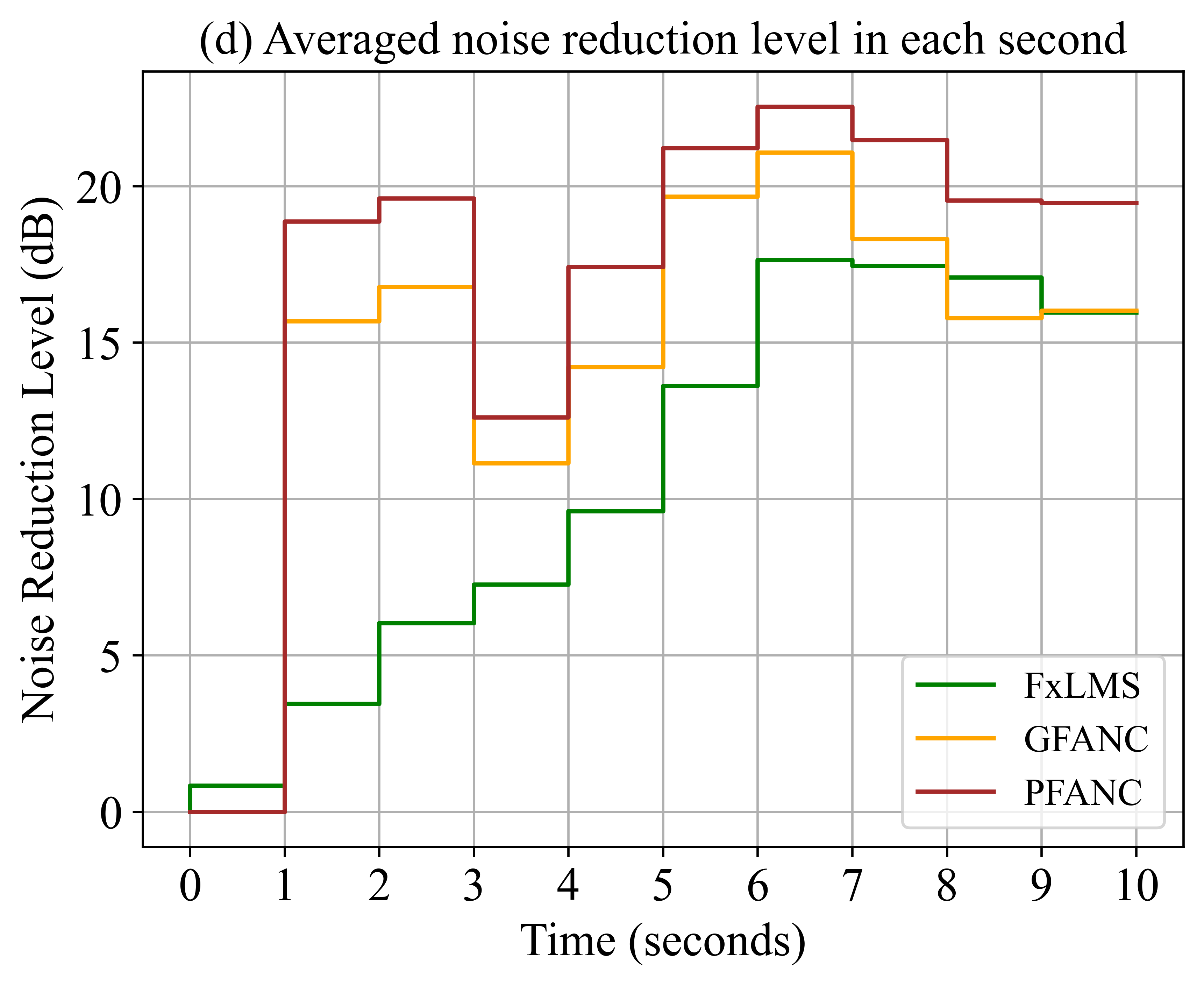}
\caption{Error signals (a)–(c) and the averaged noise reduction level per second (d) obtained with the PFANC method, the GFANC method, and the FxLMS algorithm for the traffic noise.}
\label{Fig Simulation traffic}
\end{figure}

\begin{figure}[!t]
\centering
\includegraphics[width=0.49\linewidth, height=4.5cm]{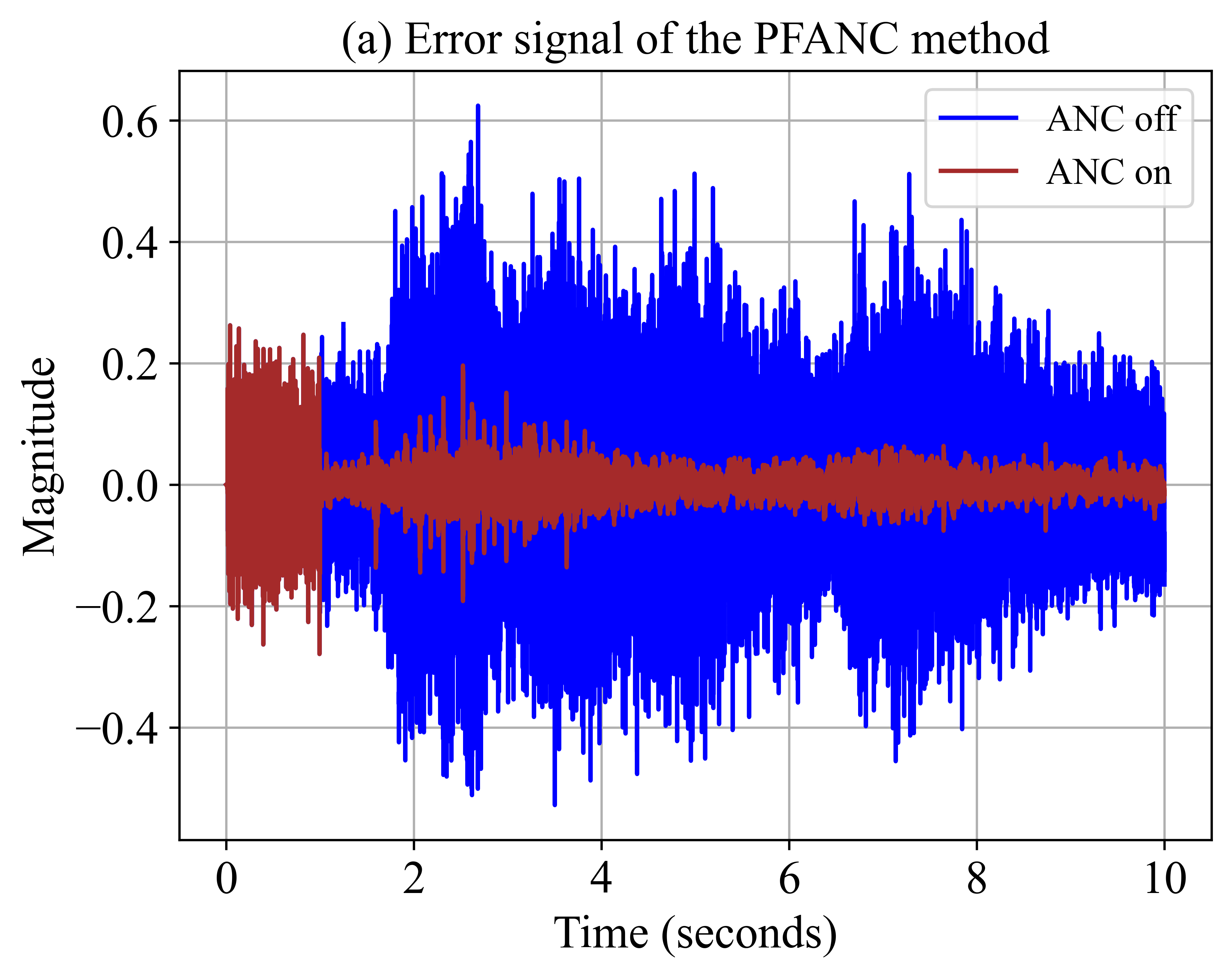}
\includegraphics[width=0.49\linewidth, height=4.5cm]{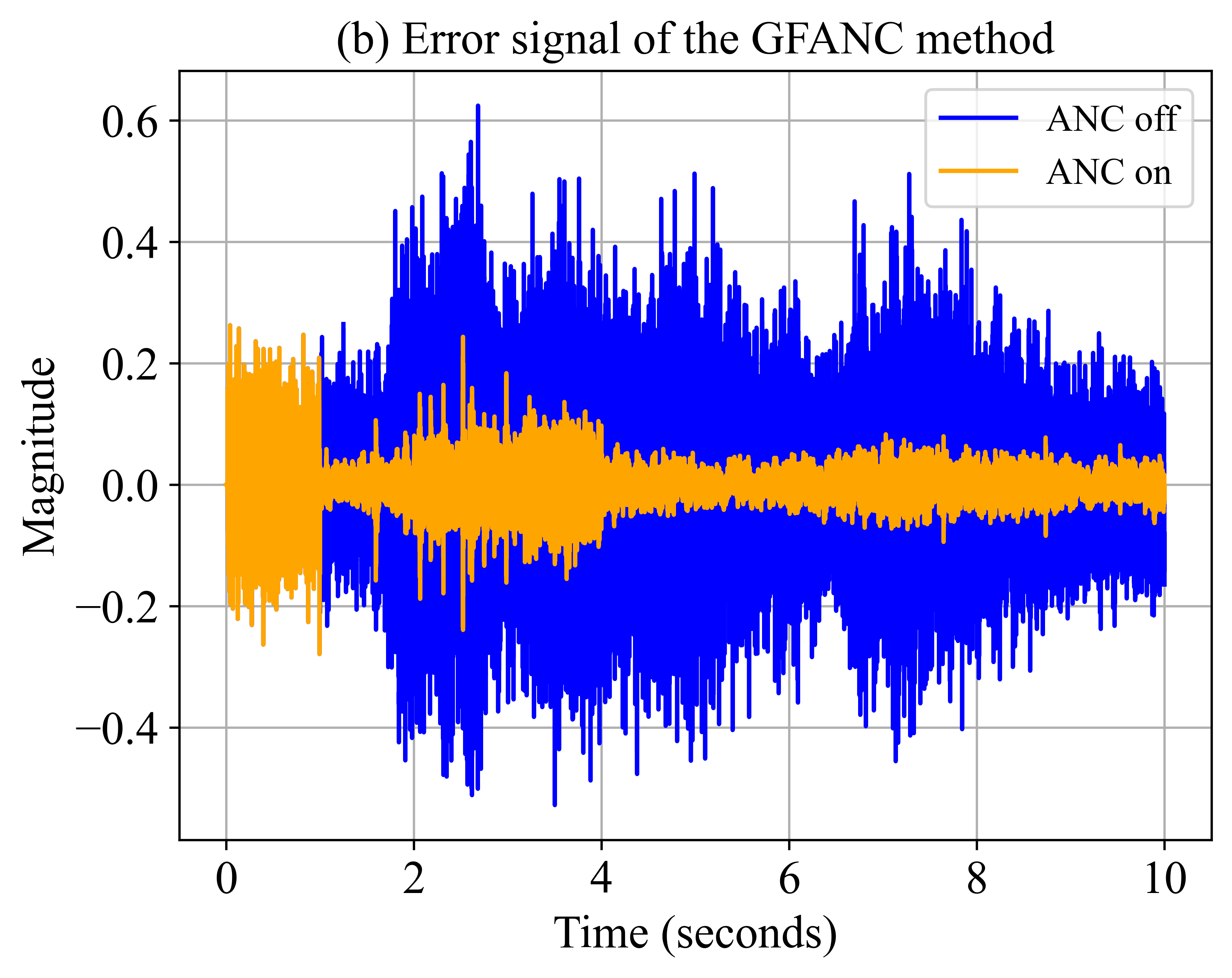}
\includegraphics[width=0.49\linewidth, height=4.5cm]{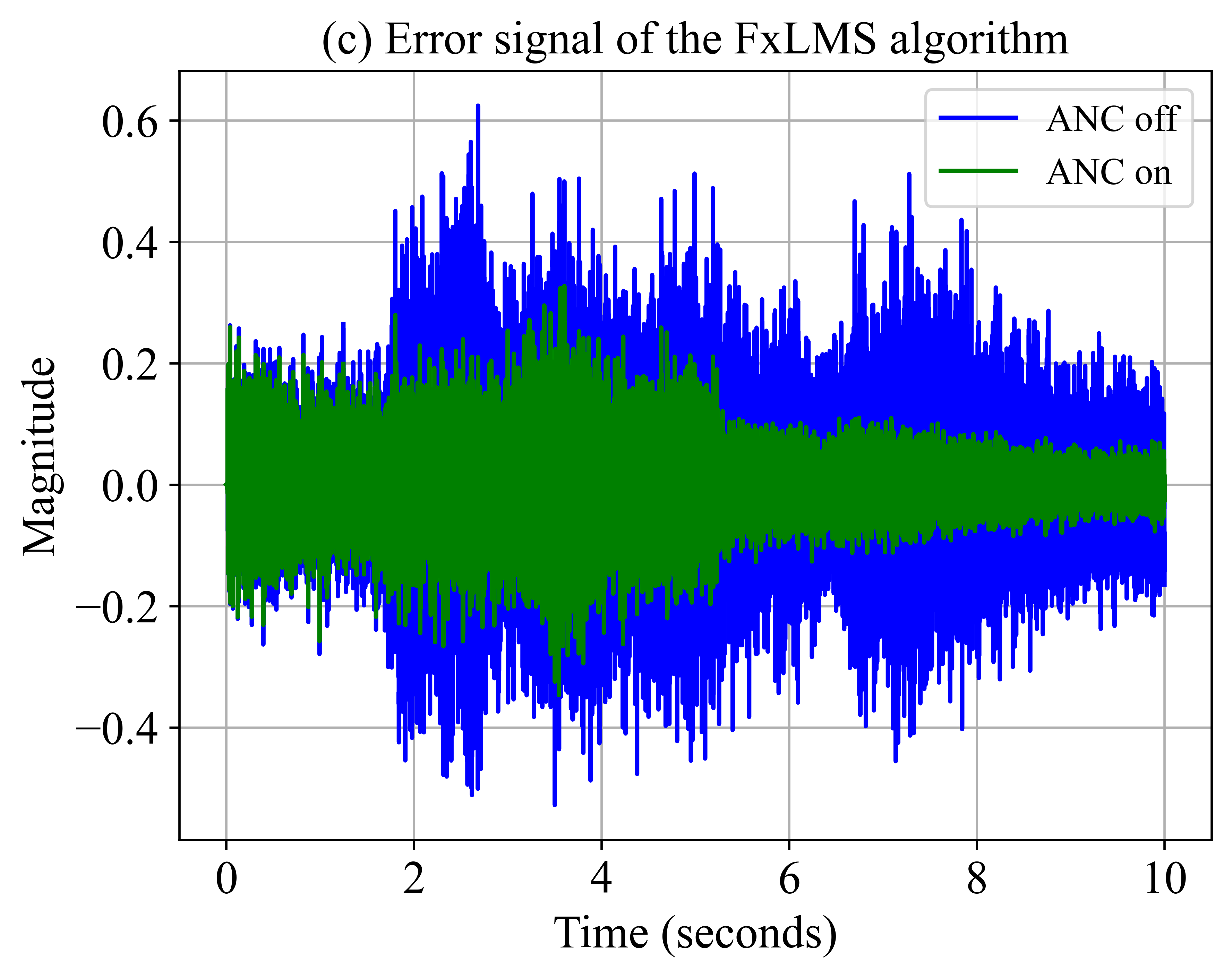}
\includegraphics[width=0.49\linewidth, height=4.5cm]{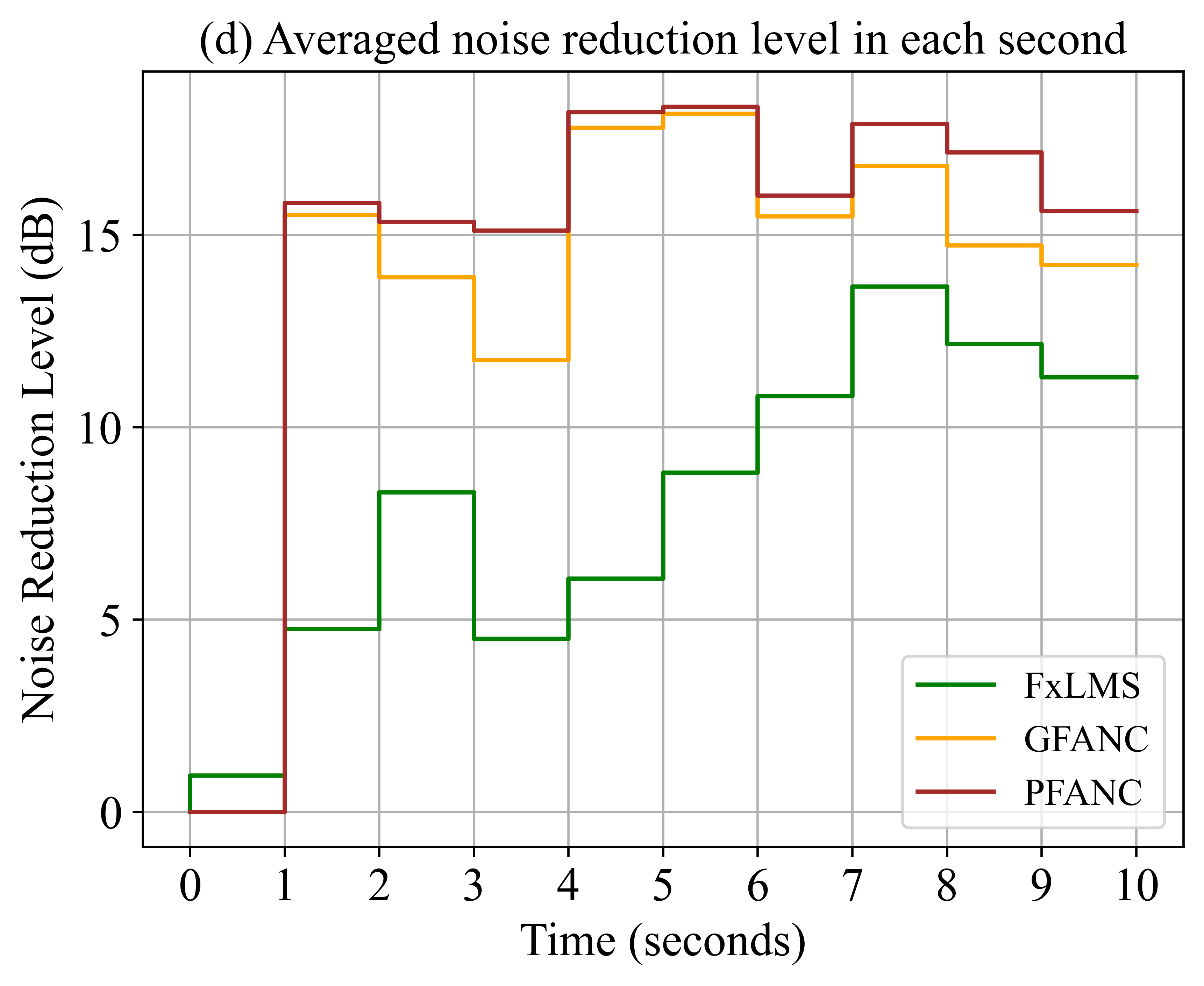}
\caption{Error signals (a)–(c) and the averaged noise reduction level per second (d) obtained with the PFANC method, the GFANC method, and the FxLMS algorithm for the trolley noise.}
\label{Fig Simulation trolley}
\end{figure}

Furthermore, the real-world noises (traffic and trolley) exhibit milder temporal variations in their spectral characteristics than the dynamic chirp signals, as indicated by their lower spectral-flux values in Figure~\ref{Fig Spectral flux}. Spectral flux~\cite{spectralflux} quantifies the frame-to-frame variation of the short-time spectrum and therefore characterizes the temporal variation of a noise signal's spectral characteristics, with higher values indicating more pronounced spectral changes across consecutive frames. As shown in Figure~\ref{Fig Simulation traffic} and Figure~\ref{Fig Simulation trolley}, the lower spectral-flux values of the traffic and trolley noises are accompanied by a smaller performance difference between PFANC and GFANC compared with the chirp cases. Even under such conditions, PFANC consistently achieves higher NR levels, typically by approximately $3$ to $4$ dB over several seconds, demonstrating its capability to track moderate temporal variations in the spectral characteristics of real-world noises.

%------------------------------------------------------------------------
%-------------------------------------------------------------------------

%------------------------------------------------------------------------
%-------------------------------------------------------------------------

\subsubsection{Comparison with GFANC-Bayes and GFANC-Kalman}
As introduced in Section~\ref{From Single-Frame to Multi-Frame Temporal Modeling}, the GFANC-Bayes and GFANC-Kalman methods \cite{LuoBayes,LuoKalman} extend GFANC by filtering the newly generated control filter with the filter used in the previous frame, thereby exploiting temporal correlations between adjacent noise frames. However, similar to GFANC, these approaches lack predictive capability. When the previously applied control filter becomes mismatched to a rapidly changing noise environment, the recursive filtering operation propagates and even amplifies the mismatch instead of correcting it. Consequently, estimation errors accumulate over time, leading to degraded performance under highly non-stationary conditions. This behavior is clearly observed in the linear and logarithmic chirp noises shown in Figure~\ref{Fig Simulation linear chirp} and Figure~\ref{Fig Simulation log chirp}, where both GFANC-Bayes and GFANC-Kalman exhibit limited effectiveness, particularly in the later stages of the chirp signals, providing only approximately $5$ to $7$ dB of noise reduction. In contrast, the PFANC method continues to attenuate the chirp signals throughout the sequence, with most seconds achieving more than $20$ dB of reduction. These results suggest that the smoothing-based GFANC approaches, GFANC-Bayes and GFANC-Kalman, struggle to track dynamic spectral changes due to their lack of predictive capability, whereas the proposed PFANC method can effectively anticipate and track these variations.

\subsubsection{Comparison with the FxLMS algorithm}
In this subsection, the proposed PFANC method is compared with the traditional adaptive algorithm, the FxLMS algorithm, in attenuating the real noises (traffic and trolley). The performance comparisons of the proposed PFANC method and the FxLMS algorithm on the traffic and trolley noises are illustrated in Figure~\ref{Fig Simulation traffic} and Figure~\ref{Fig Simulation trolley}, respectively. The step size of the FxLMS algorithm is set to $0.0001$. As shown in Figure~\ref{Fig Simulation traffic} and Figure~\ref{Fig Simulation trolley}, the PFANC method provides a quick response to the real noises, whereas the FxLMS algorithm still requires some time to converge. Specifically, for the traffic noise, the PFANC method achieves an NR value of more than $15$ dB after the first second, while the FxLMS algorithm requires more than $6$ seconds to converge and reach a similar NR value. For the trolley noise, the FxLMS algorithm does not reach the steady state throughout the entire noise length due to its slow convergence speed. For real dynamic noises, the PFANC method achieves a significantly faster response than the FxLMS algorithm by directly generating the control filter from pre-trained sub control filters, thereby avoiding the iterative convergence process required by adaptive filtering.

\subsection{Transferability of PFANC}
The previous section has evaluated the effectiveness of the proposed PFANC method on synthetic acoustic paths. This section assesses the transferability of the proposed PFANC method across different measured acoustic paths. We define two new ANC systems, System-A and System-B, using acoustic paths measured from the vent of a noise chamber and from an ANC window, respectively. The magnitude and phase responses of the measured acoustic paths in ANC System-A and System-B are presented in \ref{Appendix Acoustic Paths}, and they differ considerably from the synthetic ones, which are modelled as pure acoustic delays.

%------------------------------------------------------------------------------
\begin{table}[!t]
\centering
\renewcommand\arraystretch{1.1}
\caption{Overall NR values (in dB) of the PFANC and GFANC methods in different ANC systems}
\label{Table NR different acoustic paths}
\resizebox{0.7\textwidth}{!}{
\begin{tabular}{|c|c|c|c|c|}
\hline
\multirow{2}{*}{Algorithm} 
  & \multicolumn{2}{c|}{Linear chirp signal} 
  & \multicolumn{2}{c|}{Logarithmic chirp signal} \\
\cline{2-5}
  & \multicolumn{1}{c|}{System-A}
  & \multicolumn{1}{c|}{System-B} 
  & \multicolumn{1}{c|}{System-A}
  & \multicolumn{1}{c|}{System-B}\\
\hline
PFANC & 20.73 & 11.46 & 18.50 & 16.44 \\
GFANC & 14.62 & 10.49 & 7.09  & 15.74 \\
\hline
\end{tabular}
}
\begin{tablenotes}\scriptsize
\item[]\textsuperscript{*}\;$20$–$1{,}700$~Hz linear and logarithmic chirp noise.
\end{tablenotes}
\end{table}
%--------------------------------------------------------------------------------

When transferring the PFANC method from synthetic to real acoustic paths, the CRNN model trained on synthetic acoustic paths is applied directly without retraining. Meanwhile, the pre-trained broadband control filter is updated based on the real acoustic paths and then decomposed into sub control filters as described in Section~\ref{Sub Control Filters}. The sub control filters are system-specific and depend on the acoustic paths; however, the outputs of the CRNN rely only on the input noises. This setup allows the PFANC method to be easily transferred across different application scenarios.

%------------------------------------------------------------------------
\begin{figure}[!t]
\centering
\includegraphics[width=0.325\linewidth, height=3.5cm]{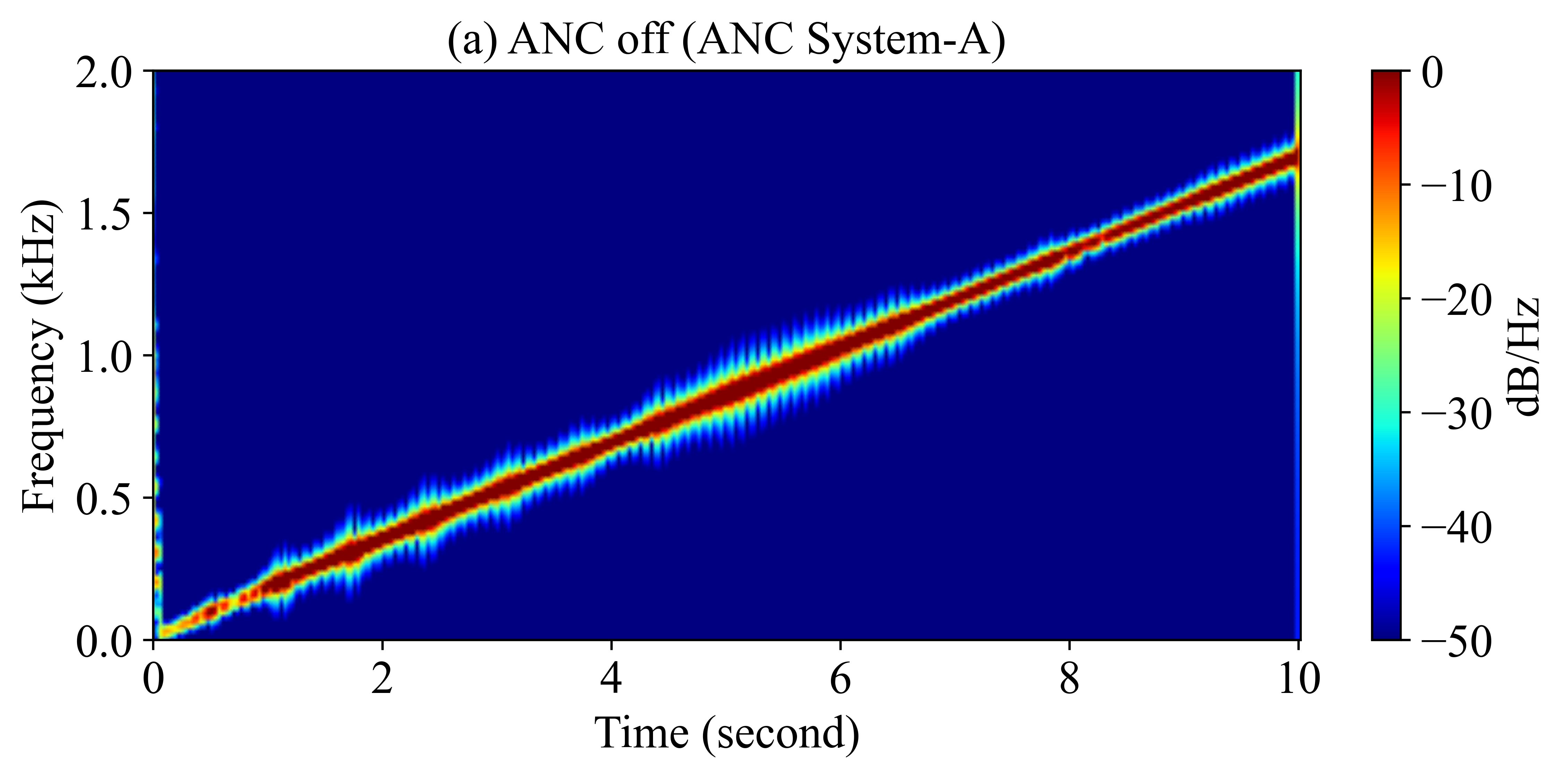}
\includegraphics[width=0.325\linewidth, height=3.5cm]{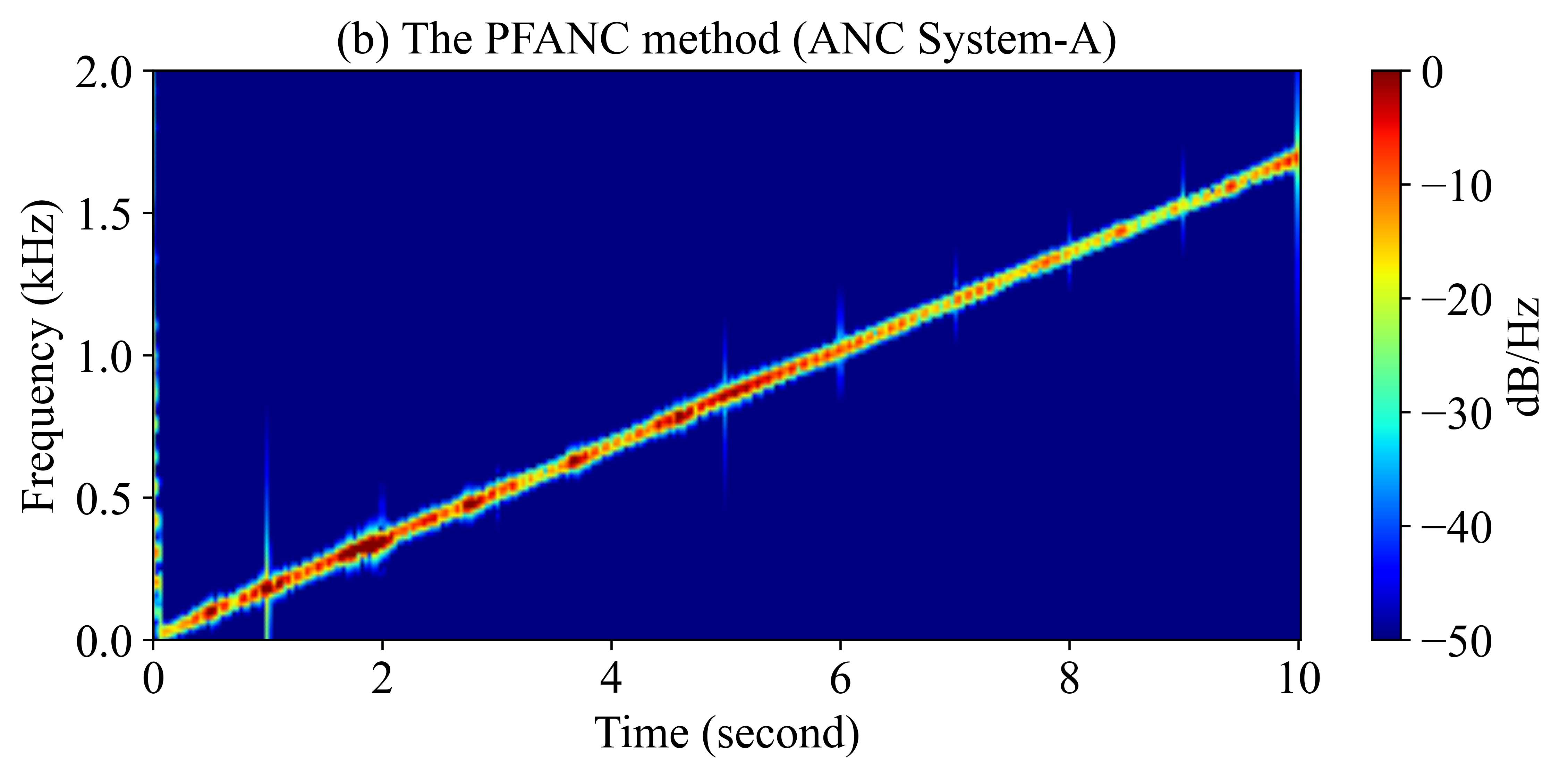}
\includegraphics[width=0.325\linewidth, height=3.5cm]{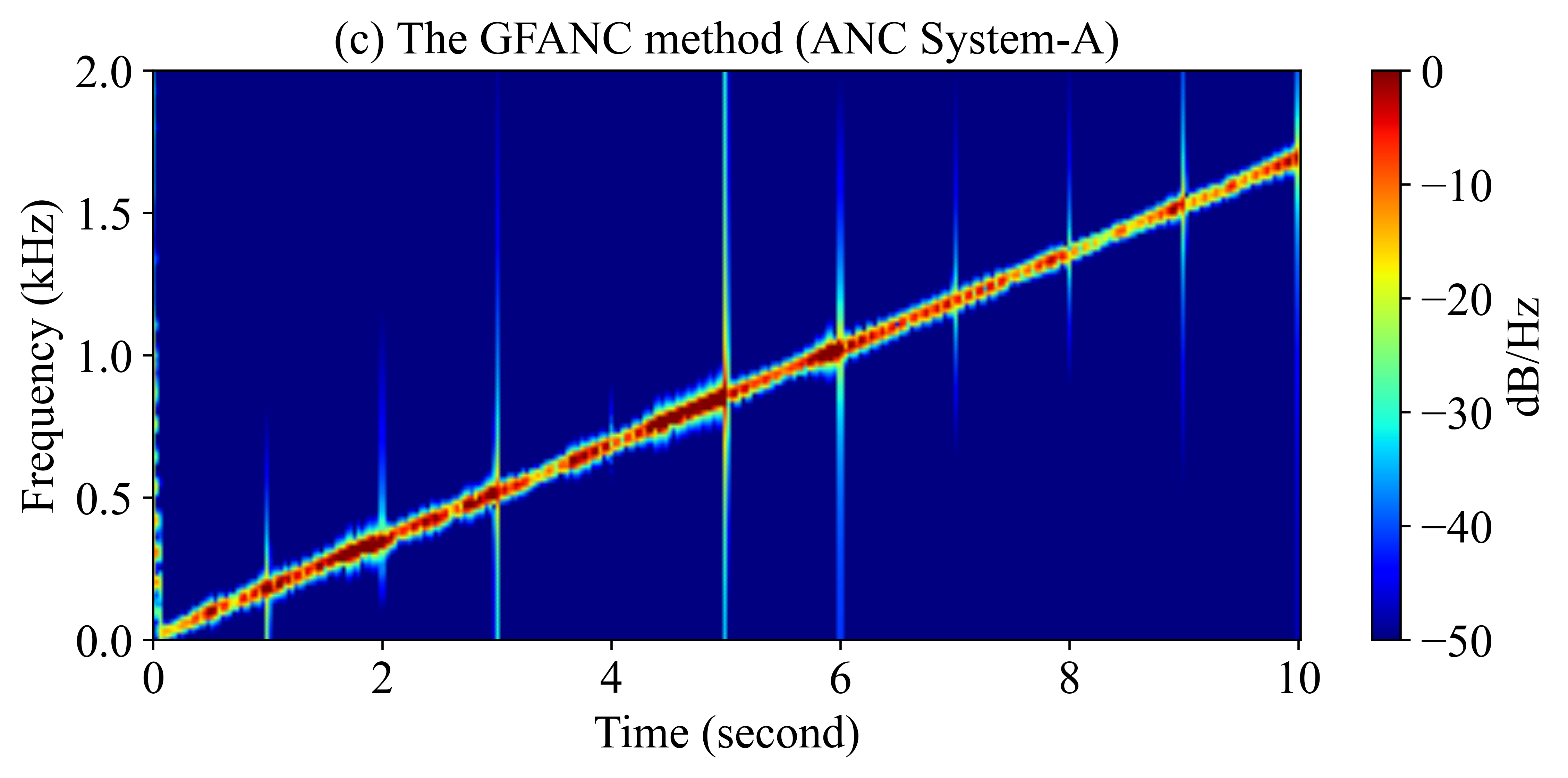}
\includegraphics[width=0.325\linewidth, height=3.5cm]{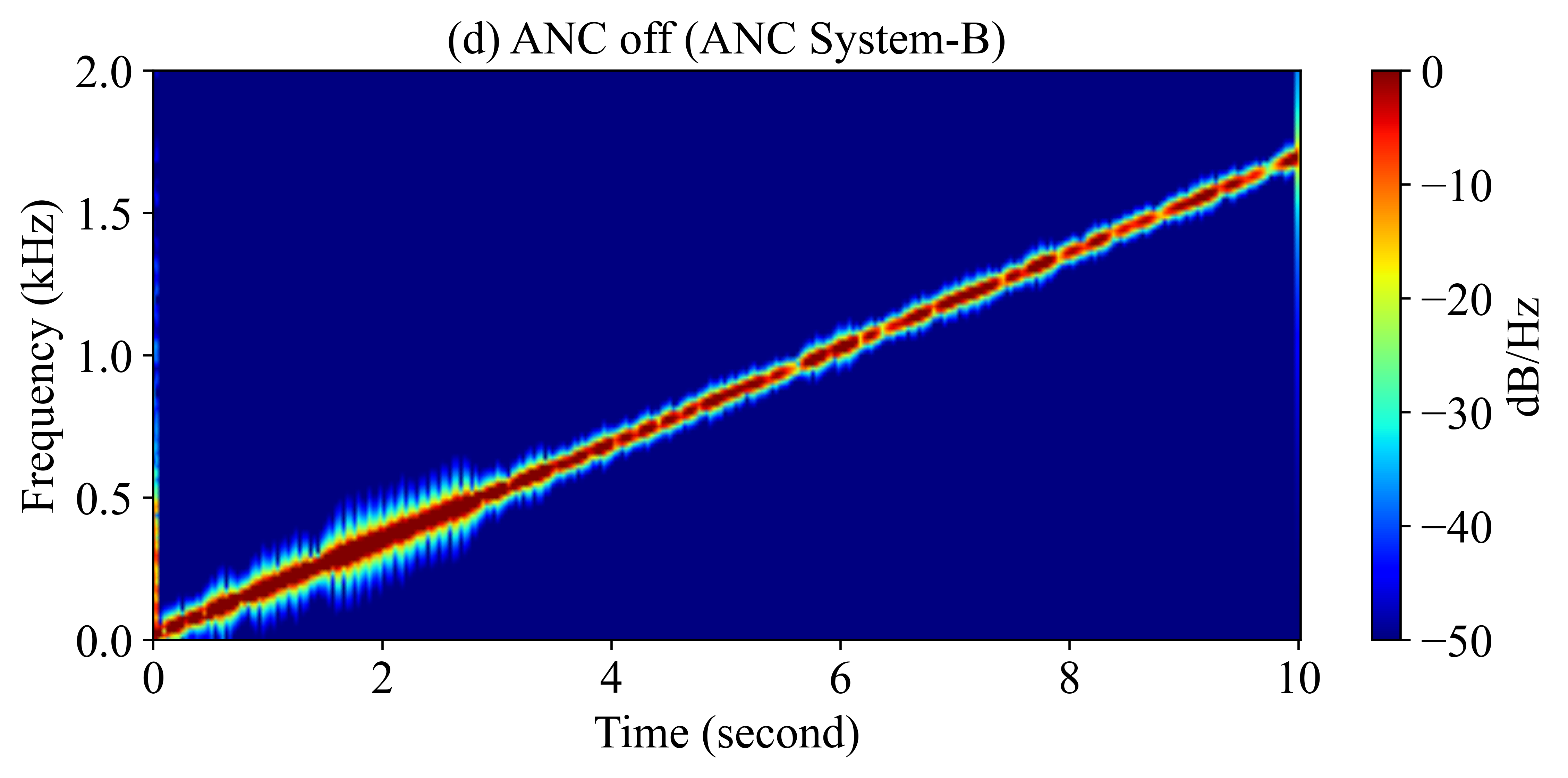}
\includegraphics[width=0.325\linewidth, height=3.5cm]{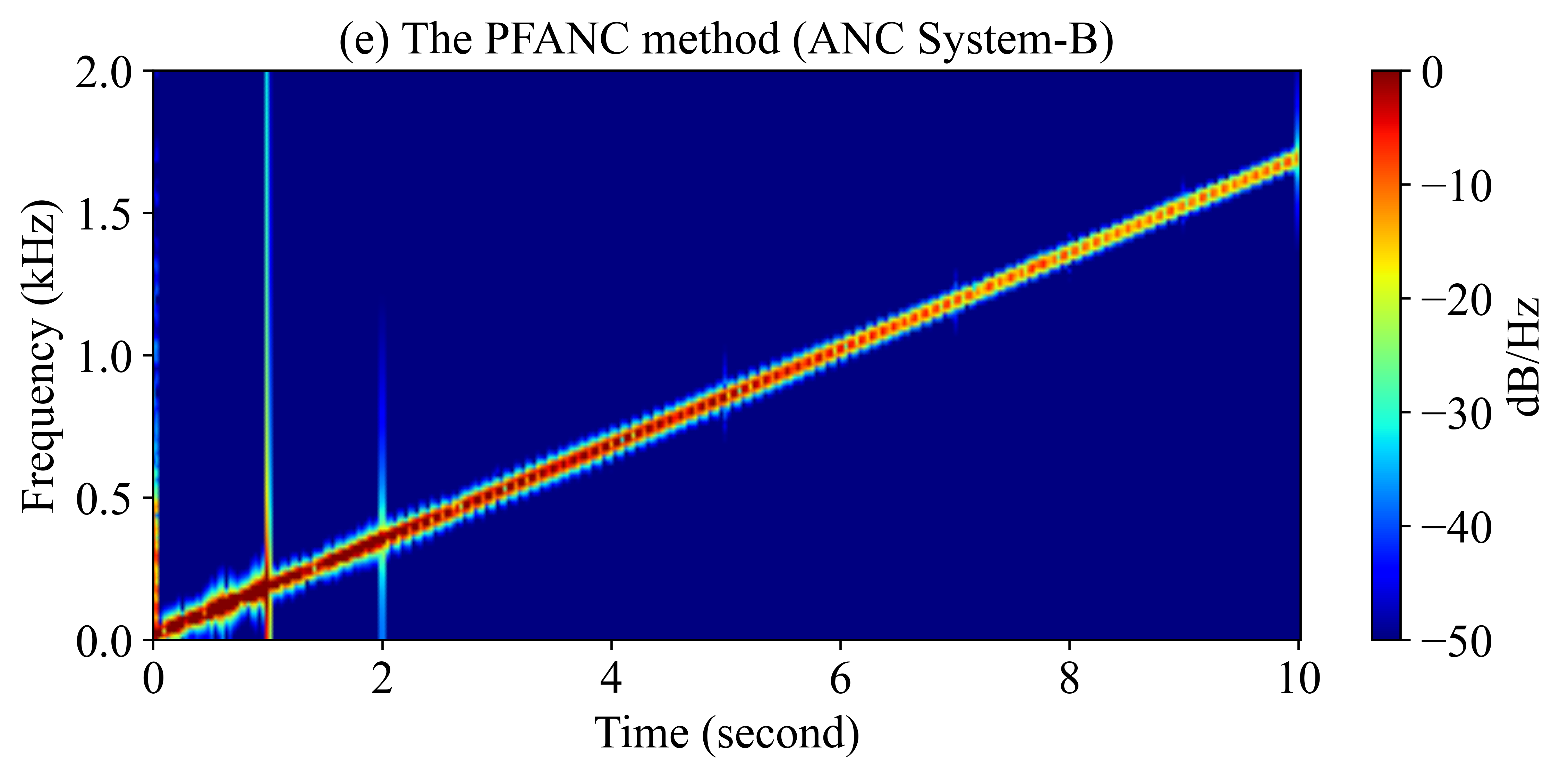}
\includegraphics[width=0.325\linewidth, height=3.5cm]{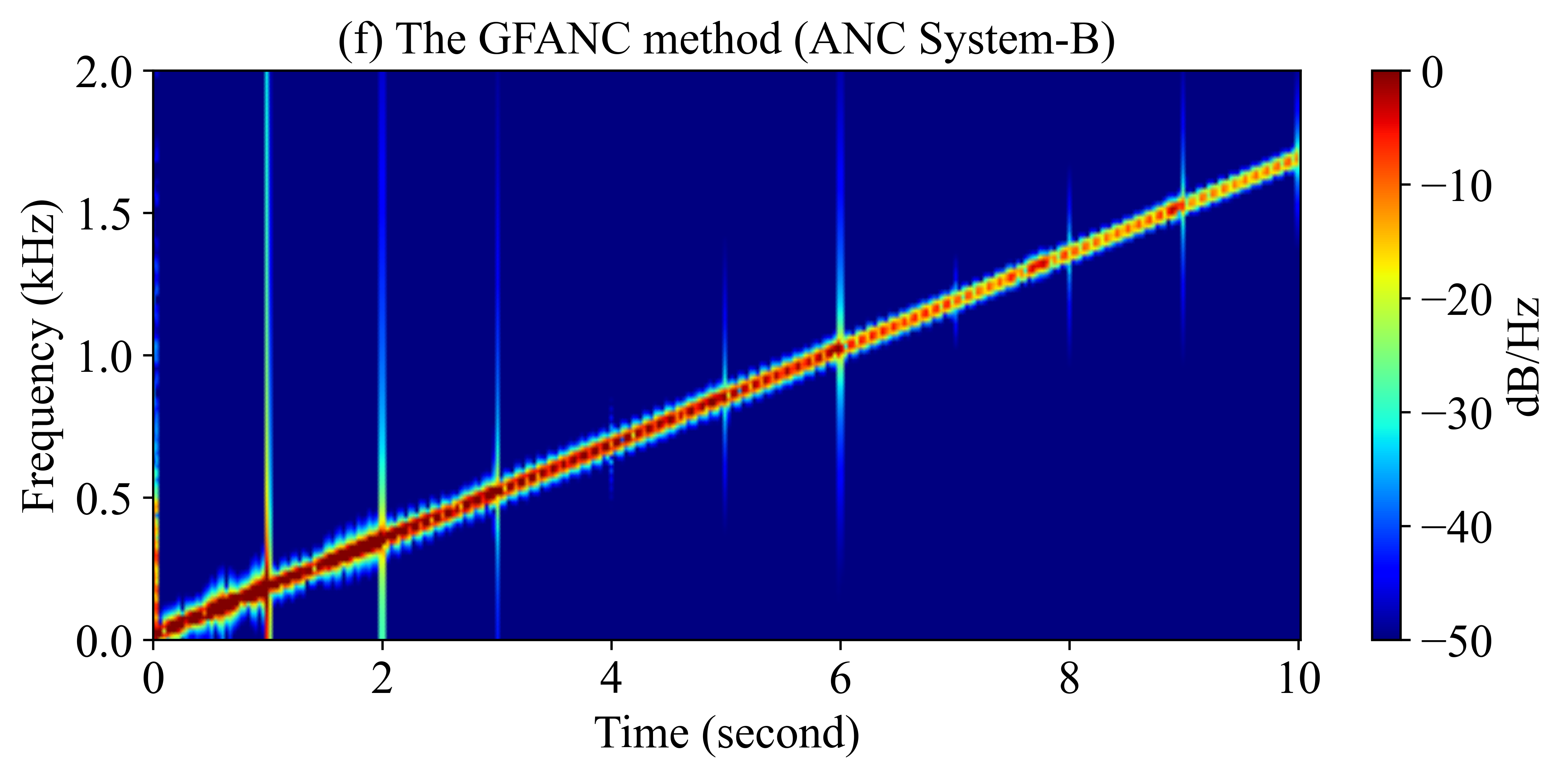}
\caption{Spectrograms of the error signal for the $20$–$1{,}700$~Hz linear chirp noise: (a)–(c) results from ANC System-A and (d)–(f) results from ANC System-B.}
\label{Fig Real Path Linear Chirp}
\end{figure}
%-------------------------------------------------------------------------

%------------------------------------------------------------------------
\begin{figure}[!t]
\centering
\includegraphics[width=0.325\linewidth, height=3.5cm]{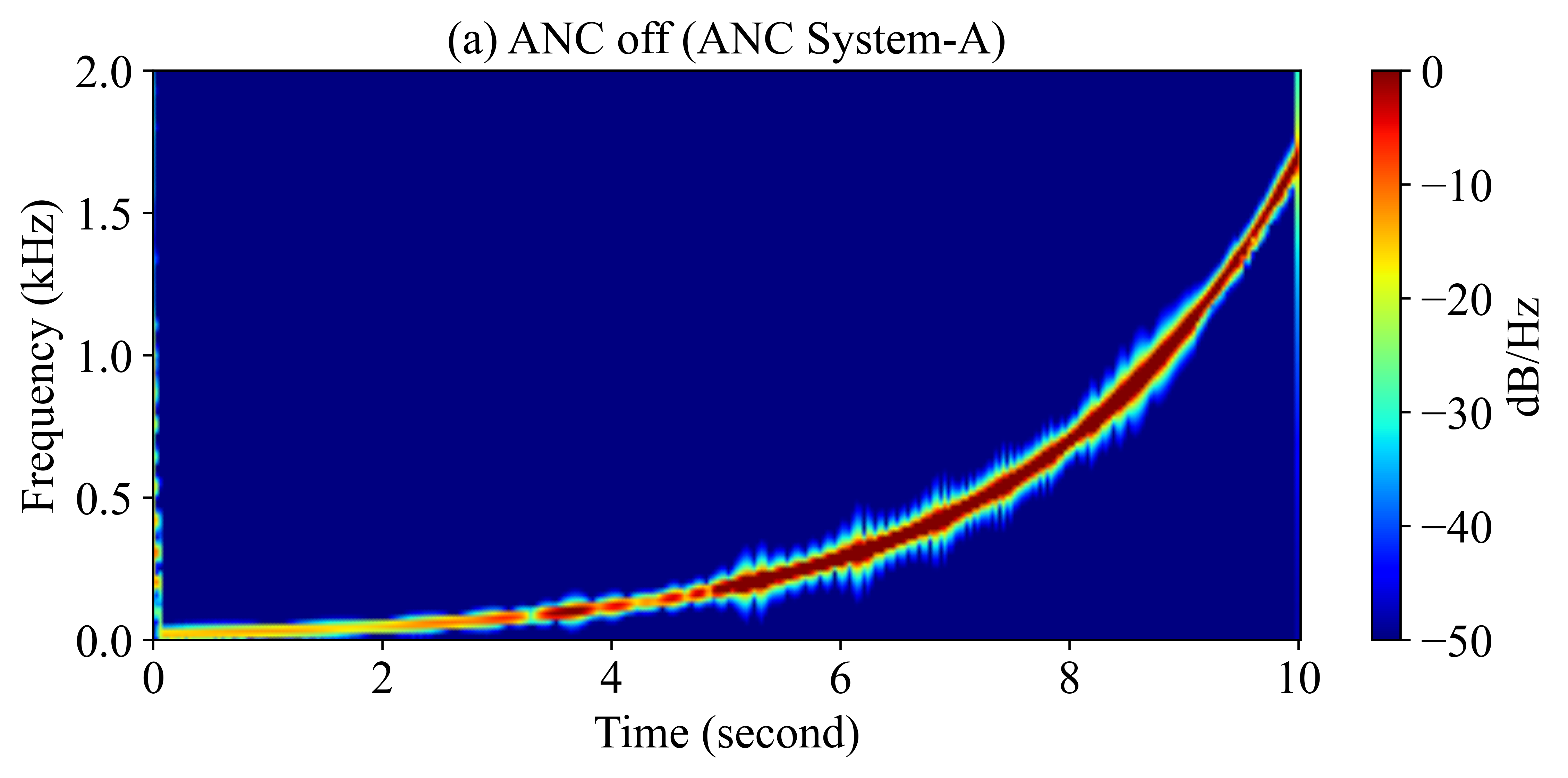}
\includegraphics[width=0.325\linewidth, height=3.5cm]{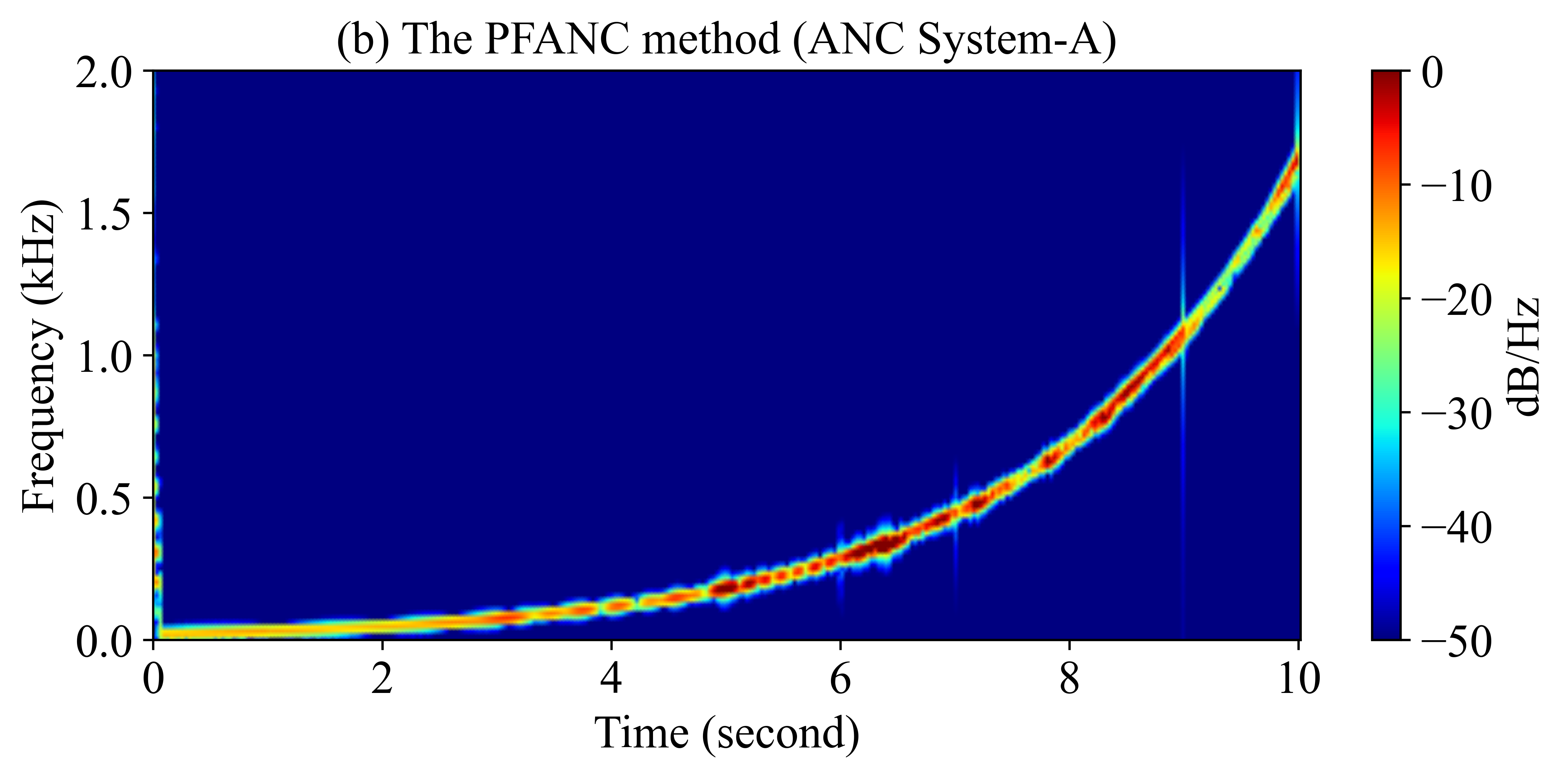}
\includegraphics[width=0.325\linewidth, height=3.5cm]{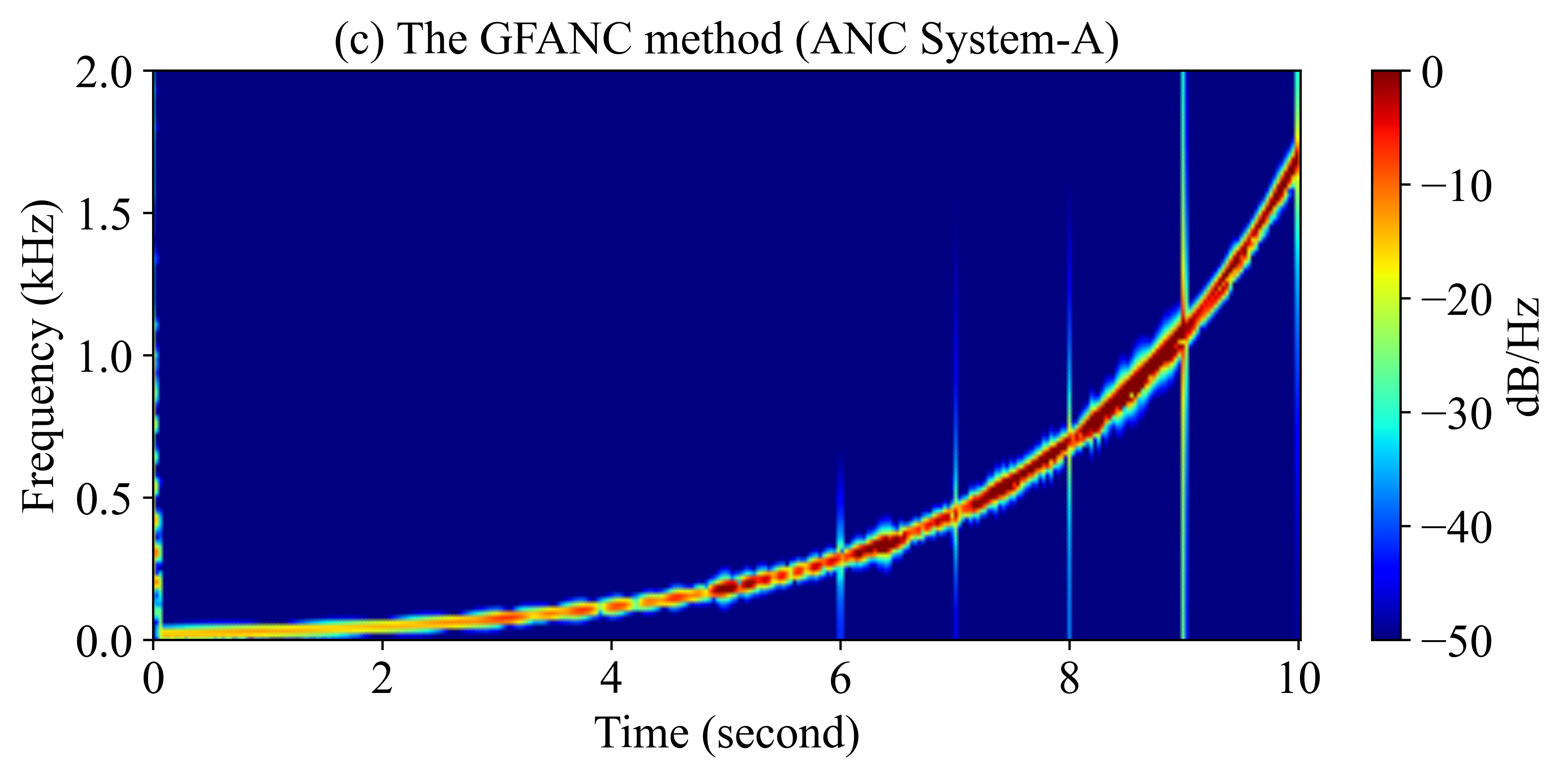}
\includegraphics[width=0.325\linewidth, height=3.5cm]{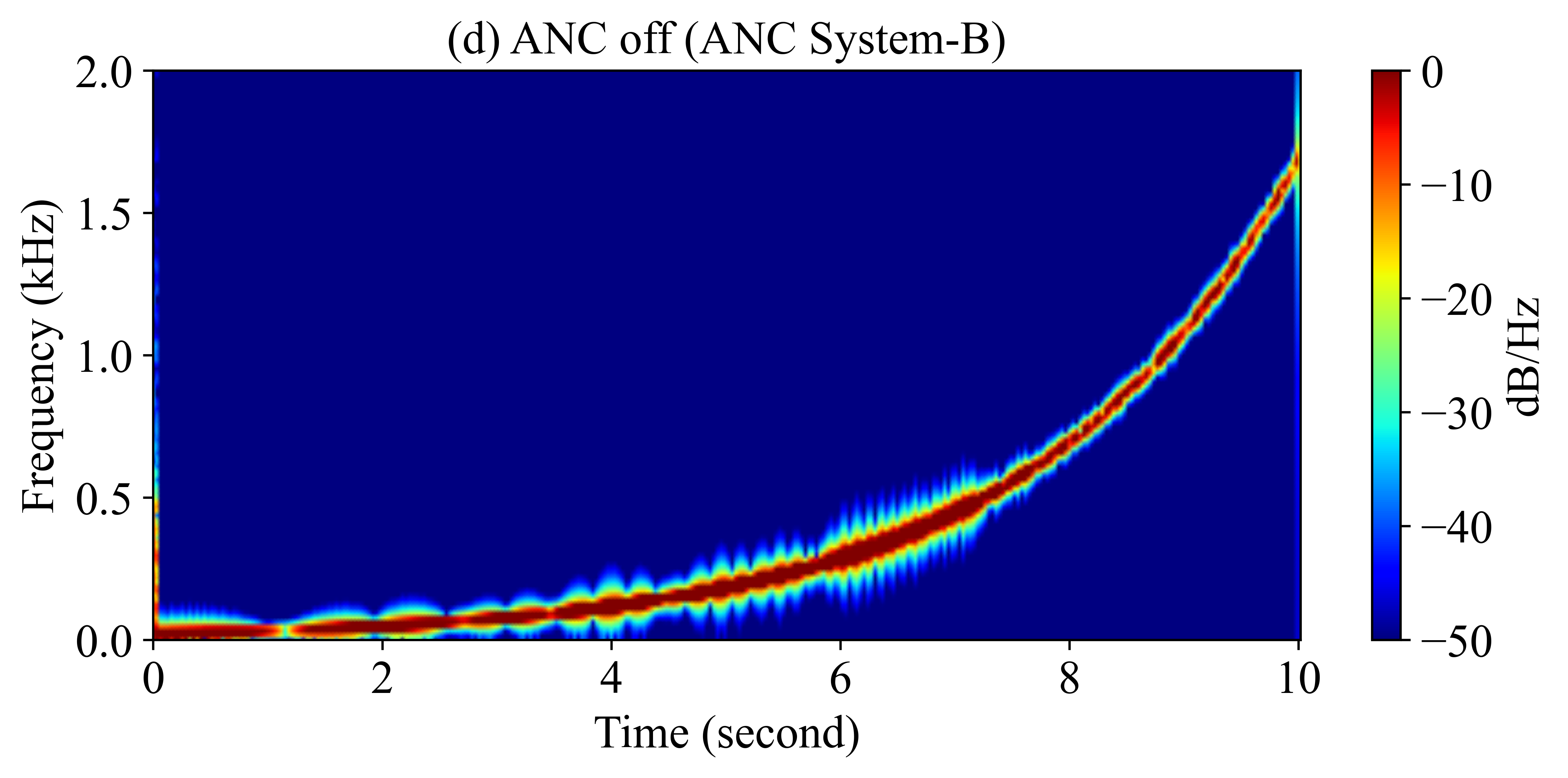}
\includegraphics[width=0.325\linewidth, height=3.5cm]{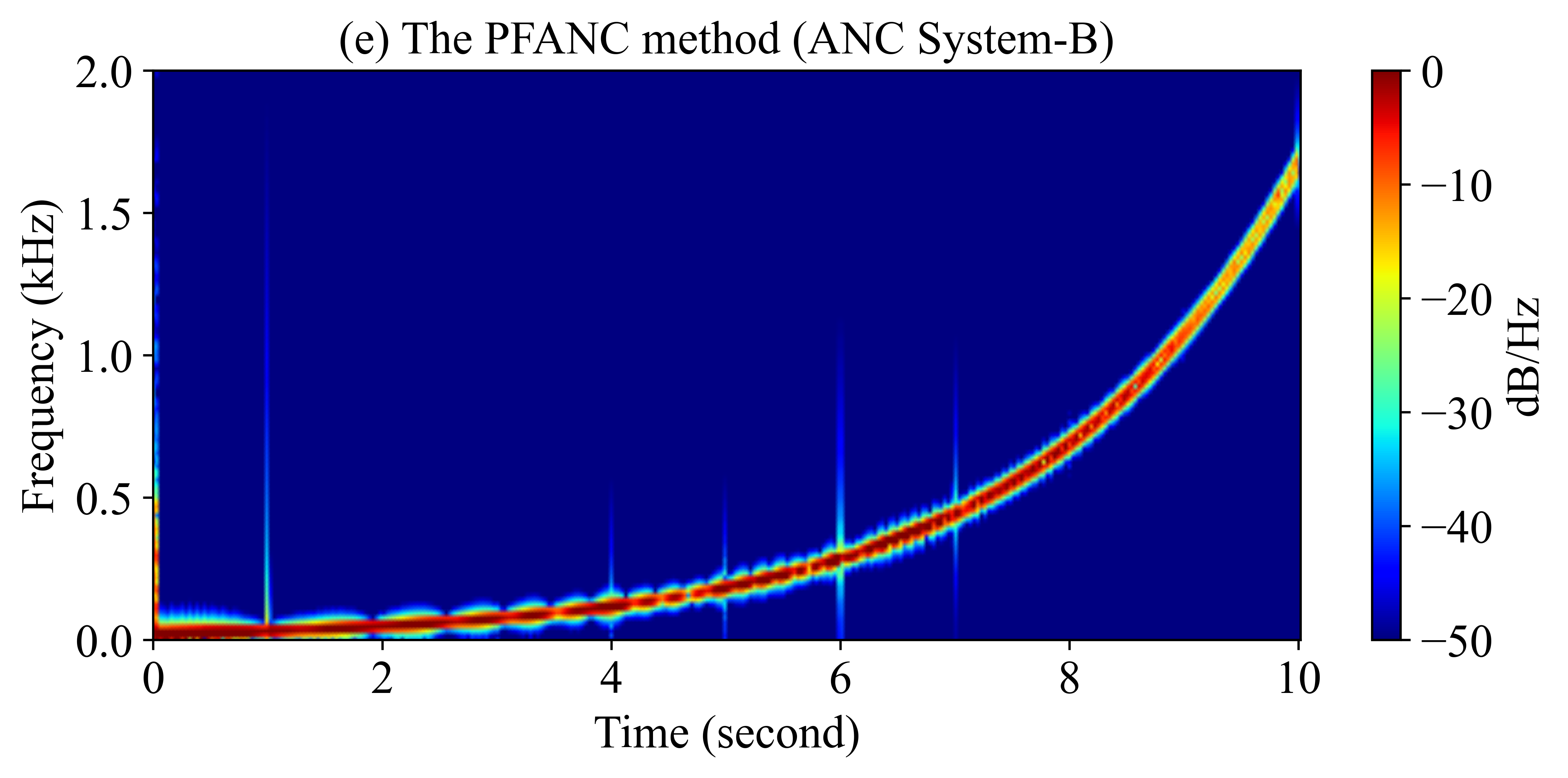}
\includegraphics[width=0.325\linewidth, height=3.5cm]{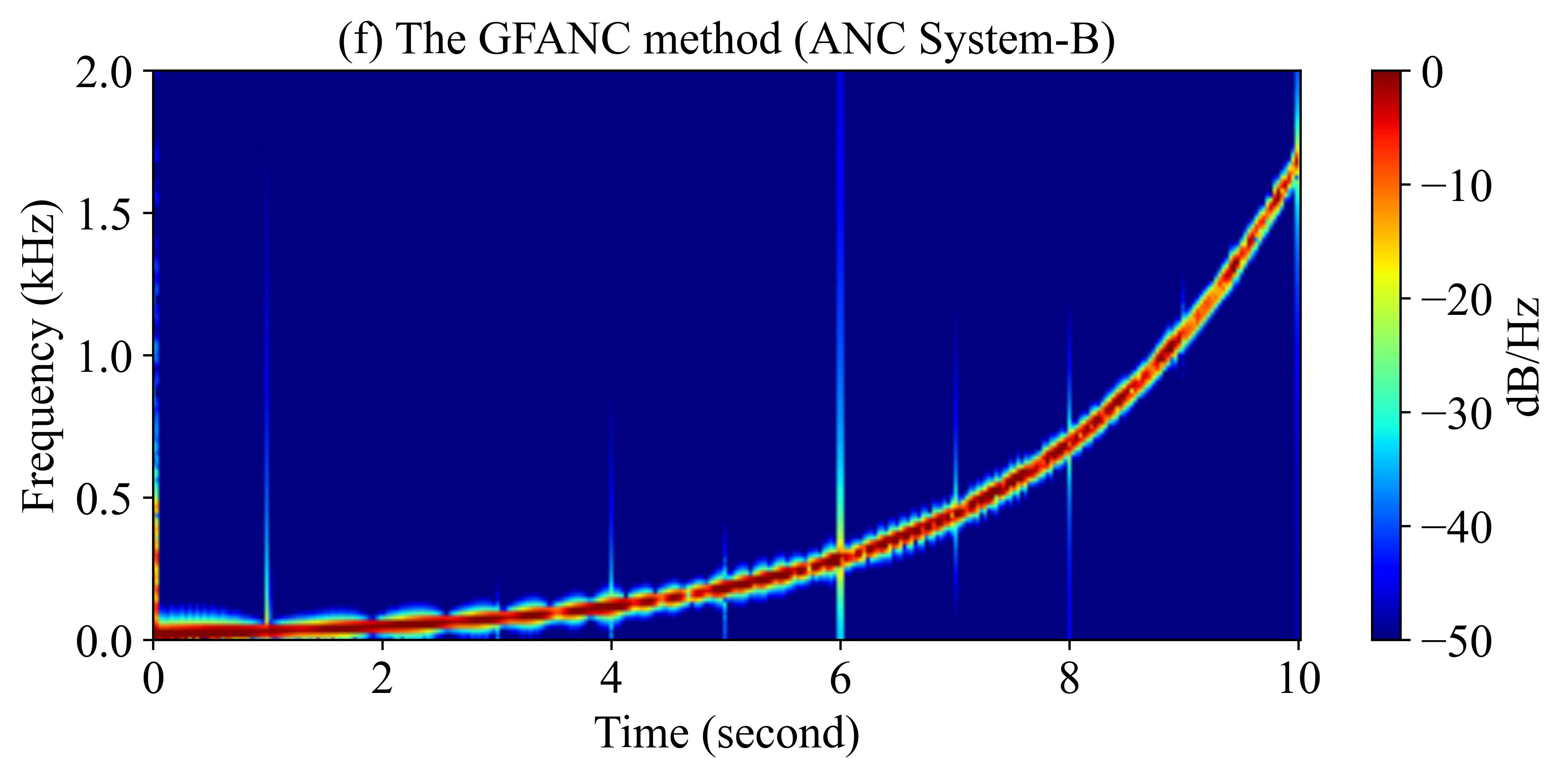}
\caption{Spectrograms of the error signal for the $20$–$1{,}700$~Hz logarithmic chirp noise: (a)–(c) results from ANC System-A and (d)–(f) results from ANC System-B.}
\label{Fig Real Path Log Chirp}
\end{figure}
%-------------------------------------------------------------------------

The overall NR values over the entire noise length for the PFANC and GFANC methods in different ANC systems are summarised in Table~\ref{Table NR different acoustic paths}. This measure is different from the averaged NR value calculated in each second. As shown in Table~\ref{Table NR different acoustic paths}, in ANC System-A, the PFANC method achieves a more than $6$ dB higher NR value than the GFANC method, whereas the improvement in ANC System-B is very small. This discrepancy primarily reflects the differences in the acoustic paths of the two systems. The detailed noise reduction performances for the linear and logarithmic chirp noises are illustrated in Figure~\ref{Fig Real Path Linear Chirp} and Figure~\ref{Fig Real Path Log Chirp}. For the linear chirp noise (Figure~\ref{Fig Real Path Linear Chirp}), the main noise frequency components differ between System-A and System-B due to their distinct acoustic path characteristics, yet the PFANC method effectively suppresses the dominant components in both cases. In contrast, the GFANC method exhibits weaker tracking ability, as evidenced by intermittent attenuation. For the logarithmic chirp noise (Figure~\ref{Fig Real Path Log Chirp}), the PFANC method consistently outperforms the GFANC method in both systems, particularly in the late portion of the noise, which aligns with the results on synthetic acoustic paths in Figure~\ref{Fig Simulation log chirp}. Furthermore, the overall NR values in System-B are lower than those in System-A because the noise energy in the first second is much higher in System-B, and neither the PFANC nor the GFANC method provides attenuation during this initial period. As discussed above, the results from System-A and System-B indicate the good transferability of the PFANC method, enabling its deployment across diverse acoustic environments.

%-----------------------------------------------------------------------
\section{Conclusion}\label{Conclusion}
Different from the existing GFANC method and its variations, this paper proposes the PFANC approach that enables a proactive control paradigm. In the PFANC method, a CRNN exploits temporal correlations across multiple noise frames to predict the optimal control filter for the next frame. The theoretical analysis shows that, for dynamic noises, predicting the next-frame control filter from multiple noise frames yields a lower prediction error than using only the current frame. Simulation results demonstrate that GFANC and its variants, including GFANC-Bayes and GFANC-Kalman, exhibit limited tracking capability for noises with rapidly varying spectral characteristics due to their lack of predictive modeling, particularly in the later portion of the logarithmic chirp noise. In contrast, the proposed PFANC method achieves superior noise reduction for both chirp signals and real-world dynamic noises and responds faster than the FxLMS algorithm. Moreover, PFANC exhibits good transferability across different acoustic paths. These results demonstrate the effectiveness of predictive ANC in tracking temporal variations in noise spectral characteristics. Future work will further investigate more complex dynamic scenarios involving simultaneous variations in both the frequency characteristics and spatial direction of the noise source.

%=====================================================
\appendix
\section{Measured Acoustic Paths}
\label{Appendix Acoustic Paths}
\begin{figure}[tp]
\centering
\includegraphics[width=0.45\linewidth, height=4cm]{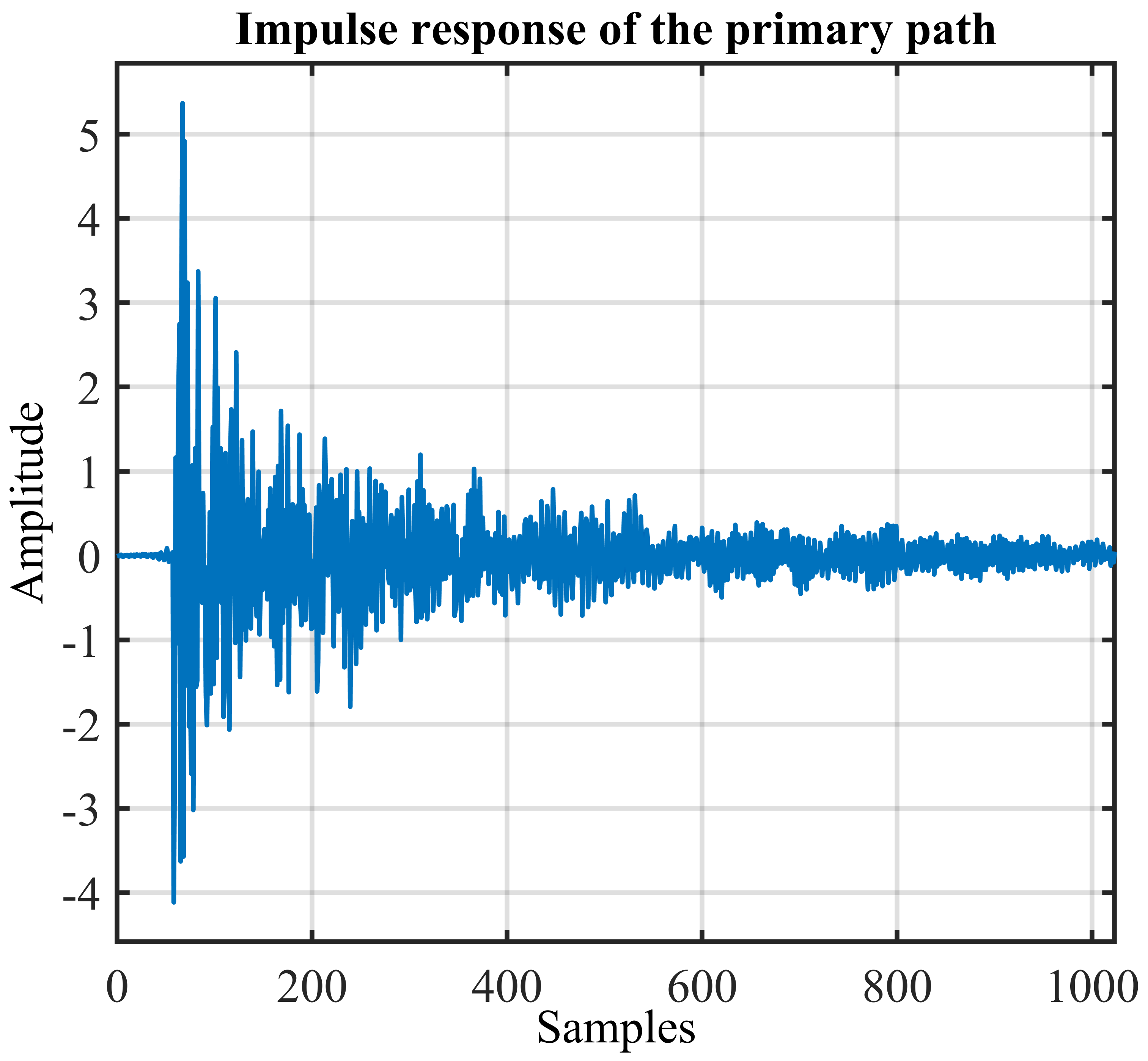}
\includegraphics[width=0.45\linewidth, height=4cm]{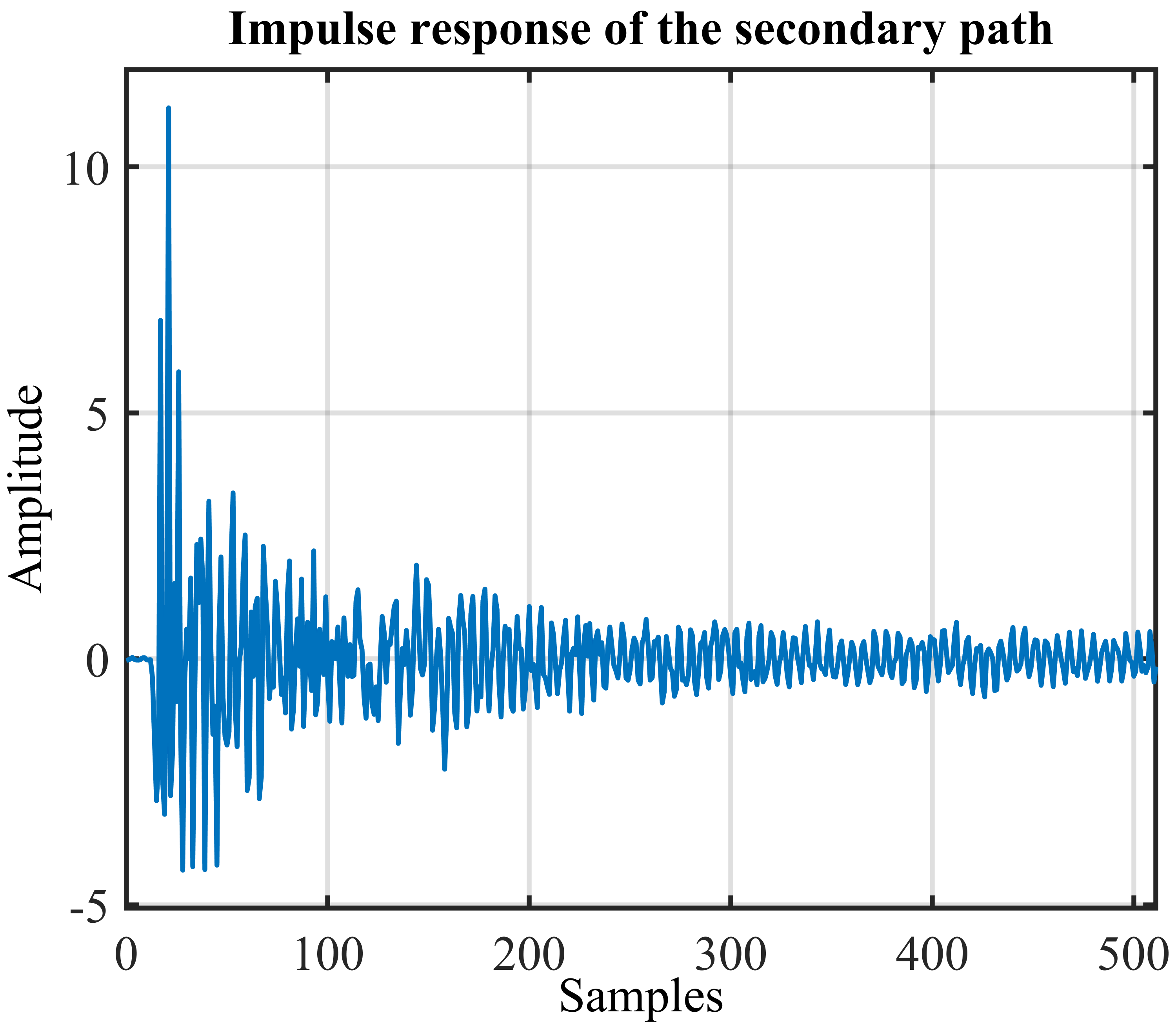}
\includegraphics[width=0.45\linewidth, height=4cm]{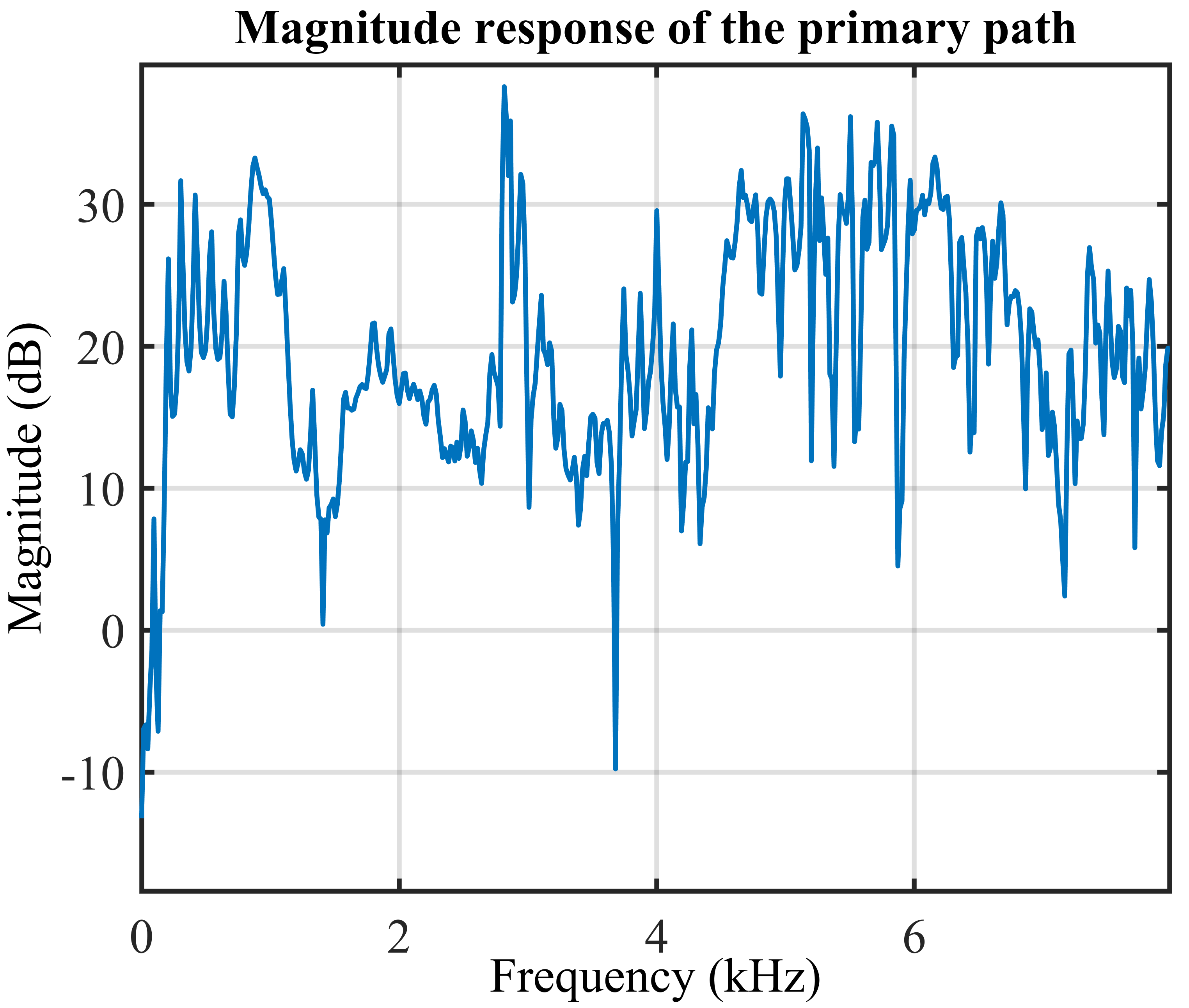}
\includegraphics[width=0.45\linewidth, height=4cm]{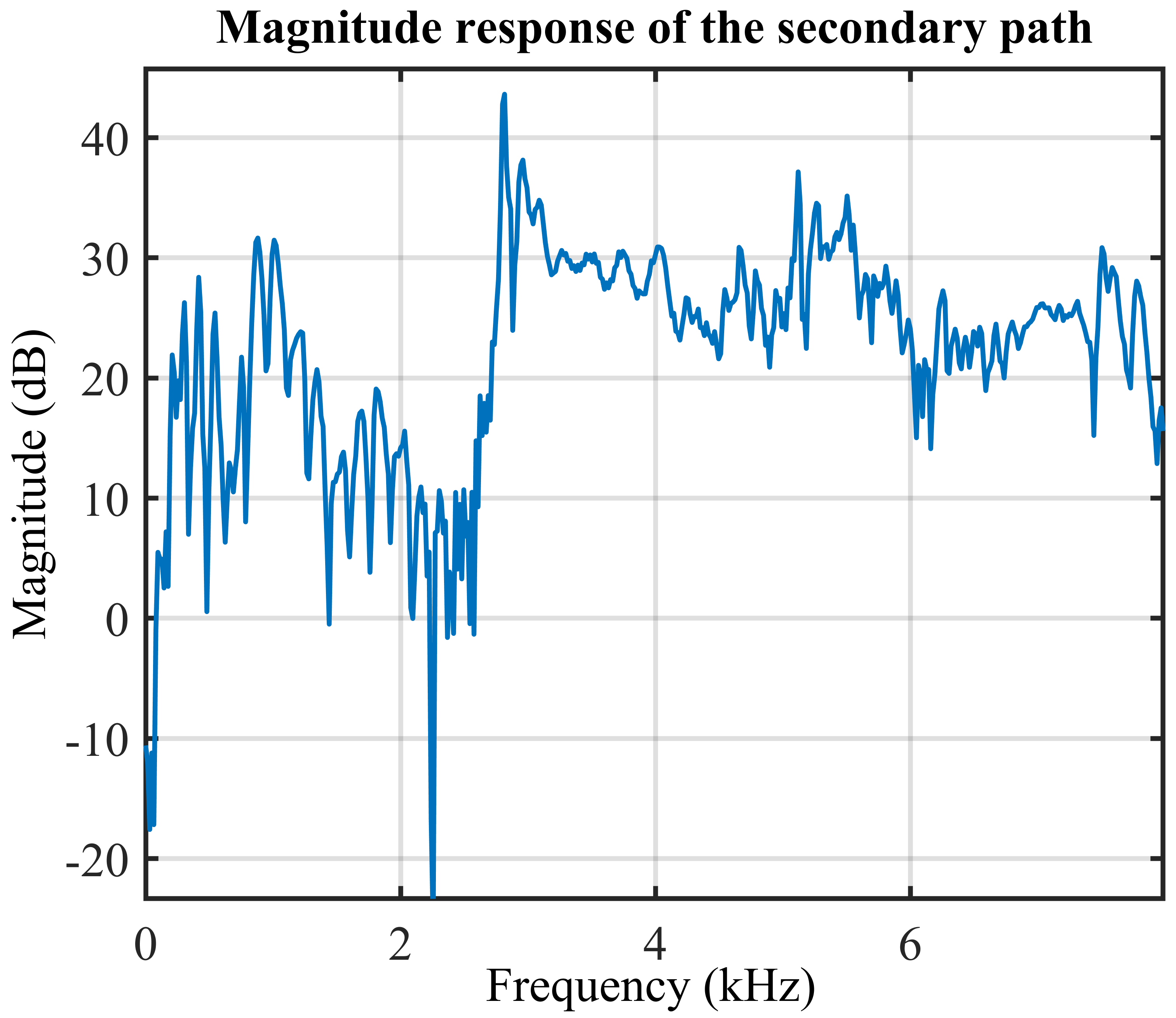}
\includegraphics[width=0.45\linewidth, height=4cm]{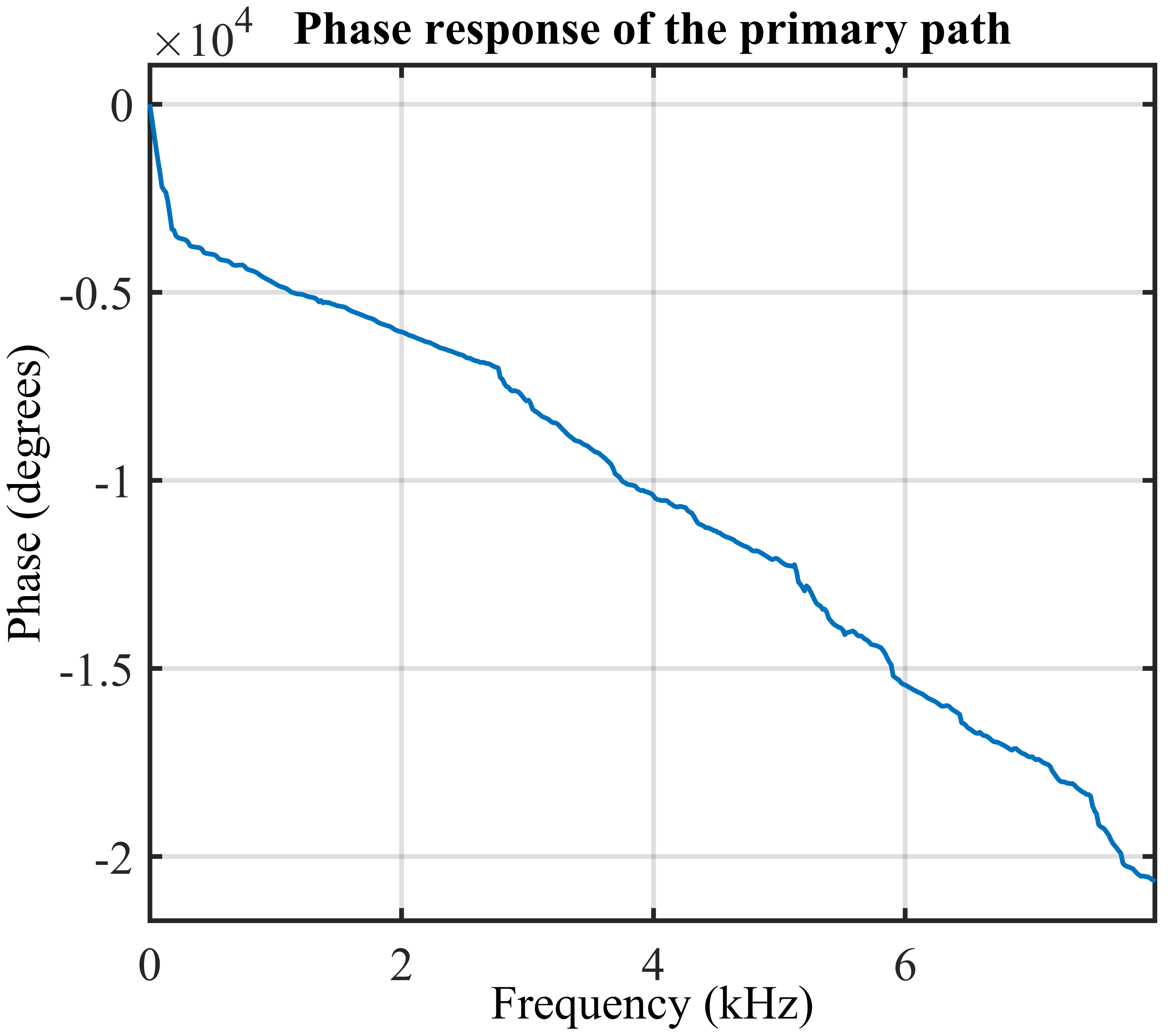}
\includegraphics[width=0.45\linewidth, height=4cm]{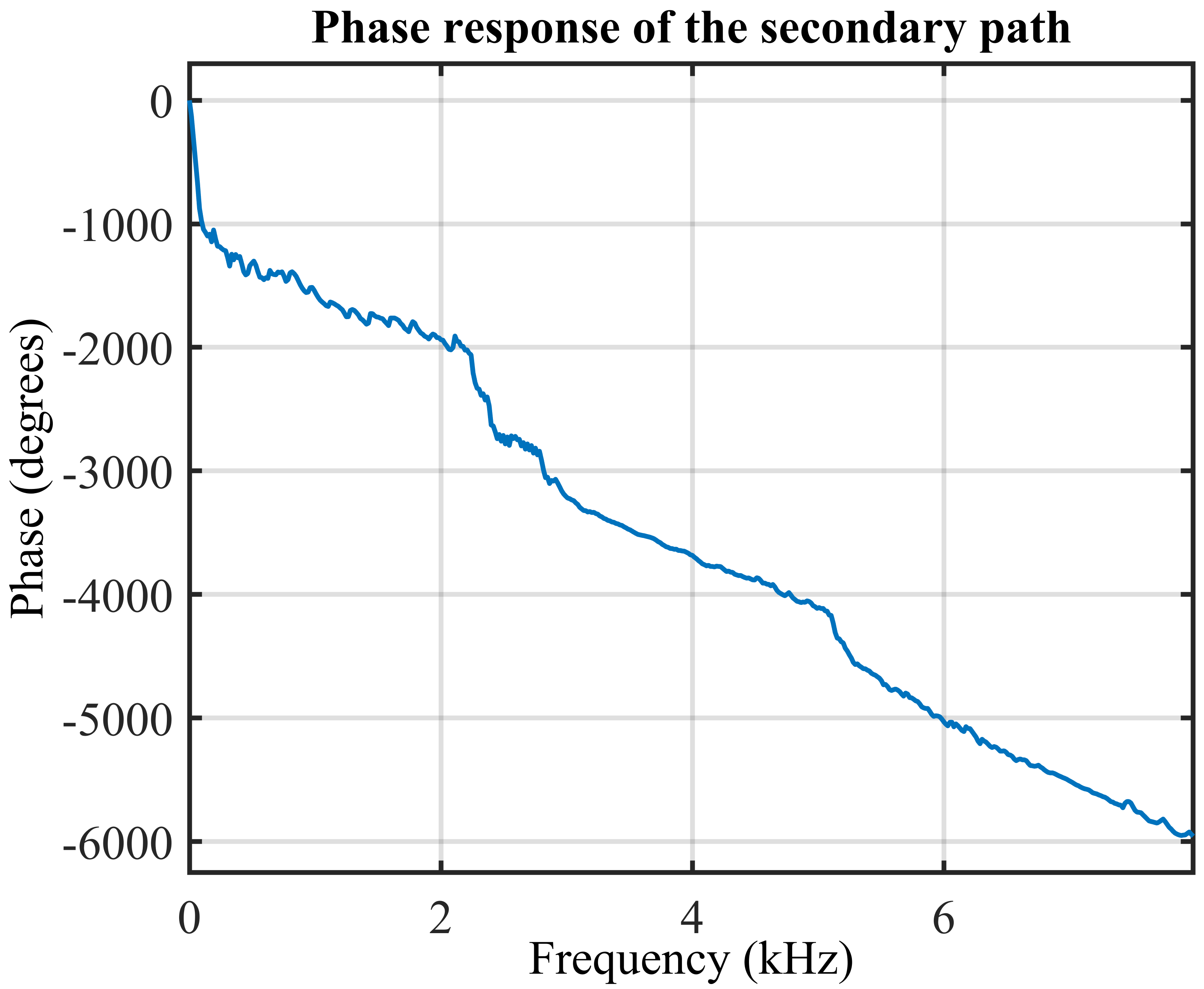}
\caption{The impulse responses, magnitude responses, and phase responses of the primary and secondary paths measured from the vent of a noise chamber (ANC System-A).}
\label{Fig duct path}
\end{figure}

\begin{figure}[tp]
\centering
\includegraphics[width=0.45\linewidth, height=4cm]{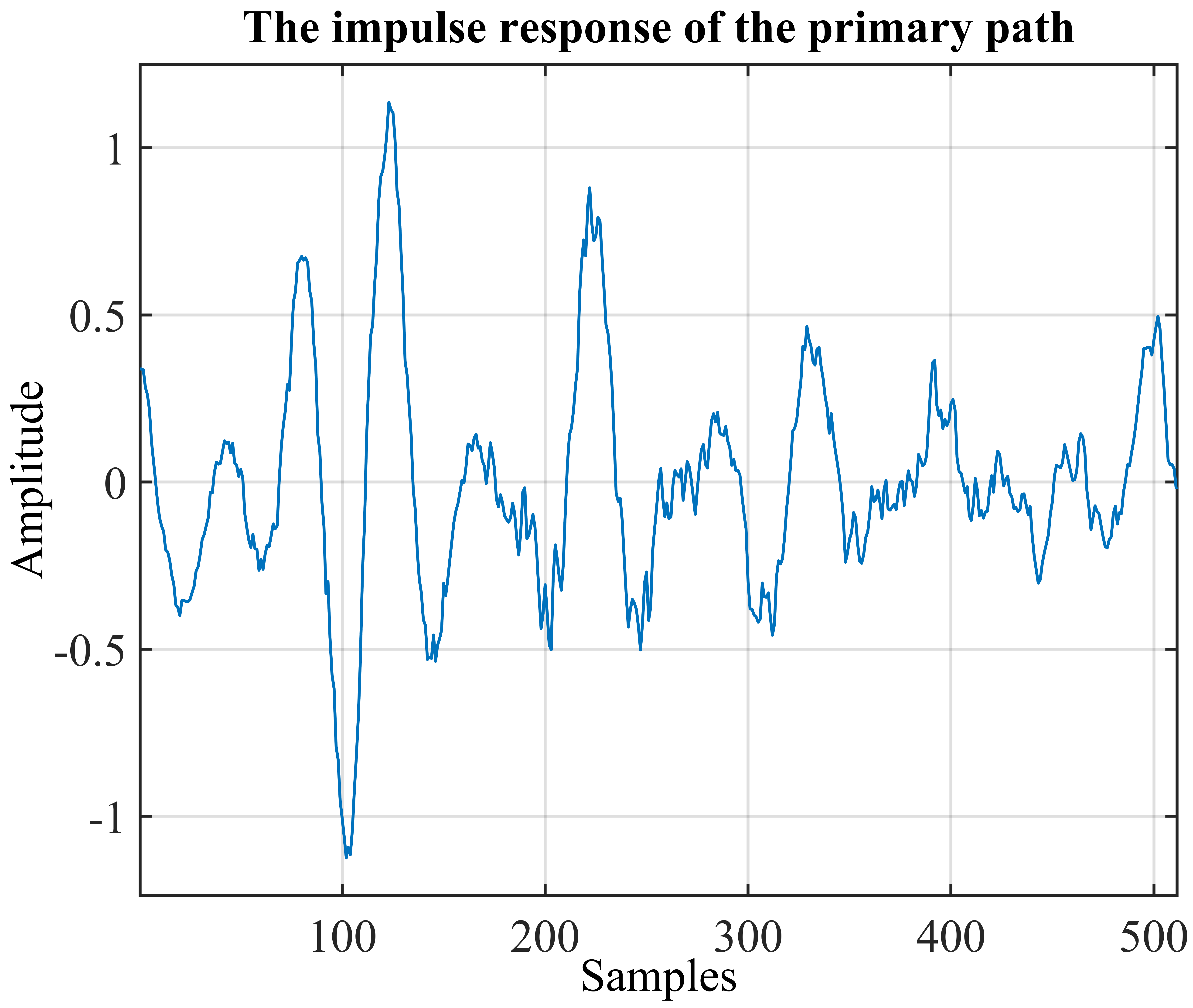}
\includegraphics[width=0.45\linewidth, height=4cm]{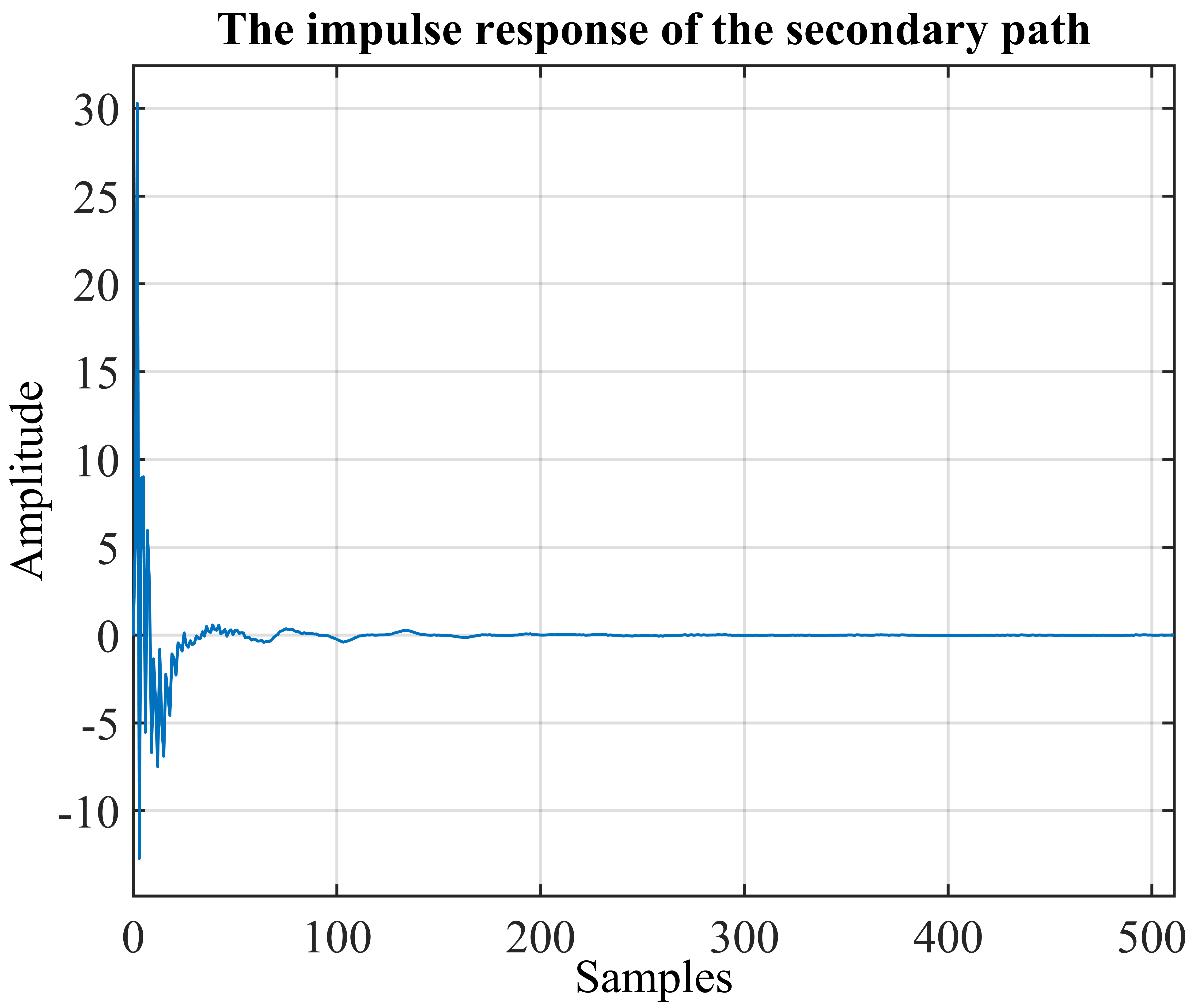}
\includegraphics[width=0.45\linewidth, height=4cm]{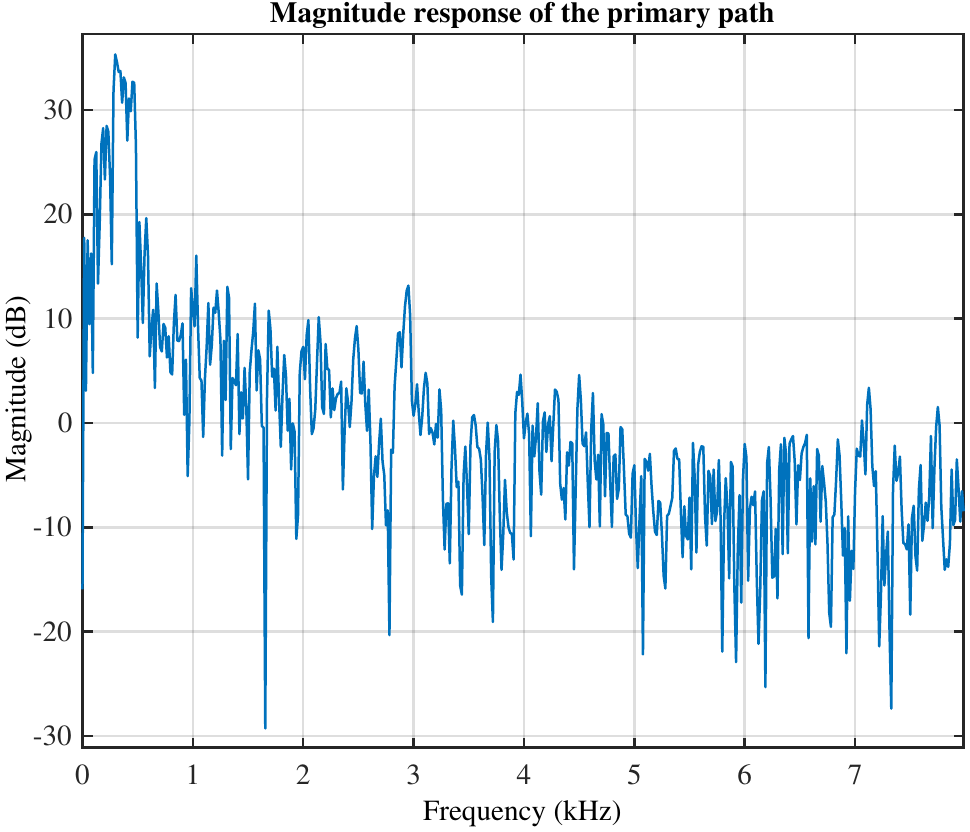}
\includegraphics[width=0.45\linewidth, height=4cm]{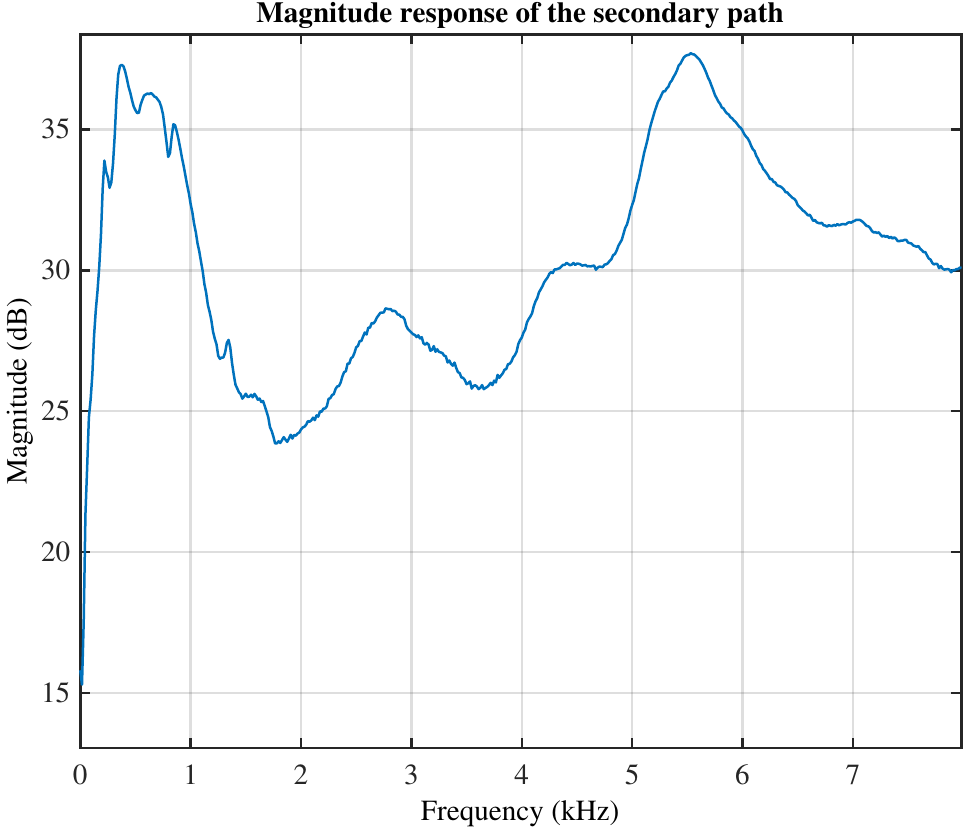}
\includegraphics[width=0.45\linewidth, height=4cm]{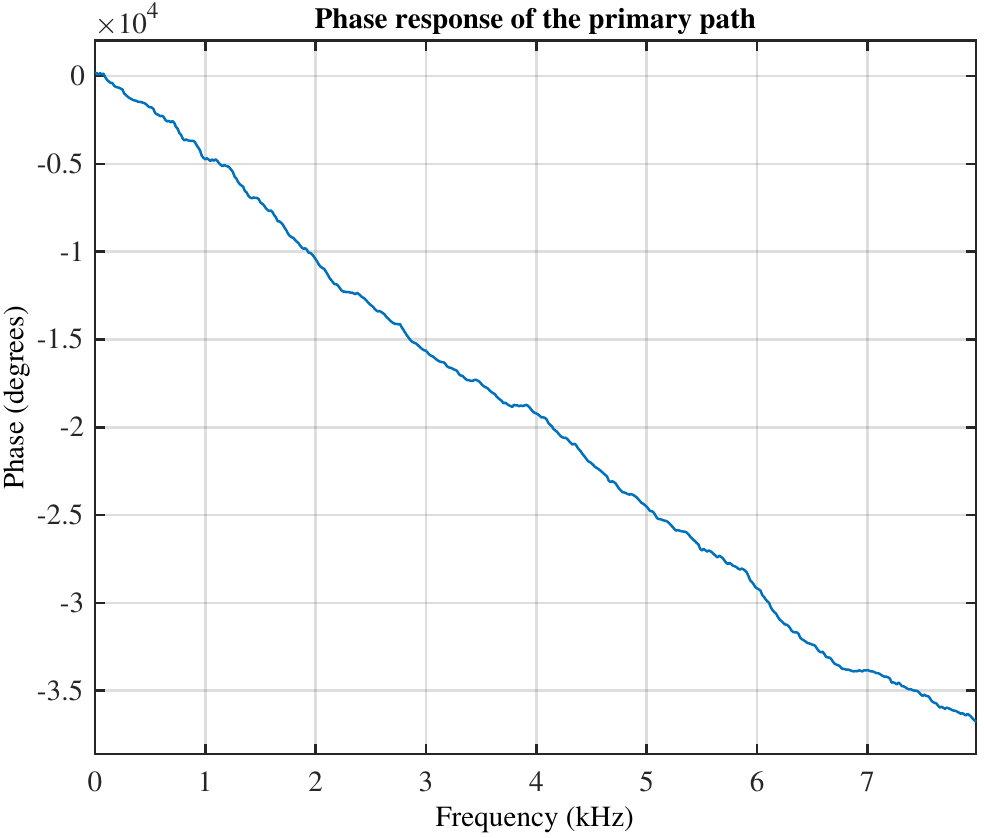}
\includegraphics[width=0.45\linewidth, height=4cm]{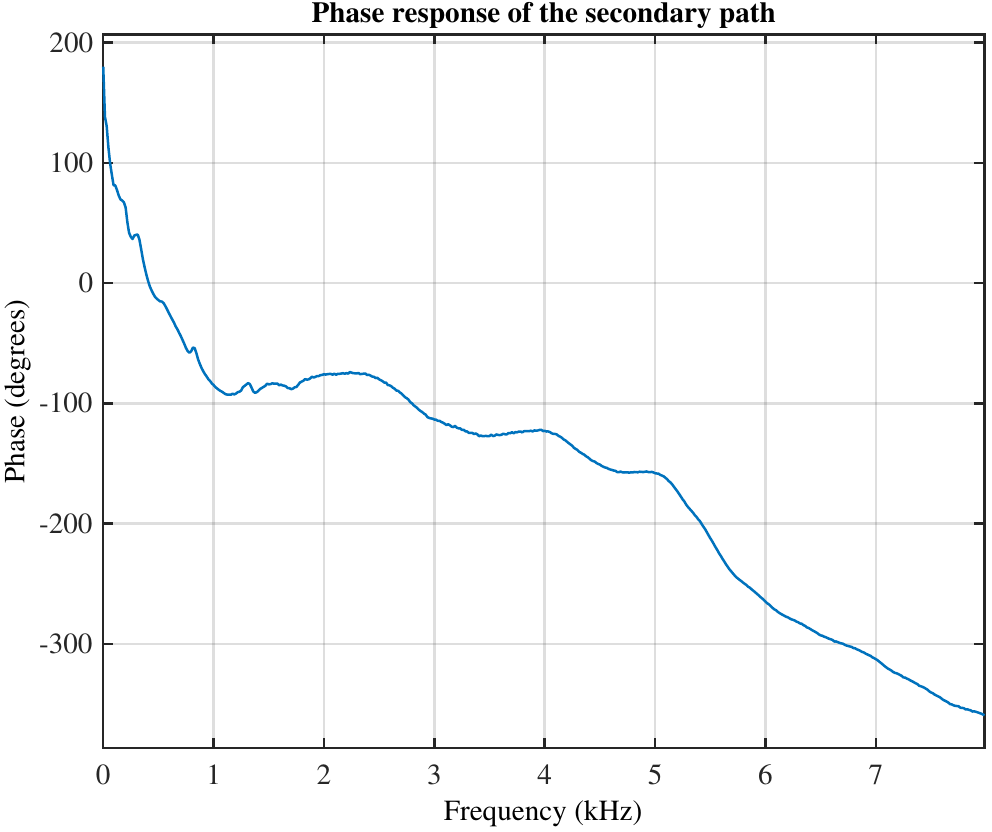}
\caption{The impulse responses, magnitude responses, and phase responses of the primary and secondary paths measured from an ANC window (ANC System-B).}
\label{Fig window path}
\end{figure}

The appendix provides the magnitude and phase responses of the acoustic paths measured in the vent of a noise chamber (ANC System-A) and an ANC window (ANC System-B), as illustrated in Figure~\ref{Fig duct path} and Figure~\ref{Fig window path}.

%------------------------------------------------------------------------------
%------------------------------------------------------------------------------

%------------------------------------------------------------------------------
%------------------------------------------------------------------------------

\clearpage
\bibliographystyle{elsarticle-num} 
\bibliography{A}
\end{document}